# Dynamic Slowdown and Spatial Correlations in Viscous Silica Melt: Perspectives from Dynamic Disorder

Shubham Kumar[1], Zhiye Tang[1,2] and Shinji Saito[1,2,*]

[1]Institute for Molecular Science, Myodaiji, Okazaki, Aichi, 444-8585, Japan

[2]The Graduate University for Advanced Studies (SOKENDAI), Myodaiji, Okazaki, Aichi, 444-8585, Japan

*Author for correspondence: shinji@ims.ac.jp

**Abstract**

The dynamic slowdown in glass-forming liquids remains a central topic in condensed matter science. Here, we report a theoretical investigation of the microscopic origin of the slowdown in amorphous silica, a prototypical strong glass former with a tetrahedral network structure. Using molecular dynamics simulations, we analyze atomic jump dynamics, the elementary structural change processes underlying relaxation. We find that the jump statistics deviate from Poisson behavior with decreasing temperature, reflecting the emergence of dynamic disorder in which slowly evolving variables modulate the jump motion. The slowdown is species-dependent: for silicon, the primary constraint arises from the fourth-nearest oxygen neighbor, while at lower temperatures, the fourth-nearest silicon also becomes relevant; for oxygen, the dominant influence comes from the second-nearest silicon neighbors. As the system is cooled, the jump dynamics become increasingly slow and intermittent, proceeding in a higher-dimensional space of multiple slow variables that reflect cooperative rearrangements of the network. Species-resolved point-to-set correlations further reveal that the spatial extent of cooperative relaxation grows differently for silicon and oxygen, directly linking their relaxation asymmetry to the extent of collective motion. Together, these results provide a microscopic framework linking dynamic disorder, species-dependent constraints, and cooperative correlations, offering deeper insight into the slowdown of strong glass-forming networks.

# I. INTRODUCTION

Understanding the microscopic mechanisms that govern the dynamics of glass-forming liquids remains a central challenge in condensed matter science.[1–4] As the temperature decreases, glass-forming liquids exhibit a dramatic increase in structural relaxation times and viscosities, often spanning many orders of magnitude, while conventional structural measures, such as the radial distribution function and static structure factor, show only minimal changes.[5–8] This apparent decoupling between structure and dynamics underscores the complexity of the glass transition and highlights the need for a microscopic understanding of the processes that govern structural relaxation.[9,10] To address this, numerous systematic investigations have focused on identifying structural features correlated with dynamic behavior, including local geometric order,[11–13] topological defects,[14–16] and microscopic frameworks such as the cage-jump model.[17–26] More recently, machine learning techniques have been applied to reveal subtle structural signatures associated with dynamic heterogeneity.[27–35] Despite these advances, a comprehensive understanding of the origin of the dynamic slowdown remains elusive.

Among glass-forming systems, silica ($SiO_2$) holds a unique position as a prototypical strong glass-former characterized by a disordered tetrahedral network formed through covalent bonds between silicon and oxygen atoms.[36–38] Unlike fragile glass-formers, which exhibit super-Arrhenius increases in relaxation times upon cooling, silica displays nearly Arrhenius behavior over a broad temperature range.[5,38–40] At the same time, oxygen and silicon show distinct relaxation dynamics, with oxygen remaining more mobile, a decoupling attributed to generalized rotational processes within the potential energy landscape.[37,41,42] This comparatively simple yet nontrivial dynamical behavior has made silica a model system for investigating the microscopic mechanisms governing structural relaxation in network-forming liquids. A closely related tetrahedral liquid, water, is known to exhibit a more complex dynamical behavior, displaying a fragile-to-strong transition upon deep supercooling.[43–45] This nuanced behavior highlights how subtle variations in network topology, bonding strength, and local structure can profoundly influence the dynamic slowdown in glass-forming liquids.[6,9,46–50]

Silica is of both fundamental and technological importance, with applications in geological processes, glass and ceramic manufacturing, optical fibers, and photonics.[51–53] Over the past few decades, both experimental[54,55] and computational studies[36,38,40,56–61] have sought to understand how structural features such as defects, ring distributions, and intermediate-range order contribute to the relaxation dynamics of silica. Despite these advances in characterizing the static structure and average dynamics, the microscopic origins of dynamic heterogeneity and intermittent relaxation in silica remain poorly understood. In many glass-forming liquids, structural relaxation proceeds through localized "jump" motions, where particles escape transient cages formed by their neighbors. Such jump events represent elementary structural rearrangements that facilitate long-time relaxation.[17–26] While jump dynamics have been studied extensively in fragile glasses and some network-forming liquids, their detailed characterization and role in the relaxation processes of silica remain limited.

It is now well established that jump events in glass-forming liquids deviate from simple Poisson statistics, especially at low temperatures.[21,22,26] These deviations arise from slowly evolving structural environments that influence the frequency and spatial distribution of jumps, a phenomenon referred to as dynamic disorder. Dynamic disorder emerges from the interplay between localized jump motions and slowly evolving collective structural fluctuations, leading to increasingly intermittent and heterogeneous jump dynamics as the temperature decreases.[62–68] In a recent study,[26] the dynamic disorder in the jump dynamics of supercooled water was investigated, demonstrating that specific slow variables, such as the displacement of the fourth nearest neighbor, control jump behavior by trapping jumping molecules within extended, low-density regions. As the temperature further decreases, the number of these slow variables grows, driving the jump dynamics into a higher-dimensional space marked by enhanced slowness and intermittency.

In this work, we aim to elucidate the microscopic origin of dynamic slowdown in amorphous silica by extending the framework of dynamic disorder in jump motion previously established for supercooled water. Our study reveals that the atomic displacements governing jump dynamics in silica differ markedly from those in water, resulting in distinct relaxation mechanisms. Specifically, silicon atom jumps are predominantly influenced by the displacement of fourth-nearest oxygen atoms, while oxygen atom jumps depend on the displacement of second-nearest silicon atoms within the network. These differences give rise to unique cooperative behaviors and increasingly intermittent jump dynamics as temperature decreases. Through point-to-set (PTS) correlation analysis, we demonstrate that this leads to differing spatial extents of cooperativity and dynamic heterogeneity between silicon and oxygen atoms. To the best of our knowledge, point-to-set correlation analysis has not been previously applied to viscous silica, and here we use it to quantify the spatial extent of cooperative rearrangements and connect them with jump dynamics. Together, these findings suggest that relaxation in silica arises from a microscopic mechanism fundamentally different from that reported for water,[26] reflecting the unique influence of its rigid tetrahedral network on cooperative dynamics in the glassy state.

The organization of the rest of this paper is as follows. In Sec. II, we discuss the theoretical and computational details employed in this study. In Sec. III, we present a detailed discussion of the results obtained from our simulations. In this section, we first discuss the relaxation dynamics, followed by changes in structural properties during jump motion and the transient nature of jump dynamics. Subsequently, we discuss the microscopic origin of dynamic disorder in jump motions. Finally, we discuss the static point-to-set correlation length in viscous silica. In Sec. IV, we summarize the present study and highlight potential directions for future research.

## II. THEORETICAL AND COMPUTATIONAL DETAILS

### A. Models and simulation details

In this study, we have carried out extensive molecular dynamics simulations of viscous silica melt using the LAMMPS simulation package,[69] employing two different models: (i) the Beest–Kramer–Santen (BKS)[70] model and (ii) the Coslovich–Pastore (CP)[71] model. The details of the models employed are given below.

### (i) The BKS model

The BKS model is a well-established empirical potential developed to simulate the structural and dynamical properties of silica.[36–38,57,70] In the BKS framework, interatomic interactions between particles $\alpha$ and $\beta$, where $\alpha, \beta \in \{\text{Si}, \text{O}\}$ are described by the following potential function:

$$\upsilon_{\alpha\beta}^{BKS}(r) = \frac{q_\alpha q_\beta e^2}{r} + A_{\alpha\beta} \exp(-B_{\alpha\beta} r) - \frac{C_{\alpha\beta}}{r^6}. \tag{1}$$

Here, $r$ denotes the interparticle distance; $q_\alpha$ and $q_\beta$ are the partial charges on species $\alpha$ and $\beta$, respectively, and $A_{\alpha\beta}$, $B_{\alpha\beta}$, and $C_{\alpha\beta}$ are empirical parameters characterizing the short-range repulsion and dispersion interactions. The partial charges assigned to the atomic species are $q_{\text{Si}}$ = +2.4 for silicon and $q_{\text{O}}$ = −1.2 for oxygen. The short-range repulsive and dispersive parameters of the BKS potential are provided in Table S1. The long-range electrostatic interactions are computed using the Ewald summation method, whereas the non-Coulombic (Buckingham and van der Waals) part of the potential is truncated and shifted at a cutoff radius of 5.5Å.[57] To avoid the finite size effect,[36] we have simulated a system of 8016 particles (2672

Si and 5344 O atoms) at a mass density of 2.37 g cm$^{-3}$. The system is simulated over a temperature range from 6100 K to 2500 K, using a time step of $\delta t = 1.0$ fs.

### (ii) The CP model

The CP model is a simplified representation of silica that captures its essential glass-forming properties without explicitly accounting for long-range electrostatic interactions. It comprises two types of particles, analogous to Si and O, in a fixed stoichiometric ratio of $N_O/N_{Si} = 2$, corresponding to the composition of SiO$_2$. The interactions between the particles are governed by spherically symmetric, short-range pair potentials, expressed as:

$$\upsilon_{\alpha\beta}^{CP}(r) = \varepsilon_{\alpha\beta}\left[\left(\frac{\sigma_{\alpha\beta}}{r}\right)^{12} - \left(1 - \delta_{\alpha\beta}\right)\left(\frac{\sigma_{\alpha\beta}}{r}\right)^{6}\right], \tag{2}$$

where $\varepsilon_{\alpha\beta}$ and $\sigma_{\alpha\beta}$ are the energy and length parameters characterizing the interaction between species $\alpha$ and $\beta$, respectively, (with $\alpha, \beta \in \{Si, O\}$) and $\delta_{\alpha\beta}$ is the Kronecker delta. All pairwise interactions are truncated at a distance of $r = 2.5\sigma_{\alpha\beta}$. The energy and length parameters are given by[71]

$$\varepsilon_{Si-O}/\varepsilon_{Si-Si} = 24.0, \qquad \varepsilon_{O-O}/\varepsilon_{Si-Si} = 1.0, \tag{3}$$

$$\sigma_{Si-O}/\sigma_{Si-Si} = 0.49, \qquad \sigma_{O-O}/\sigma_{Si-Si} = 0.85. \tag{4}$$

The mass ratio is set to $m_O/m_{Si} = 0.57$. All quantities are expressed in reduced units, with the unit of length defined as $\sigma_{Si-Si}$, temperature as $\varepsilon_{Si-Si}/k_B$, and time as $\sqrt{m_{Si}\sigma_{Si-Si}/\varepsilon_{Si-Si}}$. The model parameters are systematically optimized to suppress both demixing and crystallization, thereby favoring the formation of a structurally disordered, tetrahedrally coordinated network, akin to that observed in amorphous silica.[48,71,72] The simulations are performed at a number density of $\rho = 1.655$, using 1000 silicon and 2000 oxygen particles. This setup corresponds to a physical density of approximately 2.37 g cm$^{-3}$, consistent with the density used in the BKS model for amorphous silica.[38,71] The system is simulated over a temperature range from $T = 0.60$ to $T = 0.29$, using a time step of $\delta t = 0.001$ in reduced units.

For both models, the system is first equilibrated in the canonical (NVT) ensemble at each temperature using a Nosé–Hoover thermostat,[73] followed by production runs in the microcanonical (NVE) ensemble. Periodic boundary conditions are applied in all three spatial directions. The equations of motion are integrated using the velocity Verlet algorithm.[74] The production runs are carried out for durations at least five times longer than the characteristic structural relaxation time at each temperature, ensuring proper statistical sampling. Trajectory data are saved at intervals of 10 $\delta t$. We also examined the ensemble dependence of our results by performing analyses in the NVT ensemble with the Nosé–Hoover thermostat and found no significant differences in the statistics of jump dynamics.

## B. Identification of jump motion

To characterize the intermittent particle rearrangements commonly referred to as "jumps" or "hops," we employ the following hop function that quantifies sudden displacements of individual particles over time. For each particle $i$ of species $\alpha$, the hop function $h_i^\alpha(t)$ is defined by comparing its positions averaged over two consecutive time intervals before and after the current time $t$, and is given by[19,26,75]

$$h_i^\alpha(t) = \sqrt{\left\langle \left(\mathbf{r}_i^\alpha(t) - \left\langle \mathbf{r}_i^\alpha(t)\right\rangle_B\right)^2 \right\rangle_A \left\langle \left(\mathbf{r}_i^\alpha(t) - \left\langle \mathbf{r}_i^\alpha(t)\right\rangle_A\right)^2 \right\rangle_B}, \tag{5}$$

where $\mathbf{r}_i^\alpha(t)$ is the position vector of particle $i$ of species $\alpha$ at time $t$. $\left\langle X(t) \right\rangle_A$ and $\left\langle X(t) \right\rangle_B$ denote time averages over intervals $A$ and $B$, each of duration $\Delta t/2$, immediately preceding and following time $t$, respectively, and are defined as $\left\langle X(t) \right\rangle_A = \frac{1}{\Delta t/2} \int_{t-\Delta t/2}^{t} X(\tau) d\tau$ and $\left\langle X(t) \right\rangle_B = \frac{1}{\Delta t/2} \int_{t}^{t+\Delta t/2} X(\tau) d\tau$. Thus, $h_i^\alpha(t)$ measures the extent to which the particle shifts its average position between these adjacent intervals.

The choice of the time window $\Delta t$ is critical for accurately detecting jump events. It must be sufficiently long to average out rapid vibrational motions within the particle cage, yet short enough to resolve distinct, intermittent jumps. In this study, we select $\Delta t$ based on the characteristic timescale at which the mean squared displacement (MSD), $\left\langle \delta r^2(t) \right\rangle$, transitions from ballistic to diffusive behavior.[19–22,76] The MSD for species $\alpha$, is defined as

$$\left\langle \delta r_\alpha^2(t) \right\rangle = \frac{1}{N_\alpha} \sum_{i=1}^{N_\alpha} \left\langle \left|\mathbf{r}_i^\alpha(t) - \mathbf{r}_i^\alpha(0)\right|^2 \right\rangle, \tag{6}$$

where $N_\alpha$ is the number of particles of the species $\alpha$, $\mathbf{r}_i^\alpha(t)$ is the position vector of particle $i$ of species $\alpha$ at time $t$. The time window $\Delta t$ is chosen as the time when the logarithmic derivative of the MSD, $d\log\left\langle \delta r_\alpha^2(t) \right\rangle / d\log t$, reaches its minimum value, indicating the point of minimum diffusivity [Fig. S5].[21,22,26,77] This approach ensures that the hop function captures genuine jump motions while minimizing the contribution from localized vibrations. In previous studies, it has been observed that the results are largely insensitive to the precise choice of Δ$t$, provided it lies within the plateau region of the mean squared displacement.[26,76] This indicates that moderate variations in the averaging window do not compromise the reliable identification of jump events. The specific Δ$t$ values used in the present study are summarized in Table S3.

To distinguish localized vibrational motions, i.e., cage dynamics, from jump events, we analyze the cumulative distribution of the hop function $h_i^\alpha(t)$, representing the fraction of particle displacements exceeding a given threshold.[29,76] Typically, this distribution exhibits a sharp decline up to a threshold value $h^*$, followed by a slower, approximately exponential tail corresponding to less frequent but larger jumps. In our study, the threshold $h^*$ is found to depend on temperature, the type of species considered, and the interaction potential, particularly because one model uses reduced units while the other employs real units. The detailed variation of $h^*$ across these parameters is discussed later in the text. This underscores the necessity of carefully determining $h^*$ for each specific condition to accurately identify jump motions.

## C. Dynamic disorder in jump events: Survival probability and rate fluctuations

After identifying jump events via the hopping threshold $h^*$, a fundamental question arises: how long do particles remain confined within their local cages before making a jump? Reflecting the stability of a particle's local environment, this residence time plays a key role in the slow structural relaxation behavior of glassy systems.

To quantify this, we analyze the residence time distribution of the cage state $\psi_{cage}(t)$, which gives the probability density that a particle remains in the cage state (defined by $h < h^*$) for a duration $t$ before jumping. The nature of the residence time distribution $\psi_{cage}(t)$ indicates whether the jump dynamics follow a stochastic process characterized by Poisson statistics, exhibiting a single exponential timescale, i.e., $\psi_{cage}(t) = ke^{-kt}$, or deviate from this behavior, reflecting a broad spectrum of waiting times for jump motion due to dynamic disorder.

From $\psi_{cage}(t)$, we define the residence probability $C_R(t)$, representing the probability that a particle remains trapped in the cage for at least time $t$:[67]

$$C_R(t) = \int_t^\infty \psi_{cage}(\tau)\,d\tau. \qquad (7)$$

This cumulative measure provides a clearer picture of how particle confinement decays over time. To further characterize cage persistence, we calculate the survival probability $C_S(t)$, which normalizes $C_R(t)$ by the average residence time $\langle t \rangle$, and represents the fraction of particles remaining in the cage beyond time $t$:[67,68]

$$C_S(t) = \frac{1}{\langle t \rangle} \int_t^\infty C_R(\tau)\, d\tau. \tag{8}$$

Together, these probabilities allow us to probe the dynamics of jump events and the temporal evolution of local particle confinement, thereby shedding light on the mechanisms underlying structural relaxation in supercooled liquids and glasses. After defining the residence and survival probabilities, it is important to quantify how regularly or irregularly particles jump from their cages. This is captured by the randomness parameter $R$, defined as[78,79]

$$R = \frac{\langle t^2 \rangle - \langle t \rangle^2}{\langle t \rangle^2}, \tag{9}$$

where $\langle t^n \rangle = \int t^n \psi_{cage}(t)\, dt$ is the $n$th moment of the residence time distribution $\psi_{cage}(t)$. While the form of $\psi_{cage}(t)$ already suggests whether jump dynamics follow Poisson-like behavior, the randomness parameter ($R$) offers a compact, quantitative descriptor of this behavior. In particular, deviations of $R$ from unity reflect departures from exponential waiting times and thus serve as a sensitive metric for dynamic disorder in the system.

To further elucidate how dynamic disorder influences particle jumps, the survival probability $C_S(t)$ can be examined within the framework of a time-dependent rate process. In this framework, the decay of the survival probability $C_S(t)$ is described by a first-order differential equation as follows:[62]

$$\frac{dC_S(t)}{dt} = -k(t) C_S(t), \tag{10}$$

where $k(t)$ is the time-dependent jump rate that generally fluctuates due to changes in the local environment. The solution to this differential equation, accounting for rate fluctuations, takes the form

$$C_S(t) = \left\langle \exp\left(-\int_0^t k(\tau)\, d\tau\right) \right\rangle, \tag{11}$$

where the angular brackets denote an ensemble average over all realizations of the rate function. Depending on the characteristic timescale of fluctuations in $k(t)$, two limiting behaviors emerge.

**(i) Fast fluctuation limit**

In this case, the environmental fluctuations that influence the jump rate $k(t)$ occur on timescales much shorter than the average cage residence time. As a result, the system effectively experiences a temporally averaged jump rate, and the survival probability reduces to a simple exponential decay,[62]

$$C_{fast}(t) = \exp(-k_{fast}t), \tag{12}$$

where $k_{fast}$ denotes the average jump rate. To estimate $k_{fast}$, we consider a trajectory of total duration $T_{tot}$, divided into $N_{tot}$ discrete time steps of size $\Delta t$ (so that $T_{tot} = N_{tot}\Delta t$). Suppose a total of $N_J + 1$ jump events are detected during this period, where the first jump is taken to occur at the time of origin for convenience. The average rate is then obtained as[26,67,68]

$$k_{fast} = \frac{N_J}{(N_{tot}\Delta t)}. \tag{13}$$

This expression corresponds to the flux-over-population approach, where the number of jump events (flux) is normalized by the total time spent in the cage state (population). The fast limit thus provides a baseline for comparing the observed survival probability against the idealized Poisson behavior.

### (ii) Slow fluctuation limit

When the environmental fluctuations modulating the jump rate $k(t)$ occur on timescales much longer than the average cage residence time, the system can be viewed as residing in different quasi-static substates. In this slow-fluctuation regime, the survival probability $C_{slow}(t)$ takes the form of a multi-exponential function weighted by the distribution of $k$,[62]

$$C_{slow}(t) = \langle \exp(-kt) \rangle_k. \tag{14}$$

By defining the jump rate $k_i$ for each substate $i$, $C_{slow}(t)$ can be expressed as a weighted sum over these substates,

$$C_{slow}(t) = \sum_{i=1}^{N} p_i \exp(-k_i t), \tag{15}$$

where $N$ is the total number of substates, and $p_i$ represents the relative population of substate $i$. To estimate the relative population $p_i$ and the escape rate $k_i$ for each substate, we discretize the total observation time into $N_{tot}$ intervals of duration $\Delta t$. Let $N_{tot,i}$ denote the number of time steps during which the system resides in substate $i$, and $N_{J,i}$ represent the number of jump events ($h \geq h^*$) detected within that substate. Then, the relative population and jump rate of substate $i$ are given by $p_i = N_{tot,i}/N_{tot}$ and $k_i = N_{J,i}/(N_{tot,i}\Delta t)$, respectively.[26,67,68] Thus, the survival probability in its slow fluctuation limit can be expressed as

$$C_{slow}(t) = \sum_{i=1}^{N} \frac{N_{tot,i}}{N_{tot}} \exp\left(-\frac{N_{J,i}}{N_{tot,i}\Delta t}t\right). \tag{16}$$

Each substate represents a distinct local environment or structural configuration that modulates the jump dynamics. In contrast to the fast fluctuation limit, where rapid transitions between substates lead to a single, averaged jump rate $k_{fast}$, the slow limit reflects a distribution of rates due to persistent structural heterogeneity. Consequently, $C_{slow}(t)$ deviates from a simple exponential and captures the multi-timescale nature of relaxation in disordered systems. This formulation reveals the explicit dependence of the survival probability on the choice of a slow variable used to define the substates.

To explore how different slow variables influence particle jump dynamics, we compute $C_{slow}(t)$ using various structural descriptors such as local coordination, interatomic distances, and angular correlations. For instance, when using the distance between a central jumping atom (say Si) and its $n$th nearest neighbor (e.g., $r_{SiOn}$ or $r_{SiSin}$), we partition the range of this variable into uniform bins, each treated as a distinct substate. Similarly, angular descriptors such as Si–Si–Si or O–Si–O bond angles are discretized into bins to define substates over which the survival probability is analyzed. From the trajectory, we evaluate the residence time in each substate and the number of jumps observed therein. This allows us to determine the substate populations $p_i$ and the corresponding rates $k_i$, which together define the multi-exponential form of $C_{slow}(t)$. This analysis provides valuable insights into the structural variables most closely linked to jump events, thereby helping to uncover the underlying causes of dynamic heterogeneity within the system.

### D. Kullback-Leibler (KL) divergence analysis

Based on the analysis of jump events and their associated structural substates, it is essential to further characterize how the local environment evolves as a particle approaches the hopping threshold, $h^*$. While the survival probability captures the timescale and heterogeneity of jump dynamics, analyzing changes in structural distributions provides a microscopic perspective on the mechanisms driving these events.

Hence, we focus on the probability distribution $P(x, h)$ of a relevant structural variable $x$, e.g., the distance between a jumping atom and its $n$th nearest neighbor along the jump coordinate $h$. As $h$ approaches the threshold $h^*$, the distribution $P(x, h)$ evolves, reflecting the progressive reorganization of the local environment in anticipation of a jump. To quantify these changes, we employ the Kullback–Leibler (KL) divergence,[80] which compares the distribution at an intermediate jump coordinate $h$ with that at the jump threshold $h^*$:

$$D(h) = \int dx\, P(x,h) \log \frac{P(x,h)}{P(x,h^*)}. \tag{17}$$

This metric captures the extent to which the structural environment deviates from its final configuration at the onset of a jump. For comparison, we also compute the KL divergence between the equilibrium distribution $P^{eq}(x)$ and the distribution at $h^*$:

$$D_{eq} = \int dx\, P^{eq}(x) \log \frac{P^{eq}(x)}{P(x,h^*)}. \tag{18}$$

The normalized quantity $D(h)/D_{eq}$, thus, characterizes the progression of the structural variable $x$ as the system approaches a jump, in a manner analogous to a time-correlation function. This framework enables us to identify which structural features undergo the most significant transformations, providing deeper insight into the microscopic origins of dynamic heterogeneity in disordered systems.

### E. Static point-to-set corrections

The analysis of survival probabilities in the slow fluctuation regime reveals structural variables that strongly modulate jump dynamics and underlie dynamic disorder. These variables reflect the cooperative nature of cage rearrangements: the probability of a jump is not determined solely by an individual particle's environment but by collective fluctuations involving its neighbors. While this underscores the cooperative dynamics of supercooled liquids, where local rearrangements depend sensitively on broader spatial correlations, it raises a fundamental question: *how far do these cooperative motions extend in space, and how do we quantify their spatial extent as the system becomes increasingly sluggish upon cooling?* Addressing this question is essential to a microscopic understanding of dynamic heterogeneity and the slowdown in glass-forming systems.

To address this question, we evaluate the static point-to-set (PTS) correlation length,[81–90] which measures the spatial extent of amorphous structural correlations. Unlike standard pairwise structural descriptors, which often exhibit little temperature dependence in supercooled liquids, PTS correlations probe hidden many-body orders that become increasingly relevant as the dynamics slow down. This idea originates from the Random First-Order Transition (RFOT) theory,[1,91–94] which proposes that the growing relaxation time in glass-forming systems is driven by the expansion of cooperatively rearranging regions (CRRs).[95] Within this framework, the structural relaxation time $\tau_\alpha$ is governed by an activated scaling relation:[81,86,90]

$$\tau_\alpha(T) = \tau_0 \exp\left[\frac{\Delta_0 \xi_{PTS}^\psi(T)}{T}\right], \tag{19}$$

where $\xi_{PTS}$ is the PTS correlation length, $\psi$ is an exponent characterizing the energy barrier scaling, $\Delta_0$ and $\tau_0$ are the characteristic energy and microscopic time scales, respectively. This suggests that the sluggish dynamics observed at low temperatures are fundamentally tied to the size of regions over which particles must reorganize collectively.

To estimate $\xi_{PTS}$, we adopt a protocol that probes how the structure of a region is constrained by its frozen surroundings, thereby quantifying the spatial extent of amorphous order. This approach follows the original construction by Bouchaud and Biroli,[81,85] and involves measuring the configurational response of a subset of particles to external pinning. We implement this idea using a cubic cavity with frozen boundaries.[87] In this setup, a central cubic cavity of side length $2d$ is embedded within a three-dimensional periodic simulation box. All particles outside the cavity are permanently frozen at time $t = 0$, while the particles inside are allowed to evolve under standard Newtonian dynamics. A schematic illustration of the geometry and the associated measurement setup is provided in Fig. S29 (a). For each cavity, the system is first equilibrated at the target temperature in the NVT ensemble using a Nosé–Hoover thermostat, and the subsequent production run is then performed in the NVE ensemble. The overlap functions are then computed within the central region of the cavity to minimize boundary effects. Specifically, the central $5^3$ grid cells are used for evaluating the collective overlap function $Q(t)$, defined as:[87,88]

$$Q(t) = \frac{\sum_i \langle n_i(t) n_i(0) \rangle}{\sum_i \langle n_i(0) \rangle}, \tag{20}$$

where $n_i(t) \in \{0, 1\}$ is the occupation number of the $i$th cubic cell at time $t$, with a cell volume of approximately $\upsilon = 0.375 \sigma_{Si}^3$. Cells containing pinned particles are excluded from the sum. A value of $Q(t) = 1$ indicates perfect memory of the initial configuration, whereas values approaching the random overlap $Q_{rand} \approx \rho_0 \upsilon$ (with $\rho_0$ being the average particle number density) signify loss of memory due to relaxation. The long-time limit $Q_\infty(d) = \lim_{t \to \infty} Q(t)$ is estimated from the steady-state value of $Q(t)$, with simulations run for durations of 100 to 1000 times the bulk $\alpha$-relaxation time $\tau_\alpha$, depending on both temperature and the degree of confinement, to ensure full decorrelation. To obtain statistically reliable estimates, these long-time values are further averaged over 25–30 independent realizations for each value of $d$.

To extract $\xi_{PTS}$, the excess overlap $\tilde{q}(d)$ is fitted to a compressed exponential form:[85,88]

$$\tilde{q}(d) = Q_\infty(d) - Q_{rand} = A\exp\left[-\left(\frac{d-a}{\xi_{PTS}}\right)^\eta\right], \qquad (21)$$

where $A$ and $\eta$ are fitting parameters, and $a = 1$ accounts for the minimal cavity size containing roughly a single particle. This procedure provides a robust estimate of $\xi_{PTS}$. While random pinning has also been proposed as an alternative means of evaluating $\xi_{PTS}$, its practical application is more challenging, since the overlap decay at low pinning fractions is difficult to fit reliably, and threshold-based criteria are typically required.[87,89,96–98] For this reason, we focus on the results obtained from the cavity geometry in the present work.

## III. RESULTS AND DISCUSSION

### A. Structural and dynamical characterization

We first calculated the structural and dynamical properties of silica using the BKS and CP models across a broad temperature range to verify consistency with established data.[37,38,71,72] The radial distribution functions $g_{SiSi}(r)$, $g_{SiO}(r)$, and $g_{OO}(r)$ exhibit well-defined peaks corresponding to nearest-neighbor correlations with modest temperature dependence, confirming stable short-range order (Fig. S1, panels (a-c) for the BKS model and panels (d-f) for the CP model). In the silica network, silicon atoms are tetrahedrally coordinated by oxygen atoms in the first shell and connected to other silicon atoms in the second shell via bridging oxygens. Oxygen atoms are coordinated by silicon in the first shell and by oxygen in the second. To gain further insight into the structural properties, we computed the angular distribution functions $f_{\alpha\beta\gamma}(\theta)$, which represent the probability distribution of angles formed at a central particle of species $\beta$ with neighbors of species $\alpha$ and $\gamma$, where $\alpha, \beta, \gamma \in \{Si, O\}$. The distributions $f_{SiSiSi}(\theta)$, $f_{OSiO}(\theta)$, $f_{OOO}(\theta)$, and $f_{SiOSi}(\theta)$ similarly show subtle temperature-dependent changes (Fig. S2), indicating that both models exhibit comparable structural evolution upon cooling.

We then evaluated the dynamical behavior by computing the self-intermediate scattering function $F_s(k,t)$, and mean squared displacement (MSD) for silicon and oxygen atoms. The self-intermediate scattering function is defined as

$$F_s(k,t) = \frac{1}{N_\alpha}\left\langle\sum_{j=1}^{N_\alpha}\exp\left[i\mathbf{k}\cdot\left(\mathbf{r}_j^\alpha(t) - \mathbf{r}_j^\alpha(0)\right)\right]\right\rangle, \qquad (22)$$

where $N_\alpha$ is the number of atoms of species $\alpha$, $\mathbf{r}_j^\alpha(t)$ is the position vector of atom $j$ of species $\alpha$ at time $t$, and $\mathbf{k}$ is the wavevector. For the BKS model, $F_s(k,t)$ was calculated at wavevectors $k = 1.7$ and $2.8$ Å$^{-1}$, corresponding approximately to the first and second peaks of the static structure factor $S(k)$. For the CP model, wavevectors $k = 5.0$ and $8.0$ (in reduced units) were chosen to probe comparable structural features. The $\alpha$-relaxation times $\tau_\alpha$ were extracted from $F_s(k,t)$ as the time at which it decays to $e^{-1}$. Diffusion coefficients $D_\alpha$ were obtained from the long-time limit of the MSD [Eq. (6)] via $D_\alpha = \lim_{t\to\infty}\langle \delta r_\alpha^2(t)\rangle/6t$. Temperature-dependent $F_s(k, t)$ and MSD for silicon and oxygen in both models are shown in Figs. S3 and S4.

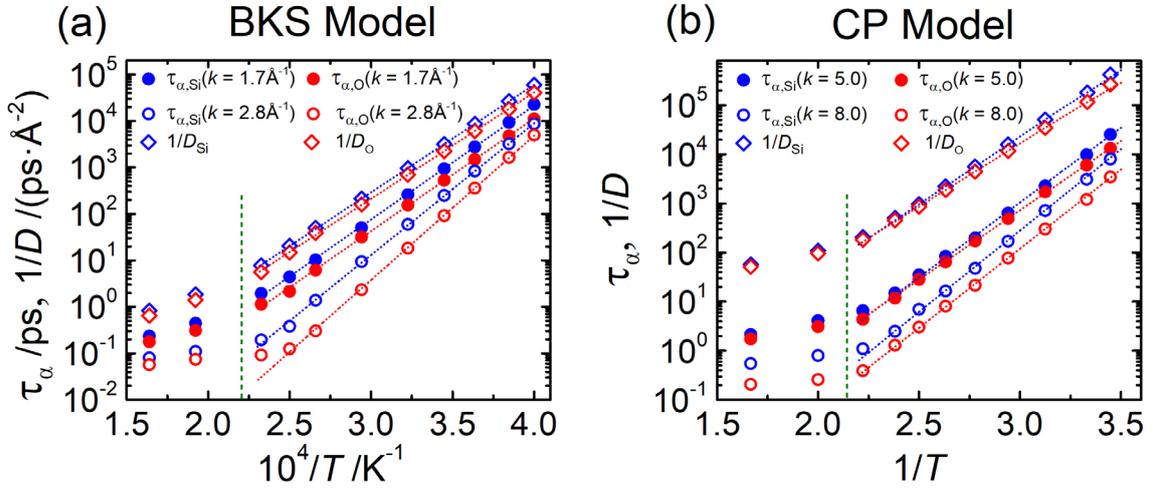

**FIG. 1.** Temperature dependence of the structural relaxation time $\tau_\alpha$ and the inverse diffusion coefficient $1/D$ for (a) the BKS and (b) the CP silica models. Data for silicon are shown in blue, while those for oxygen are shown in red. Circles denote $\tau_\alpha$: filled and open circles correspond to $k = 1.7$ and $2.8$ Å$^{-1}$ for BKS and $k = 5.0$ and $8.0$ (reduced units) for CP. Open diamonds represent $1/D$. Dotted lines are Arrhenius fits, and the vertical dashed line marks the onset of strong (Arrhenius) behavior.

Fig. 1 shows the $\alpha$-relaxation times $\tau_\alpha$ and inverse diffusion coefficients $1/D_\alpha$ for silicon and oxygen in both models. The BKS model exhibits Arrhenius-like behavior below $T = 4300$ K, whereas the CP model shows similar behavior below approximately $T = 0.45$. Activation energies, $E_a$, were determined by fitting the temperature dependence of $\tau_\alpha$ and $D_\alpha$ to an Arrhenius relation of the form $\sim \exp(\pm E_a/k_B T)$, where $k_B$ is the Boltzmann constant. The extracted values are summarized in Table S2 and show good agreement with previous studies.[38,71] The activation energies obtained here ($\sim$3.5-5.5 eV $\approx$ 320-510 kcal mol$^{-1}$) are an order of magnitude larger than those reported for supercooled water in its strong regime ($\sim$0.2-0.3 eV ≈ 4.6-6.9 kcal mol$^{-1}$)$^{99-101}$. This difference arises from the highly rigid, covalently bonded Si–O network in silica, where structural relaxation must overcome strong topological constraints of the tetrahedral network, in contrast to the easily reconfigurable hydrogen-bond network of water.

As established in Ref. [71], the energy unit of the CP model corresponds to approximately 7000 $K$ $\left(\varepsilon_{Si-Si}/k_B \approx 7000 K\right)$, enabling an approximate temperature mapping to the BKS model based on static structural properties. Specifically, radial distribution functions at a reduced temperature $T = 0.39$ in the CP model were shown to closely match those near $T \approx 2750$ $K$ in the BKS model. However, as noted in Refs. [71,72], this correspondence is limited to static structural features; dynamic behaviors differ markedly due to distinct interaction potentials and timescales. Consequently, a simultaneous one-to-one mapping of temperature-dependent structural and dynamic quantities between the two models is not feasible. In this study, we analyzed both models to ensure the robustness of our findings; however, the main text primarily focuses on the results from the BKS model for clarity, with corresponding CP results provided in the Supplementary Material.

## B. Local structural changes during jump motions

As outlined in the Introduction, structural relaxation in glass-forming liquids occurs via localized jump events where particles escape transient cages formed by their neighbors. To identify such events, we calculated the hop function $h$ [Eq. (5)]$^{19,26}$ for individual silicon and oxygen atoms, as described in Sec. II.B. Fig. 2 depicts the representative time evolution of $h$ for individual silicon and oxygen atoms in the BKS model at temperatures $T$ = 5200, 3760, and 2500 K, with analogous data for the CP model shown in Fig. S6.

The hop function $h$ exhibits long-lasting small fluctuations associated with vibrational motion within a cage, punctuated by sudden large jumps corresponding to transitions between cages. Both the frequency of such transitions and the magnitude of the associated changes in $h$ decrease with lowering temperature, reflecting slower structural relaxation. Additionally, oxygen atoms consistently show larger values of $h$ compared to silicon atoms, likely due to their smaller size and differing local environment. The threshold $h^*$ used to distinguish between cage and jump states is determined from the cumulative probability distribution of $h$ (Fig. S7).$^{26,76}$ For $T \leq 4000$ K, we set $h^* = 2.0$ Å$^2$ for silicon and 4.0 Å$^2$ for oxygen to ensure consistent identification of jump events in the strong regime, with all threshold values summarized in Table S4. Notably, at low temperatures, the threshold for oxygen is at least twice that of silicon in both models, reflecting the enhanced displacements permitted by its coordination environment.

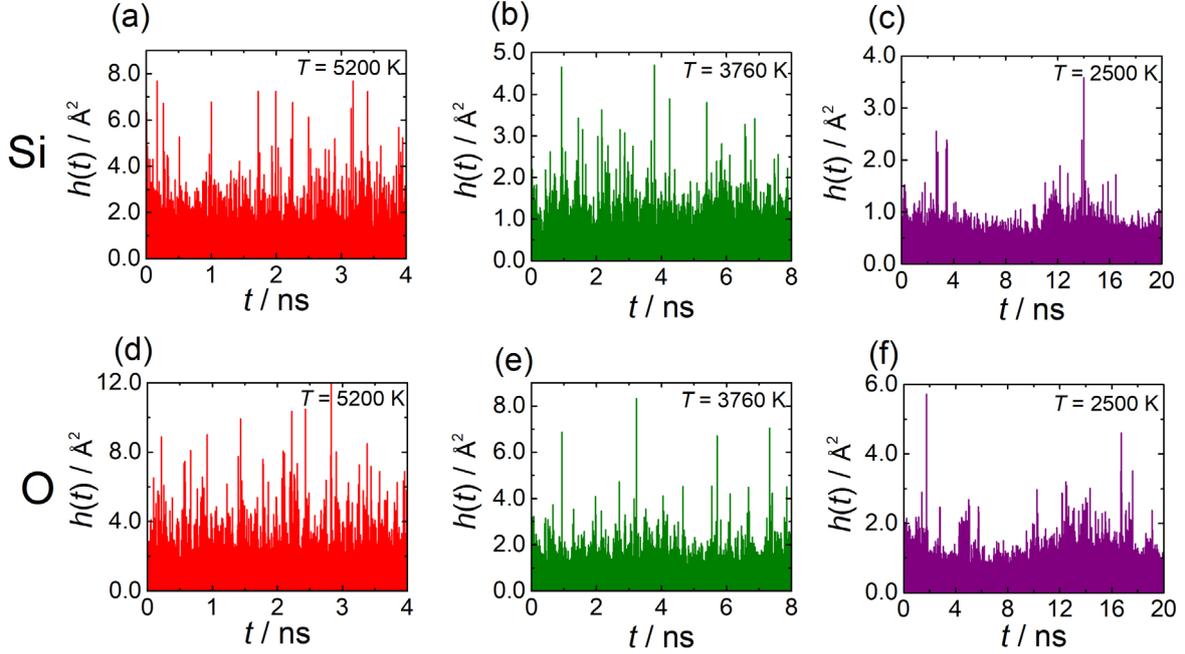

**FIG. 2.** Time evolution of the hop function $h(t)$ for individual Si (top row) and O (bottom row) atoms in the BKS model at $T = 5200$ K, 3760 K, and 2500 K. The trajectories illustrate the increasingly intermittent nature of jump events upon cooling.

To characterize the microscopic structural changes associated with jump events, we employed the hop function $h$ as an order parameter and computed various thermodynamic and structural properties as functions of $h$ for both silicon and oxygen atoms across the temperature range studied. We calculated the free energy profile along $h$, $\Delta F(h) = -k_B T \ln P(h) + C$, where $P(h)$ is the probability distribution of $h$ and $C$ is chosen such that the minimum of $\Delta F(h)$ is zero. This normalization allows direct comparison of the barrier heights across different temperatures and atom types. Figs. 3(a) and 3(b) show the free energy profiles as a function of the hop function $h$ for silicon and oxygen atoms, respectively, at representative temperatures. At high temperatures, the barrier for jump events ($h > h^*$) is low, reflecting facile structural relaxation. As temperature decreases, the barrier height increases significantly, indicating that jump events become increasingly constrained in the low-temperature viscous regime. For silicon, the dimensionless barrier height is $\Delta F(h^*)/(k_B T) = 5.06$ at $T = 4000$ K and 10.58 at $T = 2500$ K, corresponding to absolute barriers of 1.74 and 2.28 eV, respectively, i.e., an increase of about 0.54 eV. For oxygen, the barrier is $\Delta F(h^*)/(k_B T) = 6.63$ at $T = 4000$ K and 11.42 at $T = 2500$ K, corresponding to 2.28 and 2.46 eV, respectively, an increase of only 0.18 eV. The larger increase for silicon reflects its stronger involvement in the tetrahedral network, where cage breaking requires more cooperative motion and a higher energetic cost than for the more mobile oxygen atoms, consistent with earlier analyses of species-dependent relaxation in silica.[37,41,42]

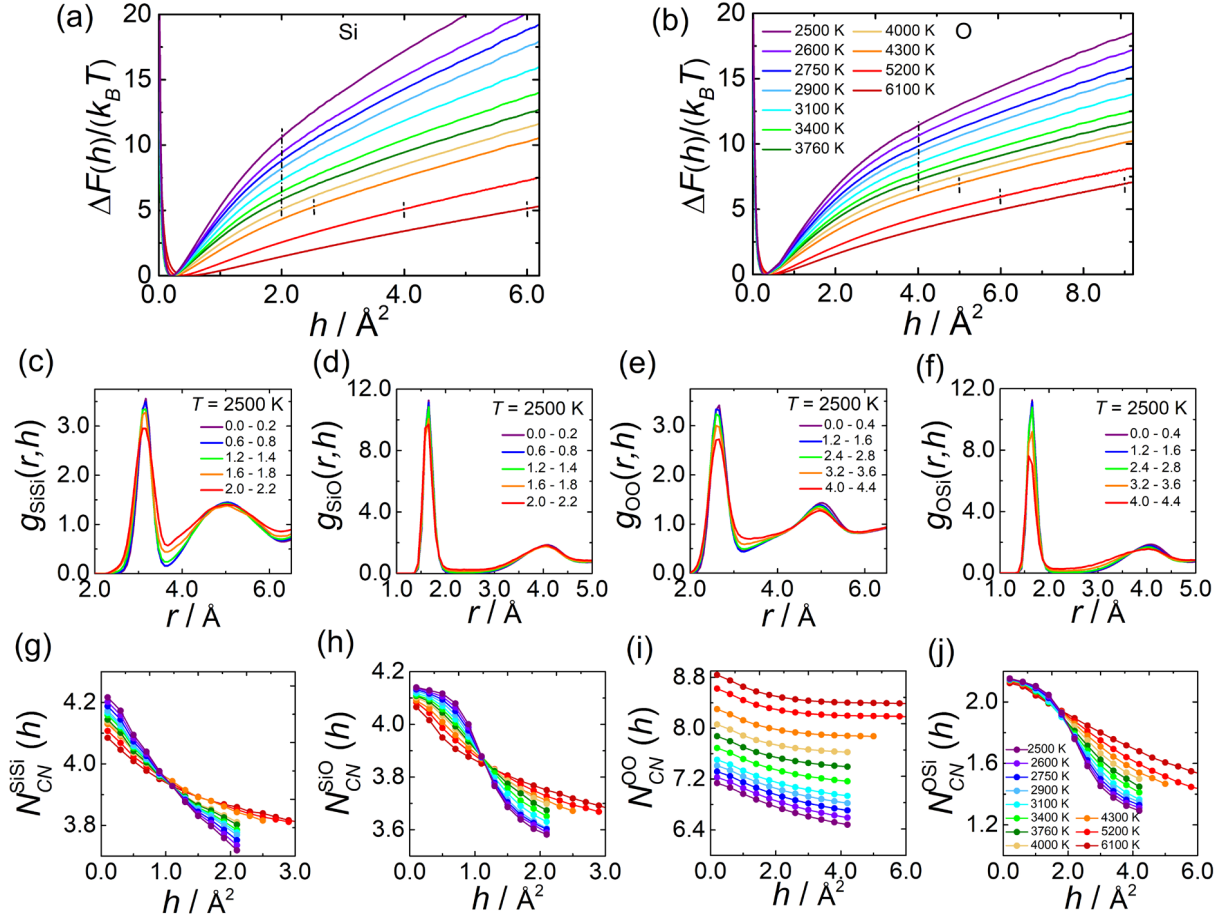

**FIG. 3.** (a)-(b) Temperature dependence of free-energy profiles $\Delta F(h)/(k_B T)$ for Si and O, showing the progressive growth of the barrier height with cooling (legend in (b)). Vertical black dashed-dotted lines indicate the hopping threshold $h^*$ at the corresponding temperatures. (c)-(f) Partial radial distribution functions $g_{\alpha\beta}(r, h)$ at $T = 2500$ K between a jumping atom $\alpha = \{Si, O\}$, resolved for specific ranges of $h$, and its surrounding neighbors $\beta = \{Si, O\}$. Increasing $h$ leads to a systematic reduction of peak heights and a shallowing of the minima. (g)-(j) Temperature dependence of the coordination numbers $N_{CN}^{\alpha\beta}$ for jumping Si and O atoms, showing a progressive loss of neighbors with increasing $h$ (legend in (j)).

To elucidate structural changes along the jump coordinate, we analyzed the partial radial distribution functions $g_{\alpha\beta}(r, h)$ for the jumping atoms $\alpha = \{Si, O\}$, and their neighboring species $\beta$ = {Si, O}, resolved by the hop function value $h$. Figs. 3(c)-3(f) show the partial radial distribution functions (RDFs) at the lowest temperature studied ($T$ = 2500 K), while representative results at other temperatures are provided in Fig. S9. Across all partial pairs ($g_{SiSi}$, $g_{SiO}$, $g_{OO}$, and $g_{OSi}$), increasing $h$ results in a systematic reduction in the heights of the first and second peaks, along with a shallowing of the first minimum. These trends indicate a progressive weakening of local structural order and a shift towards more disordered configurations as particles exhibit larger jump displacements.

To quantify changes in local coordination along the jump coordinate, we calculated the partial coordination numbers $N_{CN}^{\alpha\beta}$ by integrating $g_{\alpha\beta}(r, h)$ up to the first minimum. Figs 3(g) - 3(j) show the dependence of partial coordination numbers $N_{CN}^{\alpha\beta}(h)$ on the hop function $h$ across different temperatures. For silicon atoms, $N_{CN}^{SiO}$ and $N_{CN}^{SiSi}$ decrease monotonically with increasing $h$, reflecting a progressive loss of oxygen atoms from the first coordination shell and silicon atoms from the second shell. At $T$ = 2500 K, the reductions in $N_{CN}^{SiO}$ and $N_{CN}^{SiSi}$ are comparable in magnitude, indicating that structural rearrangements around jumping silicon atoms involve coordinated changes in both first- and second-shell environments. For oxygen atoms, $N_{CN}^{OSi}$ decreases more significantly than $N_{CN}^{OO}$ with increasing $h$, indicating a preferential loss of nearby silicon neighbors. This disparity highlights that oxygen jump events are primarily driven by changes in first-shell silicon coordination, underscoring the key role of silicon in the local hopping dynamics of oxygen atoms. The comparison of coordination numbers at equilibrium and the jump threshold $h^*$ is provided in Fig. S12.

The hop dependence of angular distribution functions $f_{SiSiSi}(\theta)$, $f_{OSiO}(\theta)$, $f_{OOO}(\theta)$, and $f_{SiOSi}(\theta)$ was analyzed to assess changes in local bond geometry, with the results presented in Fig. S13. The $f_{SiSiSi}(\theta)$ and $f_{OSiO}(\theta)$ distributions show only modest changes with increasing $h$: $f_{SiSiSi}(\theta)$ exhibits a slight reduction in its main peak and a weak enhancement near 60º, whereas $f_{OSiO}(\theta)$ shows mild broadening at an essentially unchanged mean angle, indicating that the SiO$_4$ tetrahedra remain largely intact. In contrast, the $f_{OOO}(\theta)$ distribution shows an enhanced population at 60-75º and reduced population at large angles, indicating a local compaction of the surrounding oxygen network. However, the most pronounced change appears in the inter-tetrahedral $f_{SiOSi}(\theta)$ distribution, which develops a distinct secondary lobe at 60-75º as the system approaches the hop threshold $h^*$. This feature indicates that the bridging oxygen undergoes a rotational motion around one of its Si–O bonds, transiently bending the Si–O–Si linkage. Along with this, at the lowest temperature studied, the coordination number $N_{CN}^{OSi}$ of the hopping oxygen decreases from ≈ 2.1 to ≈ 1.3, while the coordination number of the hopping silicon $N_{CN}^{SiO}$ shows a corresponding reduction from ≈ 4.1 to ≈ 3.6 at $h^*$, indicating that both species become transiently under-coordinated before returning to their typical coordination in the subsequent cage. These findings are related to earlier observations of two classes of elementary rearrangements in silica for both the BKS[37,41] and CP[71] models: (i) rotational rearrangements, in which oxygens permute among tetrahedral positions around nearly immobile silicons, known as the "rotational period" described by Heuer and co-workers[41], and (ii) localized defect-mediated jumps, where Horbach and Kob[37] showed that oxygen atoms contributing to large displacements are frequently one- or threefold coordinated and that silicon motion proceeds via transient three- or fivefold coordinated intermediates through a bond-exchange sequence involving slight recoil of neighboring oxygens. In our analysis, both silicon and oxygen atoms pass through transient under-coordinated states as they approach the hop threshold $h^*$, with oxygen hopping further involving a rotational distortion of the Si–O–Si bond angle as part of the bond-exchange process. We next focus on the temporal characteristics of these hopping events and their evolution with temperature.

### C. Transition in the nature of jump dynamics

To investigate the temperature-dependent evolution of jump dynamics, we analyzed the residence time distributions for the cage and jump states, identified using the threshold hop value $h^*$. Figs. 4(a) and 4(c) show that the residence time distributions for the jump state, $\psi_{jump}(t)$, are short and nearly temperature-independent for both silicon and oxygen. In contrast, the residence time distributions for the cage state, $\psi_{cage}(t)$, shown in Figs. 4(b) and 4(d), exhibit strong temperature dependence. At high temperatures ($T \geq 4300$ K), $\psi_{cage}(t)$ decays exponentially, indicating that jump events follow Poisson statistics. As the temperature decreases, $\psi_{cage}(t)$ develops a moderate non-exponential tail, reflecting prolonged cage trapping that gives rise to slow, intermittent non-Poisson dynamics and contributes to the emergence of dynamic heterogeneity.[10,102] Such intermittent dynamics are naturally expected when mobility becomes sparse at low temperatures, as described by the dynamic facilitation picture,[103,104] which attributes these dynamics to localized excitations that trigger mobility in neighboring immobile regions. It can also be interpreted within the activated barrier-crossing framework,[105,106] in which structural relaxation is governed by thermally activated events over local energy barriers, and fluctuations in barrier heights give rise to the long-tailed $\psi_{cage}(t)$.

To further quantify this transition from Poisson to non-Poisson behavior, we analyzed the randomness parameter $R$ (Eq. 9), which measures the deviation of the cage-time distribution from an ideal Poisson process.[78,79] As shown in Fig. 4(e), $R$ remains close to unity at high temperatures ($T \geq 4300$ K), consistent with exponential cage-time distributions and uncorrelated jump events. Upon cooling, $R$ steadily increases, indicating increasingly heterogeneous and temporally correlated jump dynamics. At lower temperatures, $R$ is consistently higher for silicon than for oxygen, with the difference becoming more pronounced as the temperature decreases. The larger $R$ values for silicon reflect more substantial deviations from Poisson statistics, indicating somewhat enhanced heterogeneity in cage dynamics. Both the BKS (Fig. 4) and CP (Fig. S14) models show similar residence time distributions and randomness parameters, suggesting these transitions in jump dynamics are intrinsic to the silica melt system.

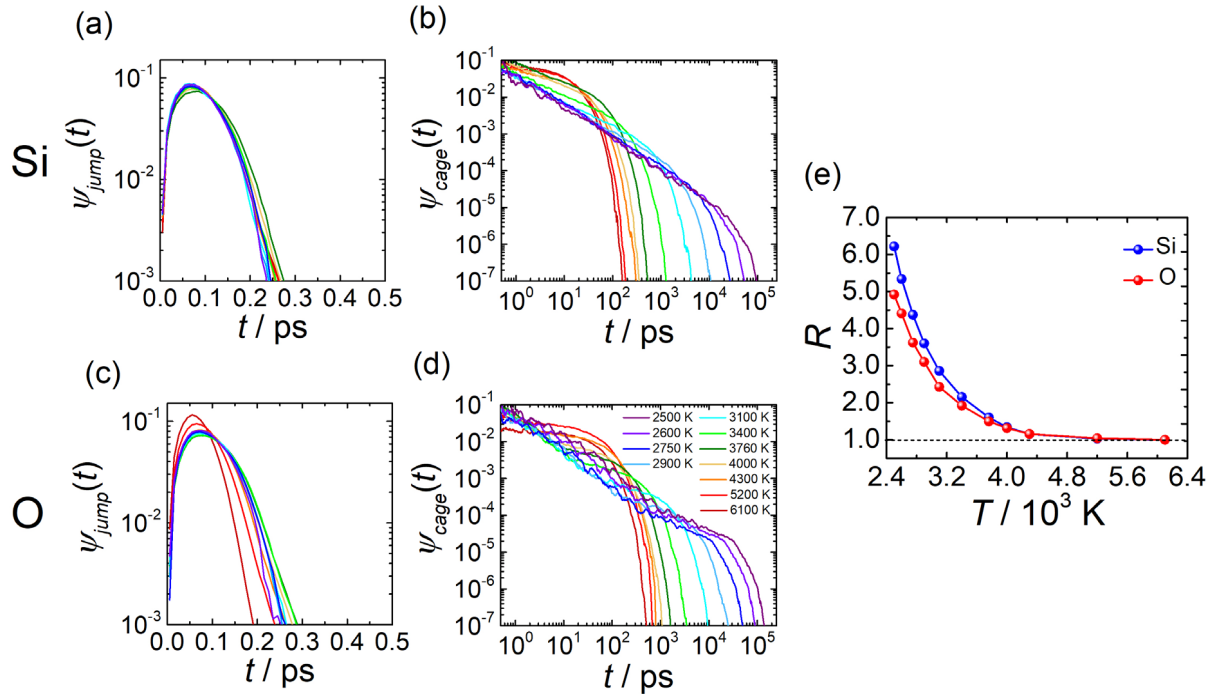

**FIG. 4.** Temperature dependence of (a), (c) jump-time distributions $\psi_{jump}(t)$ for Si and O, respectively, and (b), (d) the corresponding cage-time distributions $\psi_{cage}(t)$ in the BKS model (legend in (d)). While $\psi_{jump}(t)$ remains short and nearly temperature-independent, $\psi_{cage}(t)$ develops increasingly stretched tails upon cooling. (e) Temperature dependence of the randomness parameter $R$ for Si and O, quantifying deviations from Poisson statistics. Relatively larger $R$ values for Si indicate stronger non-Poisson behavior and enhanced heterogeneity in cage dynamics compared with O.

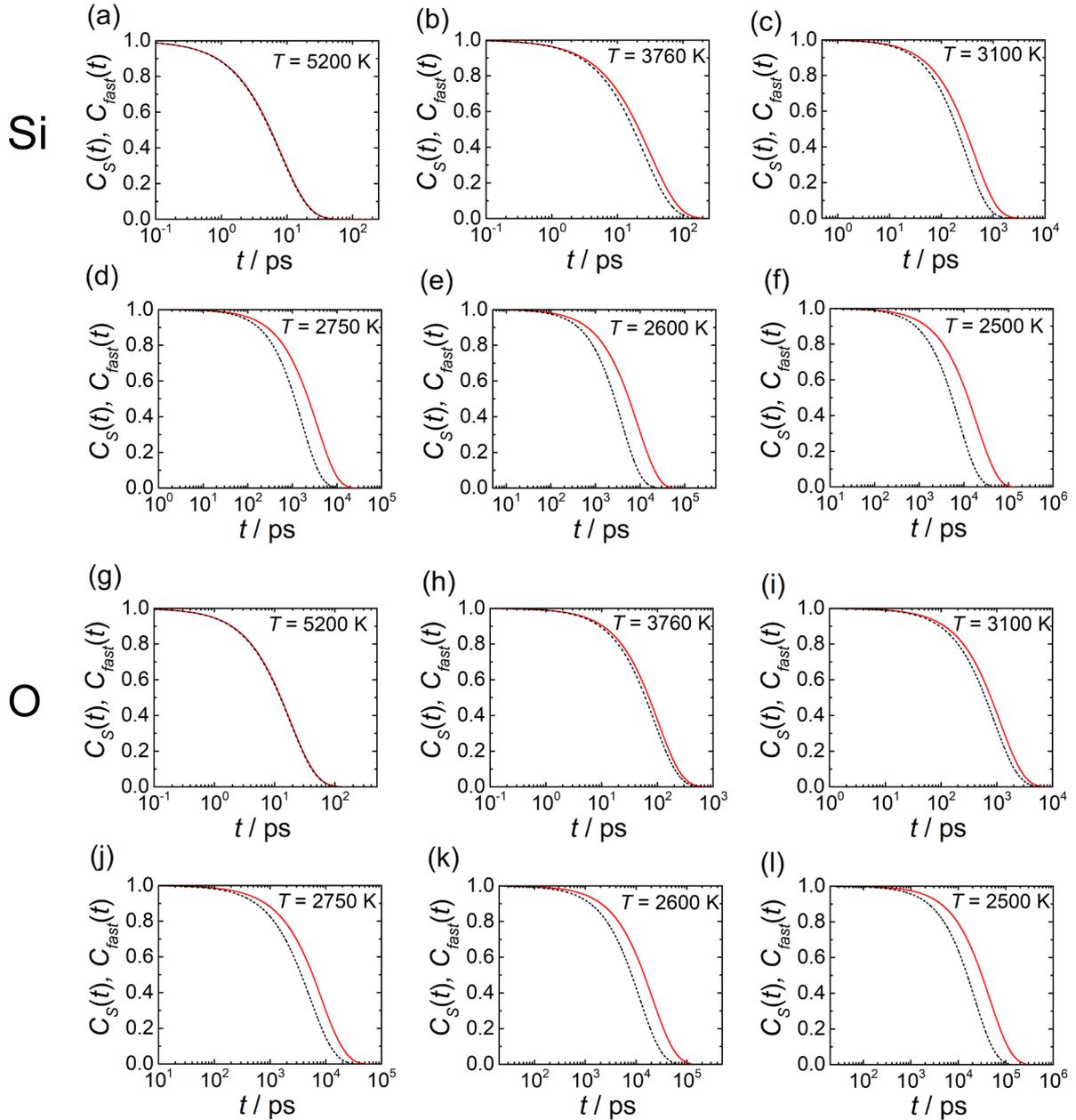

**FIG. 5.** Survival probability of the cage state, $C_S(t)$ (red), and its fast-fluctuation limit, $C_{fast}(t)$ (dashed black), for Si at (a) $T = 5200$, (b) 3760, (c) 3100, (d) 2750, (e) 2600, and (f) 2500 K in the BKS model. Panels (g-l) show the corresponding results for O at the same temperatures. At high temperatures, $C_S(t)$ closely follows $C_{fast}(t)$, consistent with Poisson statistics. However, upon cooling, systematic deviations emerge, indicating the breakdown of the fast-fluctuation approximation.

To probe the temporal origin of the non-Poisson behavior, we next analyzed the survival probability $C_S(t)$, which provides a measure of the time-dependent stability of the cage state, and compared it with its fast fluctuation limit $C_{fast}(t)$. $C_S(t)$ denotes the probability that an atom remains in the cage state up to time $t$ without undergoing a jump, as defined in Eq. (8) and discussed in Sec. II.C. **Fig. 5** shows the temperature dependence of $C_S(t)$ and $C_{fast}(t)$ for both silicon and oxygen. At high temperatures ($T \geq 4300$ K), $C_S(t)$ closely follows $C_{fast}(t)$, exhibiting exponential decay consistent with Poisson statistics. This agreement indicates that the hop function $h$ captures the slowest degree of freedom in the system, while other variables decorrelate rapidly. In this regime, the jump rate $k(t)$ fluctuates on timescales much shorter than the typical cage lifetime, and is well approximated by its fast-limit value $k_{fast}$. As the temperature decreases, $C_S(t)$ begins to deviate systematically from $C_{fast}(t)$, signaling a breakdown of the fast fluctuation approximation, i.e., $k(t) \sim k_{fast}$ is no longer valid. The jump rate becomes a slowly evolving quantity, marking the onset of dynamic disorder.[26,62–65,68] In this regime, fluctuations in the local environment persist over timescales comparable to the cage lifetime, leading to temporally correlated jump dynamics. To characterize the temporal heterogeneity of the jump dynamics, we next fit $C_S(t)$ to a stretched exponential function,

$$C_S(t) = \exp\left[-(kt)^\beta\right], \qquad (23)$$

where $k$ denotes the effective jump rate, and $\beta$ captures the degree of temporal heterogeneity in the jump dynamics. The extracted parameters are shown in **Fig. 6**.

**Fig. 6(a)** depicts the survival times defined as $\tau_S = 1/k$, together with the fast-fluctuation timescales $\tau_{fast} = 1/k_{fast}$. At high temperatures, the agreement between the two confirms the validity of the fast fluctuation approximation. Upon cooling, the survival time $\tau_S$ progressively exceeds $\tau_{fast}$, indicating that rate fluctuations persist over the timescale of jump motion. Both $\tau_S$ and $\tau_{fast}$ exhibit Arrhenius temperature dependence below $T = 4300$ K, consistent with thermally activated jump dynamics. The activation energy extracted from $\tau_S$ is slightly higher than that from $\tau_{fast}$, reflecting the influence of time-dependent rate fluctuations that are neglected in the fast fluctuation limit (Table S5). Interestingly, oxygen atoms exhibit systematically larger $\tau_S$ values than silicon atoms, despite having lower activation energies. This apparent discrepancy arises from the operational definition of $\tau_S$, which depends on the hop threshold $h^*$. Due to their larger displacement amplitudes during jumps, oxygen atoms are assigned a higher threshold, resulting in longer survival times. This underscores the need to interpret kinetic observables such as $\tau_S$ in conjunction with the jump threshold and the corresponding activation barrier. The stretching exponent $\beta$, shown in Fig. 6(b), remains close to unity at high temperatures, indicating near-exponential behavior of $C_S(t)$. Upon cooling, $\beta$ decreases steadily, reflecting the emergence of temporal heterogeneity in the jump dynamics. This decrease is more pronounced for silicon than for oxygen, suggesting that silicon atoms are more affected by persistent fluctuations in their local environment. In contrast, the weaker temperature dependence of $\beta$ for oxygen indicates more localized and less temporally correlated motion.

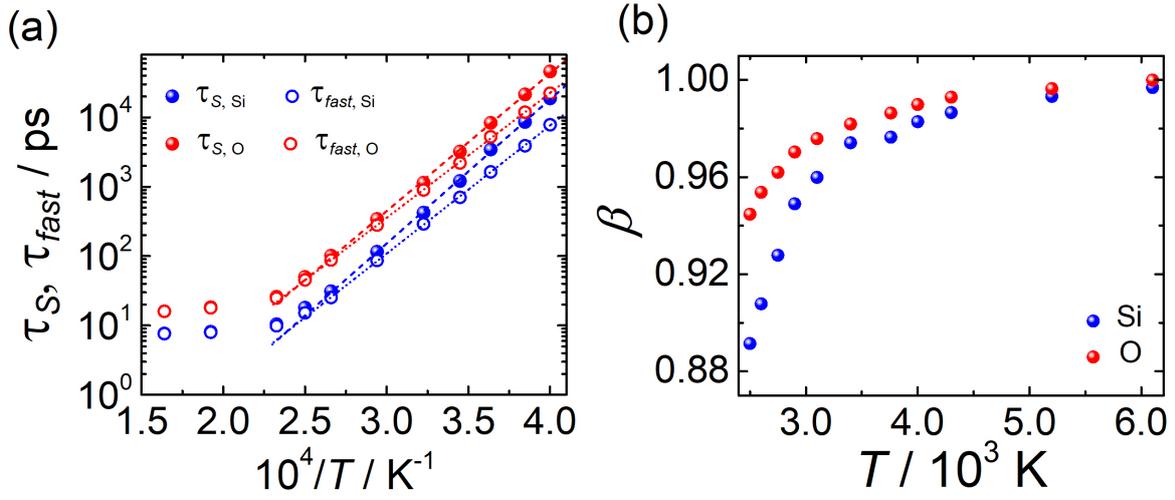

**FIG. 6.** (a) Temperature dependence of the survival time for the cage state, $\tau_S$, compared with the fast-fluctuation timescale, $\tau_{fast}$, for the BKS model. At high temperatures, the two coincide, confirming the fast-fluctuation approximation, whereas at lower temperatures, $\tau_S$ exceeds $\tau_{fast}$, showing that rate fluctuations persist over the jump timescale. Both quantities exhibit Arrhenius behavior below $T = 4300$ K, as indicated by the dashed and dotted lines. (b) Stretching exponent $\beta$ as a function of temperature. $\beta$ remains close to unity at high temperatures but decreases upon cooling, reflecting the emergence of temporal heterogeneity in jump dynamics; the reduction is more substantial for silicon than for oxygen.

The progressive divergence between $\tau_S$ and $\tau_{fast}$, together with the decrease in $\beta$, indicates that the slow dynamics become increasingly heterogeneous and cannot be represented by a single characteristic timescale. Although the hop function $h$ captures the dominant slow mode of jump dynamics at high temperatures, it becomes inadequate to describe the complexity of jump dynamics at lower temperatures. The emergence of intermittent dynamics suggests the presence of additional microscopic variables that govern rate fluctuations. This motivates a microscopic investigation of the structural and dynamical factors that contribute to dynamic disorder, which we address in the next section.

## D. Microscopic origin of dynamic disorder in jump motion

To identify the microscopic variables responsible for this dynamic disorder, we examined how the local structure around the jumping atom evolves along the hop coordinate $h$ and how these structural changes correlate with jump propensity. The analysis was carried out separately for silicon and oxygen atoms, given their distinct roles in the tetrahedral network, and further compared with water as a reference tetrahedral liquid.

### (i) For silicon

We analyzed the probability distributions of distances between jumping silicon atoms and their $n$th nearest oxygen and silicon neighbors, $P(r_{SiOn},h^*)$ and $P(r_{SiSin},h^*)$ ($1 \leq n \leq 8$), at the hop threshold $h^*$, and compared them with the corresponding equilibrium distributions, $P^{eq}(r_{SiOn})$ and $P^{eq}(r_{SiSin})$ [Figs. 7(a), 7(b), and S18]. At $h^*$, both $P(r_{SiOn},h^*)$ and $P(r_{SiSin},h^*)$ exhibit deviations from their equilibrium counterparts. For oxygen neighbors, the distributions with $n \leq 4$ shift toward longer distances, reflecting an expansion of the nearest oxygen and silicon coordination shells as local rearrangements open space for the impending jump [Fig. 7(c)], whereas those with $n \geq 5$ shift toward shorter distances. For silicon neighbors, most distributions shift toward shorter distances, indicating that atoms in the outer shells move inward to accommodate the displacements occurring closer to the jumping atom, except for the fourth-nearest silicon, $r_{SiSi4}$, whose outward shift may arise from the requirement to accommodate the expansion of $r_{SiO4}$. Among all cases, the displacement of the fourth-nearest oxygen neighbor, $r_{SiO4}$, is the most pronounced, exceeding the corresponding change for $r_{SiSi4}$, suggesting that rearrangements involving the fourth-nearest oxygen neighbor play a central role in the structural reorganization leading to a silicon jump.

We next evaluated the change in $P(r_{SiOn},h)$ and $P(r_{SiSin},h)$ toward their hop-threshold distributions $P(r_{SiOn},h^*)$ and $P(r_{SiSin},h^*)$ using the KL divergence [Eq. (17)], denoted collectively as $D_{SiXn}(h)$ [Fig. 7(d)]. Among all neighbors, $D_{SiX4}(h)$ is the largest within each sub-shell, indicating that the fourth-nearest neighbors undergo the most pronounced reorganization as a silicon atom approaches a jump. The magnitude of $D_{SiO4}(h)$ is greater than that of $D_{SiSi4}(h)$, consistent with the larger distribution shift observed for the fourth oxygen neighbor. The scaled KL divergence $D_{SiXn}(h)/D_{eq,SiXn}$, where $D_{eq,SiXn}$ denotes the KL divergence between the hop-threshold distribution $P(r_{SiXn},h^*)$ and the equilibrium distribution $P^{eq}(r_{SiXn})$, further shows that, for $n \leq 4$, the approach of $P(r_{SiOn},h)$ toward $P(r_{SiOn},h^*)$ is consistently slower than that of $P(r_{SiSin},h)$ toward $P(r_{SiSin},h^*)$ [Fig. 7(e)]. Given that oxygen atoms constitute the first coordination shell and silicon atoms occupy the second shell around a given silicon atom, this behavior indicates that the displacements of the inner oxygen neighbors occur within a local environment that has already been partially reconfigured by the faster structural adjustments of the surrounding silicon atoms.

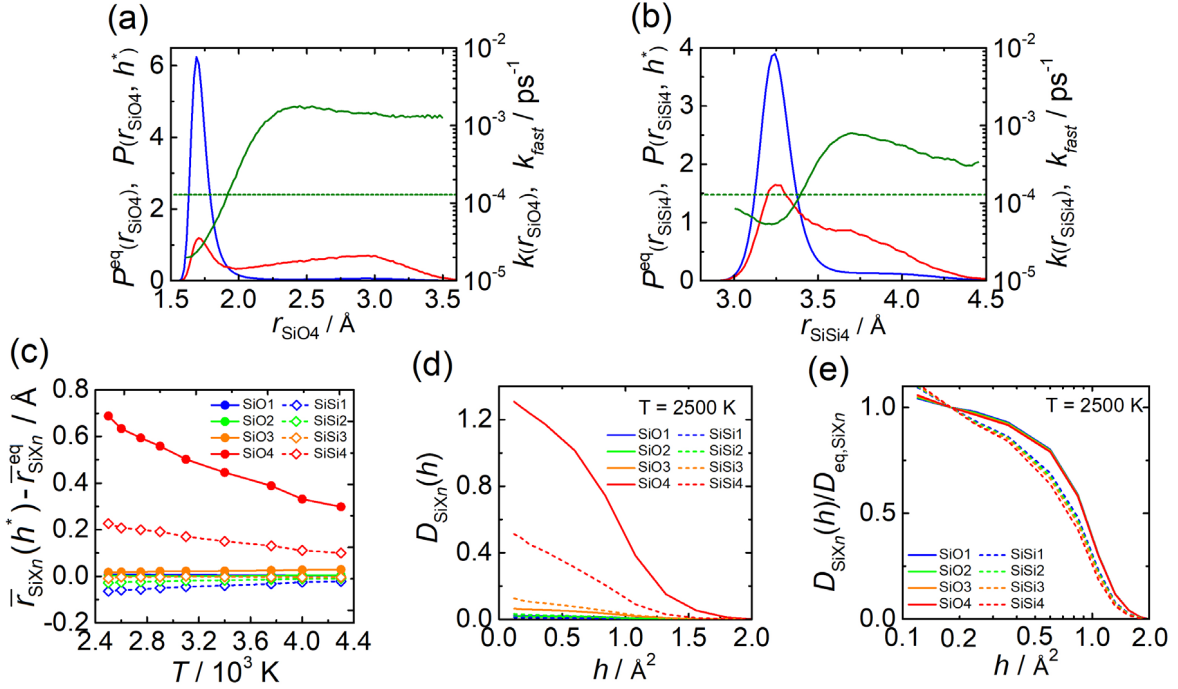

**FIG. 7.** (a) Distributions of the distance between a Si atom and its fourth-nearest O neighbor, $r_{SiO4}$, in equilibrium (blue) and for a jumping Si at the hop threshold $h^*$ (red) at $T = 2500$ K. The green curve and green dashed line represent $k(r_{SiO4})$ and $k_{fast}$, respectively. (b) Distributions of the distance between a Si atom and its fourth-nearest Si neighbor, $r_{SiSi4}$, in equilibrium (blue) and for a jumping Si at the hop threshold $h^*$ (red) at $T = 2500$ K. The green curve and green dashed line represent $k(r_{SiSi4})$ and $k_{fast}$, respectively. (c) Temperature dependence of the difference between the average distances at equilibrium and at $h^*$ for Si$Xn$. (d) KL divergence $D_{SiXn}(h)$ and (e) scaled KL divergence $D_{SiXn}(h)/D_{eq,SiXn}$ at $T = 2500$ K. In (c)-(e), $X \in \{O, Si\}$ and $n \leq 4$.

In conventional rate descriptions, such as transition state theory, the reaction coordinate is assumed to evolve on a timescale much slower than all other relevant degrees of freedom.[107] This separation of timescales results in a constant rate and an exponential survival probability. In viscous silica melts, below $T = 4300$ K, certain structural variables evolve on timescales comparable to the jump dynamics, producing pronounced rate fluctuations and deviations from exponential decay. Following the KL divergence analysis above, we examined the dependence of the jump rate on $r_{SiOn}$ and $r_{SiSin}$ for $1 \leq n \leq 8$. The most pronounced modulation is observed for $n = 4$ in the oxygen sub-shell [solid green line in Fig. 7(a)], where the ratio between the rate at the average distance $\bar{r}_{SiO4}(h^*)$ and that at the equilibrium average $\bar{r}_{SiO4}^{eq}$ is approximately 36, largely exceeding the corresponding ratios for all other $n$ [Figs. 7(a), 7(b), and Table S6]. Together with the $D_{SiXn}(h)$ analysis, these results point to $r_{SiO4}$ as the most relevant structural parameter for characterizing dynamic disorder in silicon jumps.

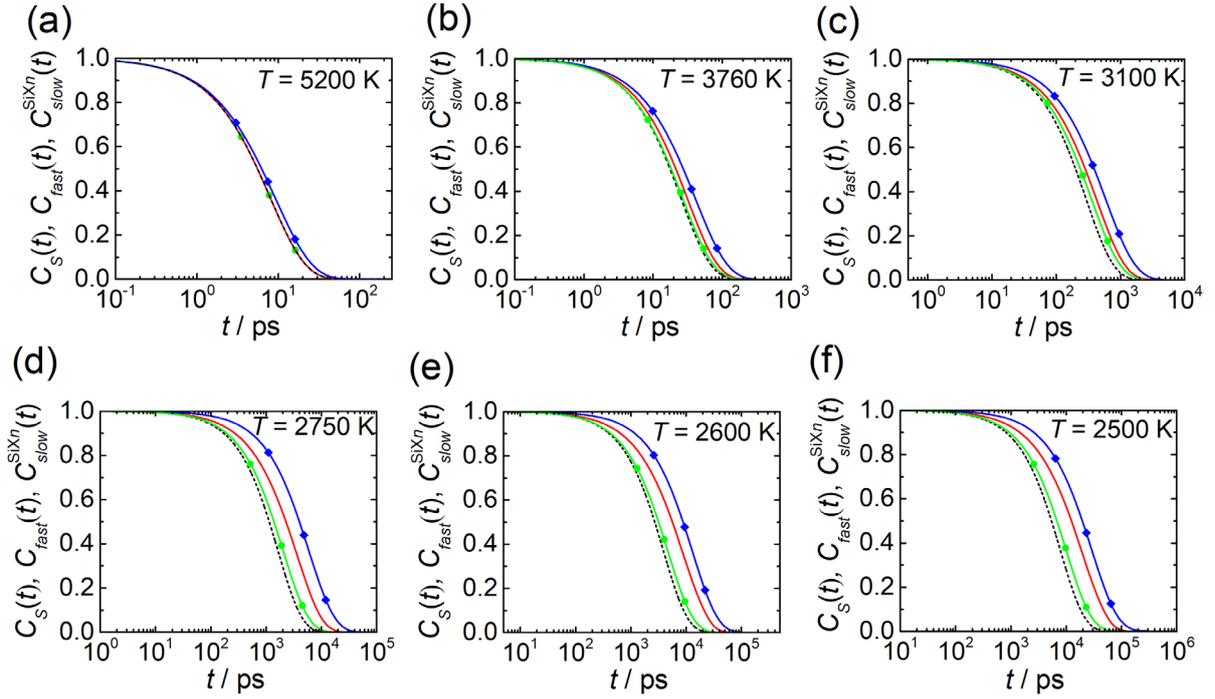

**FIG. 8.** Survival probability for the Si cage state, $C_S(t)$, its fast-fluctuation limit, $C_{fast}(t)$, and the slow-fluctuation limits $C_{slow}^{SiXn}(t)$ (with $X \in \{O, Si\}$) in the BKS model at (a) $T = 5200$, (b) 3760, (c) 3100, (d) 2750, (e) 2600, and (f) 2500 K. The red curve, black dashed curve, blue curve with squares, and green curve with circles represent $C_S(t)$, $C_{fast}(t)$, $C_{slow}^{SiO4}(t)$, and $C_{slow}^{SiSi4}(t)$, respectively.

We next examined the role of $r_{SiO4}$ as an additional slow degree of freedom by analyzing the survival probability in the slow fluctuation limit, $C_{slow}^{SiO4}(t)$, where $r_{SiO4}$ is treated as a slow variable competing with the hop function $h$ (see Sec. II.C). At higher temperatures ($T \geq 4300$ K), $C_{fast}(t)$ already provides a close approximation to $C_S(t)$, while $C_{slow}^{SiO4}(t)$ decays more slowly than $C_S(t)$ [Fig. 8(a)], indicating that $r_{SiO4}$ still fluctuates faster than the jump dynamics and that $h$ alone remains an adequate reaction coordinate. Upon cooling, $C_S(t)$ becomes progressively slower and approaches $C_{slow}^{SiO4}(t)$ [Figs. 8(b)-8(e)], signifying that the influence of $r_{SiO4}$ on the jump rate grows and that a one-dimensional description in terms of $h$ becomes increasingly inadequate. At the lowest temperature studied (2500 K), $C_{slow}^{SiO4}(t)$ remains slightly slower than $C_S(t)$ [Fig. 8(f)], demonstrating that $r_{SiO4}$ captures the dominant slow structural contribution over the range accessible in the BKS model and provides a nearly converged description of the jump dynamics. However, even at $T = 2500$ K, a small but discernible difference between $C_{slow}^{SiO4}(t)$ and the actual survival probability $C_S(t)$ indicates that $r_{SiO4}$ is not a completely slow variable, and it continues to fluctuate slightly faster than the jump dynamics.

To clarify whether the residual difference is due to the limited temperature range studied with the BKS model or reflects an intrinsic feature of the dynamics, we sought to extend the survival probability analysis to lower temperatures. However, calculations of survival probabilities and jump dynamics at such low temperatures were not feasible with the BKS model because sampling of long trajectories is computationally prohibitive. Therefore, we turned to the CP model, whose short-range potential (without long-range Coulombic interactions) makes it computationally feasible to perform survival probability and jump dynamics analysis at substantially lower reduced temperatures (for reference, $T = 0.39$ in CP corresponds approximately to 2750 K in BKS based on static structure, while the lowest CP temperature studied, $T = 0.29$, corresponds to about 2040 K in approximate mapping between the two models). For $T \geq 0.45$, the CP model exhibits the same high-temperature behavior as BKS: $C_{fast}(t)$ follows $C_S(t)$, while $C_{slow}^{SiO4}(t)$ decays more slowly [Fig. S20(a)]. With further cooling ($T = 0.42$-$0.30$), $C_S(t)$ progressively approaches $C_{slow}^{SiO4}(t)$ [Figs. S20(b)-S20(e)], indicating the growing influence of $r_{SiO4}$ on the jump dynamics, consistent with the behavior observed in BKS. At the lowest temperature studied ($T = 0.29$), $C_{slow}^{SiO4}(t)$ follows $C_S(t)$ closely, although it decays slightly faster for t $\leq 1.3\times10^2$ [Fig. 9(a)]. This residual short-time discrepancy indicates that $r_{SiO4}$ alone does not fully capture the slow structural degrees of freedom governing the jump dynamics. To address this, we examined a combined slow-variable description, $C_{slow}^{(SiO4-SiSi4)}(t)$, in which both $r_{SiO4}$ and $r_{SiSi4}$ are treated as slow coordinates alongside $h$ [Fig. 9 (b)]. This multi-variable formulation provides a better description at short times, exhibits a slower overall decay than $C_{slow}^{SiO4}(t)$, and also outperforms other combinations $C_{slow}^{(SiO4-SiSin)}(t)$ with $1 \leq n \leq 3$. The enhanced slowdown observed for the $r_{SiO4}$-$r_{SiSi4}$ pair compared to other $r_{SiO4}$-$r_{SiSin}$ pairs arises from their cooperative fluctuations, whereas correlations between $r_{SiO4}$ and $r_{SiSin}$ ($1 \leq n \leq 3$) are comparatively weak. This cooperative coupling highlights the increased dimensionality of the jump dynamics of silicon upon cooling. Nevertheless, incorporating $r_{SiO4}$, or even both $r_{SiO4}$ and $r_{SiSi4}$, alongside the hop function improves the description but still fails to reproduce the survival probability across the entire time range. A schematic illustration of this cooperative rearrangement involving both $r_{SiO4}$ and $r_{SiSi4}$ during a silicon jump is provided in Fig. S21. These findings establish $r_{SiO4}$, and at the lowest temperatures also $r_{SiSi4}$, as the structural variables most strongly linked to dynamic disorder in silicon jumps.

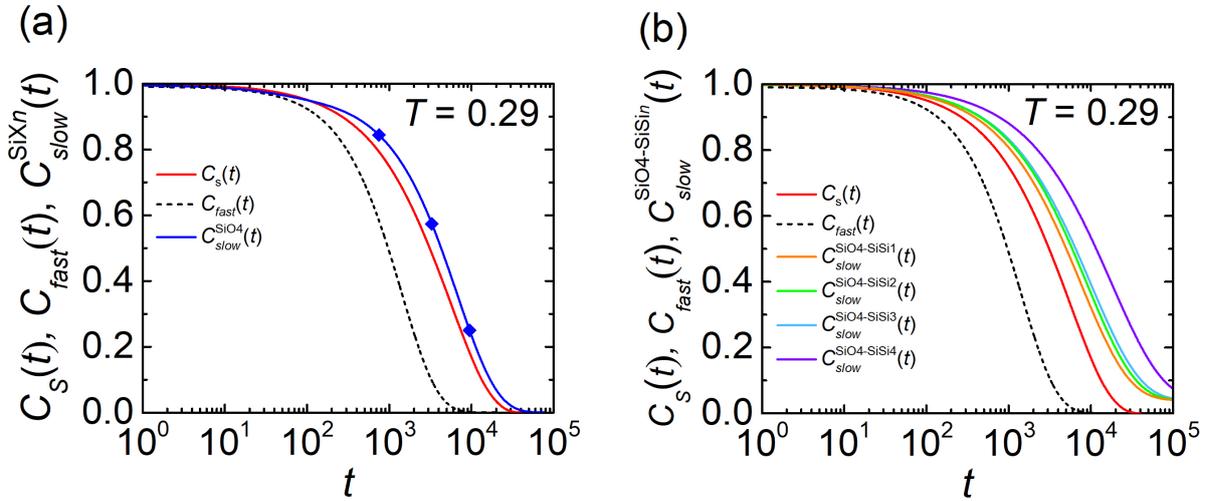

**FIG. 9.** Survival probability for the Si cage state, $C_S(t)$, and its fast- and slow-fluctuation limits at $T = 0.29$ in the CP model. The solid red and dashed black curves represent $C_S(t)$ and $C_{fast}(t)$, respectively. (a) Comparison with the slow-fluctuation limit determined by the fourth-nearest oxygen neighbor, $C_{slow}^{SiO4}(t)$ (blue with diamonds). (b) Comparison with the slow-fluctuation limits constructed from the joint variables $C_{slow}^{(SiO4-SiSin)}(t)$ with $n$ = 1, 2, 3, and 4, shown as orange, green, light blue, and violet curves, respectively.

**(ii) For oxygen**

Following the analysis of dynamic disorder in silicon jumps, we carried out a corresponding examination of oxygen jumps, focusing on the structural rearrangements of the surrounding silicon and oxygen neighbors. We analyzed the distributions of distances between a jumping oxygen atom and its $n$th nearest silicon and oxygen neighbors, $P(r_{OSin}, h^*)$ and $P(r_{OOn}, h^*)$ ($1 \leq n \leq 8$), at the hop threshold $h^*$ and compared them with the corresponding equilibrium distributions, $P^{eq}(r_{OSin})$ and $P^{eq}(r_{OOn})$ [Figs. 10(a), 10(b), and S23]. At $h^*$, the distributions for the first-shell silicon neighbors ($n \leq 2$) shift to longer distances, with the most significant change observed for $r_{OSi2}$, indicating an expansion of the immediate silicon coordination shell preceding the jump [Fig. 10(c)]. Silicon neighbors beyond the first shell ($n \geq 3$) exhibit only minor deviations from equilibrium, suggesting weaker coupling to the local cage rearrangement. The oxygen neighbor distributions show the most pronounced outward shifts for $n = 4$, 5, and 6, reflecting substantial rearrangements within the second oxygen shell that create a local environment facilitating the displacement of the second-nearest silicon.

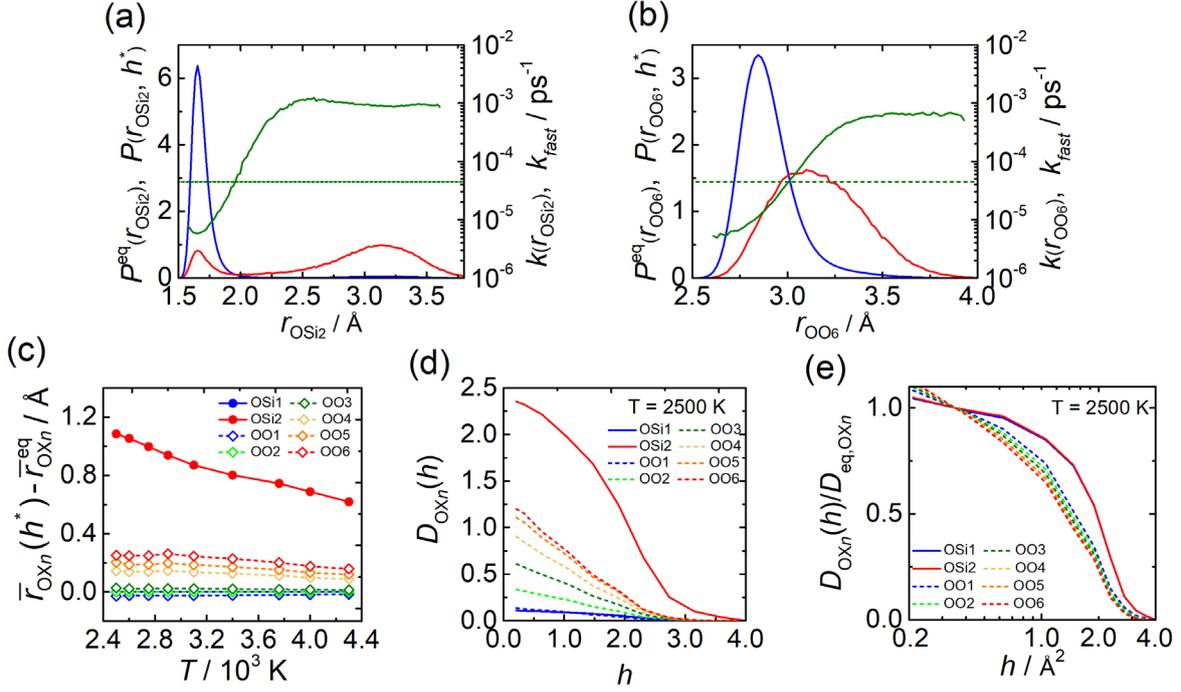

**FIG. 10.** (a) Distributions of the distance between an O atom and its second-nearest Si neighbor, $r_{OSi2}$, in equilibrium (blue) and for a jumping O at the hop threshold $h^*$ (red) at $T = 2500$ K. The green curve and green dashed line represent $k(r_{OSi2})$ and $k_{fast}$, respectively. (b) Distributions of the distance between an O atom and its sixth-nearest O neighbor, $r_{OO6}$, in equilibrium (blue) and for a jumping O at the hop threshold $h^*$ (red) at $T = 2500$ K. The green curve and green dashed line represent $k(r_{OO6})$ and $k_{fast}$, respectively. (c) Temperature dependence of the difference between the average distances at equilibrium and at $h^*$ for OX$n$. (d) KL divergence $D_{OXn}(h)$ and (e) scaled KL divergence $D_{OXn}(h)/D_{eq,OXn}$ at $T = 2500$ K. In (c)-(e), $X \in \{Si, O\}$; for X = Si, $n \leq 2$; for X = O, $n \leq 6$.

The KL divergence analysis $D_{OSin}(h)$ and $D_{OOn}(h)$ [Fig. 10(d)] shows that the second-nearest silicon neighbor, $r_{OSi2}$, undergoes the most pronounced reorganization as the oxygen atom approaches a jump. In the oxygen shell, $D_{OOn}(h)$ attains substantial values for $n = 4$, 5, and 6, consistent with the pronounced shifts in their distance distributions. The scaled KL divergence $D_{OXn}(h)/D_{eq,OXn}$ further reveals that the approach of $P(r_{OSin},h)$ toward $P(r_{OSin},h^*)$ is consistently slower than that of $P(r_{OOn},h)$ toward $P(r_{OOn},h^*)$ [Fig. 10(e)]. Since two silicon atoms constitute the first coordination shell around oxygen, this indicates that the displacements of these inner silicon neighbors occur within a local environment that is already being reorganized by the faster structural adjustments of the surrounding oxygen atoms in the second shell. The jump-rate analysis shows that modulation of the oxygen jump rate is governed predominantly by $r_{OSi2}$, for which the rate at the average distance $\bar{r}_{OSi2}(h^*)$ exceeds that at the equilibrium average $\bar{r}_{OSi2}^{eq}$ by nearly a factor of 145, substantially larger than for any other neighbor [Figures 10(a), 10(b), and Table S7]. These analyses indicate $r_{OSi2}$ as the most relevant structural parameter associated with dynamic disorder in oxygen jumps.

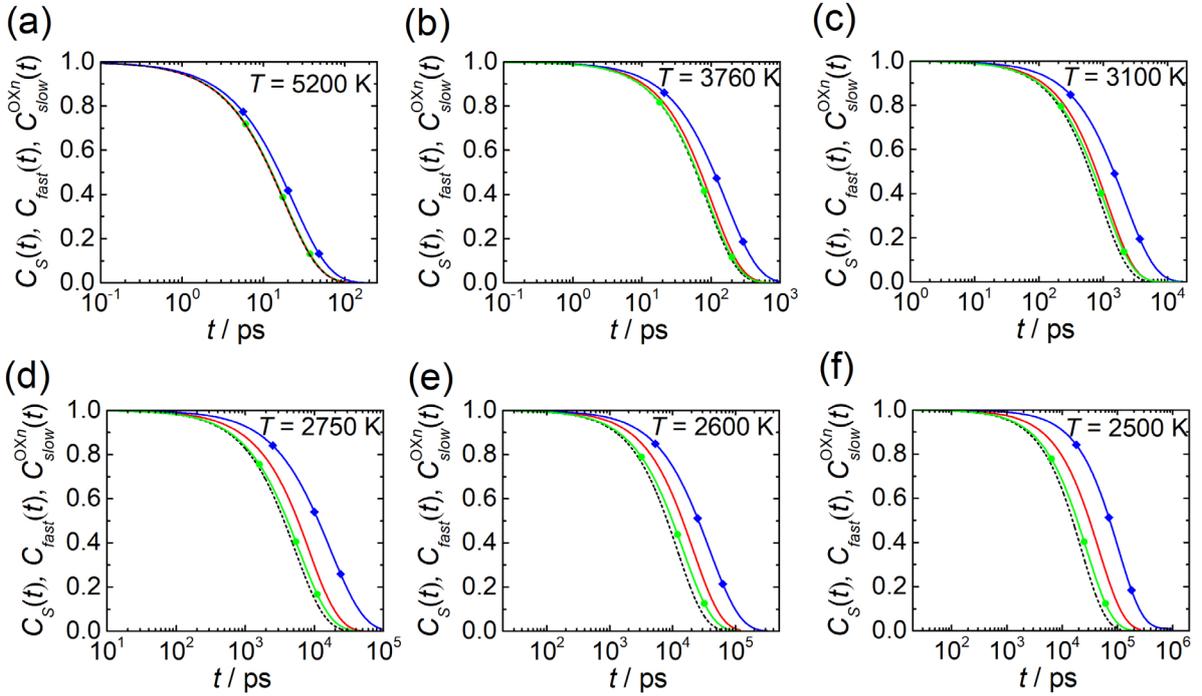

**FIG. 11.** Survival probability for the O cage state, $C_S(t)$, its fast-fluctuation limit, $C_{fast}(t)$, and the slow-fluctuation limits $C_{slow}^{OXn}(t)$ (with $X \in \{Si, O\}$) for O-centered substates at (a) $T = 5200$, (b) 3760, (c) 3100, (d) 2750, (e) 2600, and (f) 2500 K. The red curve, black dashed curve, blue curve with squares, and green curve with circles represent $C_S(t)$, $C_{fast}(t)$, $C_{slow}^{OSi2}(t)$, and $C_{slow}^{OO6}(t)$, respectively.

We next assessed the role of $r_{OSi2}$ as a potential slow variable by analyzing the survival probability in the slow fluctuation limit, $C_{slow}^{OSi2}(t)$, where $r_{OSi2}$ is treated as a competing coordinate to the hop function $h$. At higher temperatures ($T \geq 4300$ K), $C_{fast}(t)$ closely tracks $C_S(t)$, indicating that $h$ alone captures the oxygen jump dynamics. In this regime, $C_{slow}^{OSi2}(t)$ decays more slowly than $C_S(t)$ [Fig. 11(a)]. Upon cooling, $C_S(t)$ gradually shifts toward $C_{slow}^{OSi2}(t)$, highlighting the growing relevance of the second-nearest silicon neighbor in regulating oxygen jumps [Figs. 11(b)-11(e)]. At the lowest temperature studied ($T = 2500$ K), the two functions approach each other but still do not coincide Fig. 11(f)]. Similar to silicon, this residual difference may reflect the restricted temperature range of BKS. We therefore examined the CP model, which extends the analysis to lower temperatures. In CP, the overall behavior follows that of BKS: upon cooling, $C_S(t)$ progressively shifts toward $C_{slow}^{OSi2}(t)$ [Figs. S25(a)-S25(e)]. However, unlike silicon, even at the lowest temperature studied ($T = 0.29$), $C_{slow}^{OSi2}(t)$ still decays more slowly than $C_S(t)$ [Fig. S25(f)]. This indicates that no additional slow variable is required, as incorporating further neighbors only drives the slow-limit away from $C_S(t)$ (Fig. S26). These results show that oxygen jumps are effectively characterized by a two-variable description involving $h$ and $r_{OSi2}$, in contrast to silicon jumps, which require additional structural coordinates. As discussed above, the second-shell oxygens $r_{OOn}$ ($n = 4$, 5, and 6) also undergo pronounced displacement as the system approaches a jump; however, these rearrangements relax on timescales faster than the survival probability, preventing them from acting as rate-controlling variables. Taken together, these findings highlight a fundamental difference: silicon jumps involve strongly cooperative rearrangements of both the fourth-nearest oxygen and silicon neighbors, whereas oxygen jumps are governed by a more localized motion dominated by the modulation from $r_{OSi2}$. This asymmetry is consistent with potential energy landscape analyses, which attribute the decoupling of oxygen and silicon dynamics to distinct relaxation pathways on the landscape.[41] A schematic illustration of the oxygen jump mechanism is provided in Fig. S27.

In addition, we considered the hop-dependent angular distribution functions as candidate slow variables: $f_{SiSiSi}(\theta)$ and $f_{OSiO}(\theta)$ for silicon jumps, and $f_{OOO}(\theta)$ and $f_{SiOSi}(\theta)$ for oxygen jumps. Although the equilibrium and $h^*$ distributions show discernible deviations, the corresponding survival probability functions exhibit negligible separation between the fast- and slow-fluctuation limits (Fig. S28). An exception is $f_{SiOSi}(\theta)$, which produces a modest difference between the fast- and slow-fluctuation limits; however, its slow-limit function still decays faster than $C_S(t)$. This suggests that angular correlations play only a marginal role in dynamic disorder, reinforcing the notion that atomic displacements constitute the primary slow variables governing jump dynamics.

### (iii) Comparison with water

Both water and silica are tetrahedral network-forming liquids, yet the nature and microscopic origin of their dynamic disorder differ fundamentally. A key distinction lies in the degree of temporal heterogeneity in jump dynamics: silica shows only moderate heterogeneity, with $\psi_{cage}(t)$ distributions that are long-tailed but relatively narrow at low temperatures, whereas water in its fragile regime exhibits broader and more pronounced $\psi_{cage}(t)$ distributions,[26] reflecting markedly stronger dynamic heterogeneity. In supercooled water, this pronounced heterogeneity arises from distinct microscopic mechanisms. Previous analyses have shown that the displacements of the fourth-nearest oxygen neighbor, $r_{OO4}$, act as the dominant slow variable governing jump dynamics by confining molecules within extended low-density (LDL-like) regions.[26] At lower temperatures (e.g., 197 K), the fifth-nearest oxygen, $r_{OO5}$, also becomes relevant, reflecting enhanced cooperativity and the influence of LDL–HDL fluctuations on jump dynamics.[26,44]

In silica, our results reveal a clear separation of roles between species. Silicon jumps are primarily governed by the displacements of the fourth-nearest oxygens, $r_{SiO4}$. Upon cooling (e.g., $T = 0.29$), the displacement of the fourth-nearest silicon, $r_{SiSi4}$, also contributes, reflecting the emergence of additional cooperative constraints analogous to those observed in water. Jumps of oxygen atoms, however, are modulated by the displacements of the second-nearest silicon atom, $r_{OSi2}$, which remain comparatively localized and involve weaker cooperativity. *Thus, unlike water, where a single class of neighbor ($r_{OO4}$, and later $r_{OO5}$) dominates the slow variables, silica exhibits species-specific mechanisms of rate modulation.* These differences arise from the rigidity of the silicate network, which enforces asymmetric contributions of silicon and oxygen to cooperative rearrangements. These findings emphasize that while both systems share a tetrahedral backbone, the microscopic variables governing dynamic disorder and cooperativity differ markedly between water and silica. This naturally motivates a quantitative assessment of the spatial extent of such cooperative effects in silica, which we address through point-to-set (PTS) correlation analysis in the following section.

### E. Cooperative length scales from point-to-set correlations

To quantify the spatial extent of cooperative effects underlying dynamic disorder, we computed the point-to-set (PTS) correlation length $\xi_{PTS}$ from cubic cavity simulations[87] (Sec. II.E). The overlap function was analyzed in two ways: (i) considering all particles to obtain the total correlation length, and (ii) considering only silicon or oxygen atoms to resolve species-specific cooperativity (Fig. S29). In all cases, the asymptotic value of the overlap was taken only after the corresponding self-component had fully decayed, ensuring that $\xi_{PTS}$ reflects static structural correlations rather than residual self-memory. Since PTS analysis requires multiple independent equilibrations over very long timescales of partially frozen systems at low temperatures, the calculations are particularly demanding with the long-ranged BKS potential. We therefore performed them using the short-ranged CP model. Given that BKS and CP exhibit nearly identical structural and dynamical trends in our other analyses, this choice allows us to capture the essential static correlations without loss of generality.

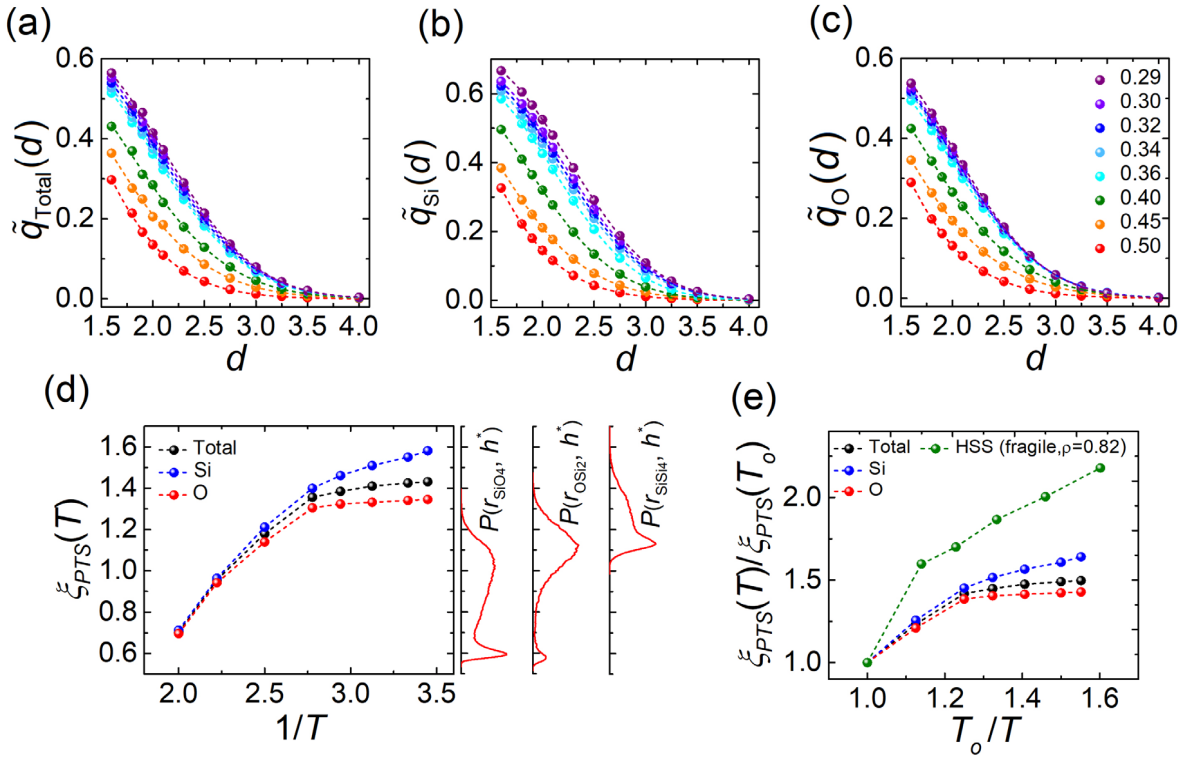

**FIG. 12.** Excess overlap $\tilde{q}(d)$ for (a) Total, (b) Si, and (c) O, respectively, at the indicated temperatures (legend in (c)); dashed curves are fits to Eq. (21). Here $d$ denotes the confining length (half-edge) of the cubic cavity. (d) Static point-to-set correlation length $\xi_{PTS}$ extracted from the fits in (a)-(c), plotted versus $1/T$ for total (black), Si (blue), and O (red); dashed lines are guides to the eye. Side panels show representative neighbor-resolved distributions at $h^*$ [$P(r_{SiO4}, h^*)$, $P(r_{OSi2}, h^*)$, and $P(r_{SiSi4}, h^*)$] for $T = 0.29$, which identify the neighbor shells progressively encompassed as $\xi_{PTS}$ grows on cooling. (e) Normalized point-to-set correlation length $\xi_{PTS}/\xi_{PTS}(T_{onset})$ versus $T_{onset}/T$, with $T_{onset} = 0.45$ for the CP model. Results for silica (this work) are compared with the harmonic soft-sphere (HSS) model in its fragile (high density, $\rho = 0.82$) regime, with HSS data taken from Ref. [90]. The growth in silica is substantially weaker than in the fragile HSS benchmark.

Figs. 12 (a)-12(c) depict the confinement-length $d$ dependence of the excess overlap, i.e., the asymptotic value of the overlap difference $\tilde{q}(d) = Q_\infty(d) - Q_{rand}$, at different temperatures for the total, silicon, and oxygen components, together with exponential fits to Eq. (21). From these fits, we obtained the correlation length $\xi_{PTS}$ as well as the stretching exponent $\eta$. The total PTS length exhibits only a modest increase upon cooling (Fig. 12 (d)), indicating little to no growth of amorphous order. This behavior is characteristic of a strong glass former and contrasts with fragile liquids, where $\xi_{PTS}$ exhibits a pronounced growth with decreasing temperature.[88,90] Analysis of the individual atomic components reveals a clear asymmetry between silicon and oxygen. For silicon, $\xi_{PTS}$ grows systematically with decreasing temperature, indicating a slight enhancement of cooperative structural correlations. In contrast, $\xi_{PTS}$ for oxygen essentially saturates at a constant value below $T = 0.36$. Thus, while silicon atoms progressively require larger cooperative rearrangements at low temperatures, oxygen dynamics remain comparatively localized. We further verified these trends using random pinning protocols, which yielded consistent results. It is to be noted that the growth of $\xi_{PTS}$ for silicon remains modest compared with fragile liquids. To highlight this difference, we normalized $\xi_{PTS}$ by its value at the onset temperature, $T_{onset} = 0.45$, and compared it with the harmonic soft-sphere (HSS) model in its fragile regime (high density, $\rho = 0.82$).[90] In contrast to the sharp increase observed in the fragile regime of the HSS model, the growth of $\xi_{PTS}$ for silicon remains substantially weaker [Fig. 12(e)]. The stretching exponent $\eta$ is shown in Fig. S30.

At high temperatures, the PTS length $\xi_{PTS}$ [Fig. 12(d)] remains confined within the first minimum of $g_{SiO}(r)$, consistent with jump events occurring largely independently of extended neighbor displacements. Upon cooling, $\xi_{PTS}$ for silicon extends beyond $r_{SiO4}$, as identified from $P(r_{SiO4}, h^*)$ [Figs. 12(d) and S17(a)], marking the onset of correlations between a jumping silicon atom and its fourth-nearest oxygen neighbor. At the lowest simulated temperature, the correlation length extends even beyond $P(r_{SiO4}, h^*)$ to include $P(r_{SiSi4}, h^*)$ [Figs. 12(d) and S17(b)], reflecting the emergence of more collective structural correlations. In contrast, for oxygen, $\xi_{PTS}$ reaches only up to $r_{OSi2}$, identified from $P(r_{OSi2}, h^*)$ [Figs. 12(d) and S22(a)], and exhibits no further increase upon cooling. This species-dependent asymmetry parallels the microscopic origin of dynamic disorder discussed earlier, with silicon dynamics becoming increasingly cooperative, whereas oxygen motion remains comparatively localized.

To directly assess the role of the slow structural variables identified earlier in governing jump motions, we performed constrained cavity simulations using configurations selected immediately before a hopping event. In each case, the particle about to hop was designated as the cavity center, and a cubic cavity of size $2d = 5.5$ was defined around it. Particles outside the cavity were frozen to mimic a fixed environment, while those inside the cavity were allowed to evolve dynamically. The cavity size was chosen to suppress extended bulk-like rearrangements while retaining sufficient space for local relaxation, including the possibility of a hop. Within this framework, we selectively immobilized specific neighbors of the central particle to isolate the effect of the slow variables. For silicon jumps, we separately froze the fourth-nearest oxygen atoms, the fourth-nearest silicon atoms, and both sets simultaneously. For oxygen jumps, we froze the second-nearest silicon atoms, the sixth-nearest oxygen atoms, and a combined set of the second silicon and sixth oxygen neighbors. To ensure that the observed effects were not simply due to the immobilization of nearby atoms, we repeated the simulations by freezing other neighbors of the same species identified in the pre-hop configuration. When two slow variables were frozen simultaneously, the comparison was made with cases where two other neighbors of the same species at comparable distances were frozen. This analysis was carried out at $T = 0.32$ and $T = 0.29$, where the role of slow variables is expected to be most pronounced. For each analysis, 120–150 independent configurations of about-to-hop particles were considered to obtain reliable averages.

We then calculated the species-resolved overlap functions under these constrained conditions to quantify the influence of the frozen neighbors on local relaxation. The results are shown in Figs. 13 (a)-13(d). For silicon, at $T = 0.32$, freezing the fourth-nearest silicon neighbor results in an overlap function that saturates at a substantially higher asymptotic value compared with cases where no neighbors or other silicon neighbors are frozen [Fig. 13 (a)]. When the fourth-nearest oxygen neighbor is frozen, the overlap function shows almost no relaxation beyond the initial drop, underscoring its dominant role in constraining the hopping event and thereby suppressing the relaxation process. Freezing both slow-variable sets together produces behavior very close to that obtained by freezing the fourth nearest oxygen neighbor alone. At $T = 0.29$, however, the difference between the two cases becomes more pronounced [Fig. 13(b)], indicating that the joint influence of the two slow variables strengthens upon cooling. For oxygen, freezing the second-nearest silicon neighbor leads to an overlap function that saturates at a markedly higher value than in cases where no neighbors or other silicon atoms are frozen [Fig. 13 (c)]. Freezing the sixth-nearest oxygen neighbor (and likewise the fourth or fifth) also enhances the long-time overlap, but the effect remains weaker than that of the second silicon neighbor. When both sets of neighbors are frozen simultaneously, the overlap function lies very close to that obtained by freezing the second silicon neighbor alone. At $T = 0.29$, the similarity between these two cases persists, underscoring that the dominant structural constraint on oxygen jumps originates from the second-nearest silicon neighbor, while the contribution of the sixth oxygen neighbor remains comparatively minor [Fig. 13 (d)].

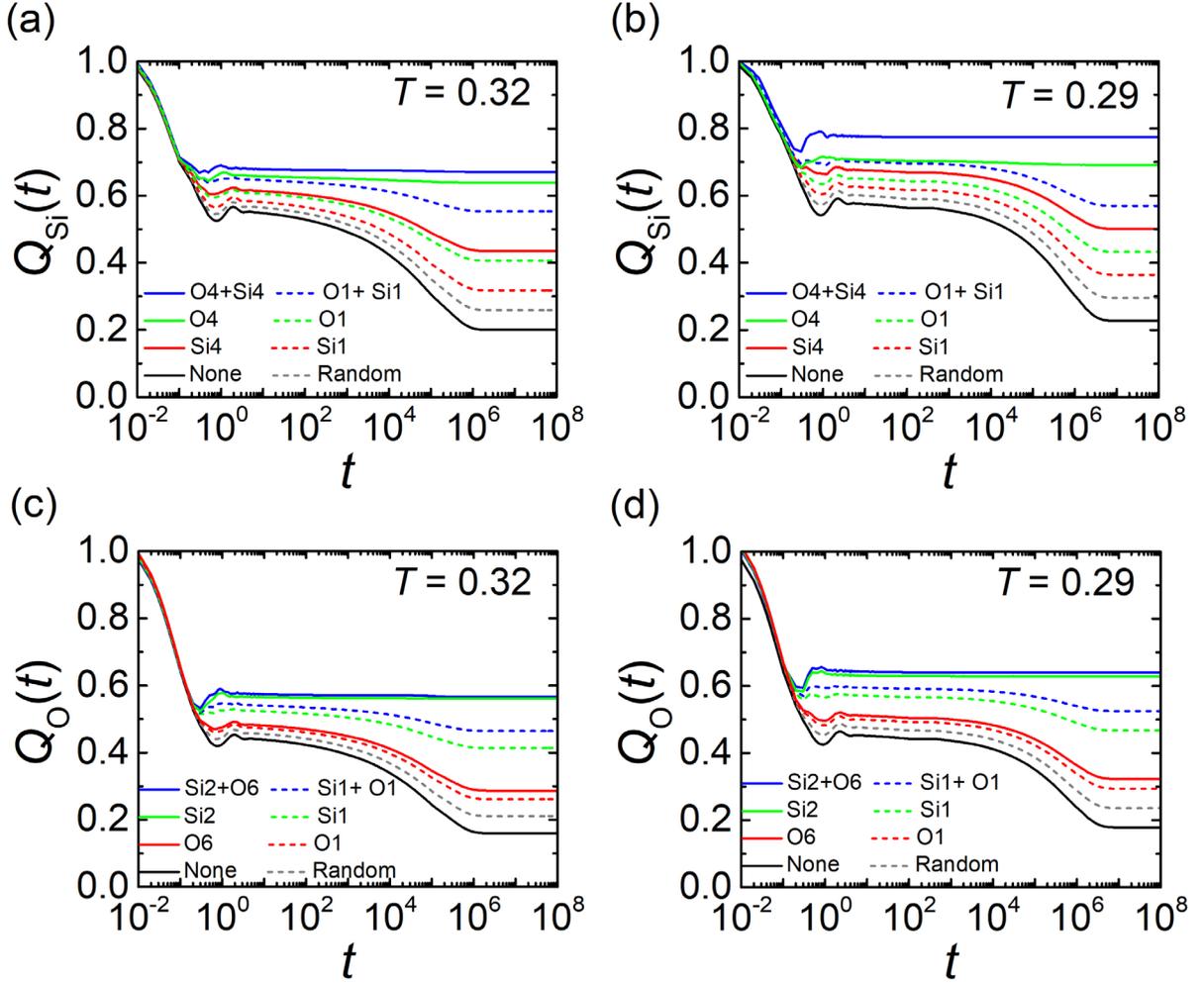

**FIG. 13.** Time-dependent overlap in cavity simulations with selective freezing. Si-centered cavity at (a) $T=0.32$ and (b) 0.29: $Q_{Si}(t)$ for the frozen neighbor sets None, O4, Si4, O4+Si4, O1, Si1, O1+Si1, and Random (size-matched control). O-centered cavity at (c) $T = 0.32$ and (d) 0.29: $Q_O(t)$ for the frozen neighbor sets None, Si2, O6, Si2+O6, Si1, O1, Si1+O1, and Random. Here, O$n$/Si$n$ denote the $n$th nearest O/Si to the tagged central atom. Random freezes a single randomly chosen particle located outside the first coordination shell. Freezing is applied at $t = 0$ (the last pre-jump frame, identified via $h^*$); only unfrozen particles are propagated thereafter. A cubic cavity with an edge length of $2d = 5.5$ is used in all panels. Each curve represents an average of 125–150 independent realizations.

These cavity simulations directly confirm that the neighbors identified as slow variables govern jump dynamics. For silicon, both the fourth-nearest oxygen and silicon neighbors impose collective constraints that intensify upon cooling, whereas for oxygen, the dominant influence arises from the second-nearest silicon atoms. These results reinforce the connection to dynamic disorder, where silicon dynamics require collective rearrangements, in contrast to the more localized motion of oxygen.

## IV. CONCLUDING REMARKS

In this work, we investigated the microscopic mechanisms of dynamic slowdown in viscous silica melt by analyzing jump dynamics within the framework of dynamic disorder. By characterizing residence-time distributions and survival probabilities, we showed that relaxation cannot be described by a single characteristic timescale but instead arises from fluctuating rates that evolve over time. Upon cooling, these rate fluctuations become increasingly heterogeneous, giving rise to dynamic disorder with a clear species dependence: *silicon dynamics are constrained by comparatively collective environments, whereas oxygen dynamics remain localized, indicating distinct microscopic routes to slowdown within the same network liquid.* These behaviors were consistently observed in both the BKS and CP models of silica, confirming that the microscopic picture established here is robust against the choice of interaction potential.

The species-dependent nature of this dynamic disorder was elucidated by identifying slow variables associated with specific fluctuations of neighboring atoms. For silicon jumps, both the fourth-nearest oxygen ($r_{OSi4}$) and silicon ($r_{SiSi4}$) neighbors become relevant at low temperatures, whereas for oxygen jumps, the dominant contribution arises from the second-nearest silicon neighbor ($r_{OSi2}$). When these neighbors were frozen within a cavity geometry, relaxation was directly suppressed, indicating that they govern the relaxation process. The total point-to-set correlation length showed only weak growth upon cooling, consistent with the strong character of silica and the nearly Arrhenius behavior of $\tau_\alpha(T)$ as described by Eq. (19). However, species-resolved analysis revealed that the cooperative length scale for silicon increased moderately with cooling, whereas that for oxygen showed little to no growth at the lowest temperatures. This contrast suggests that silicon dynamics exhibit a relatively fragile-like tendency, while oxygen remains strong-like, leading to an overall strong character for silica. These findings demonstrate that intermittent dynamics originate from specific structural constraints.

The present analysis further establishes a direct connection between the microscopic origin of dynamic disorder in silica and theoretical frameworks of the glass transition. The survival probability exhibits pronounced non-exponential behavior and becomes intermittent upon cooling, reflecting enhanced cooperativity and an increase in the effective dimensionality of the jump dynamics. These signatures closely correspond to the predictions of dynamic facilitation theory, which posits that glassy slowdown arises from localized excitations rather than uniform relaxation.[103,104] Furthermore, the modest growth of $\xi_{PTS}$ for Si atoms upon cooling is broadly consistent with the prediction of the random first-order transition (RFOT) theory, which links the slowdown of relaxation to the emergence of static amorphous order.[1,2,91,108] In contrast, the much weaker growth of oxygen and total $\xi_{PTS}$ emphasizes that a comprehensive description of silica requires species-resolved theoretical treatments of relaxation.

In conclusion, this study establishes a microscopic framework for understanding dynamic slowdown in glass-forming liquids by linking dynamic disorder in jump dynamics with the growth of structural correlations. We demonstrated that rate fluctuations, as captured by survival probabilities $C_S(t)$, originate from specific neighbor reorganizations that are also reflected in the cooperative length scale $\xi_{PTS}$. This reveals that dynamic disorder and static correlations are two complementary manifestations of the relaxation process. A comparison with supercooled water underscores the distinct microscopic routes to slowdown in tetrahedral network liquids. In silica, dynamic disorder arises from localized neighbor constraints with species-dependent cooperativity, whereas in water, it stems from the slow displacements of fourth-nearest oxygen atoms within the hydrogen-bond network.[26] The emergence of an LDL-like structure upon cooling results in slower and more intermittent jumps, due to enhanced cooperativity. Importantly, water undergoes a temperature-dependent fragile-to-strong transition[43–45,109] whose microscopic origin remains unresolved, highlighting the need to reinterpret this phenomenon through the joint framework of dynamic disorder and static correlations. A compelling future direction is to extend this framework to liquids with systematically tunable fragility,[90,110] enabling a rigorous characterization of how disorder-driven fluctuations and cooperative length scales evolve across the fragile–strong spectrum. Extending this framework to biological and materials relevant systems, such as confined water,[111–113] conformational dynamics of proteins,[66–68,114] selective transport through ion channels,[115–117] and ion conduction in glassy solid-state electrolytes[75] offers an opportunity to uncover how disorder-driven fluctuations and cooperative correlations influence relaxation and transport. Elucidating their microscopic origins would provide deeper insight into these processes and could pave the way for a unified framework to understand complex dynamics across physical and biological domains.

## SUPPLEMENTARY MATERIAL

See the supplementary material for additional details referenced in the text.

## ACKNOWLEDGEMENTS

The present study was supported by the Grant-in-Aid for Scientific Research (JP21H04676 and JP23K17361). The calculations were partially carried out using the supercomputers at the Research Center for Computational Science in Okazaki (Projects: 24 IMS-473 C193 and 25-IMS-C223).

## AUTHOR DECLARATIONS
### Conflict of Interest

The authors have no conflicts to disclose.

## DATA AVAILABILITY STATEMENT

The data that support the findings of this study are available within the article and its Supplementary Material.

# Supplementary Material

# Dynamic Slowdown and Spatial Correlations in Viscous Silica Melt: Perspectives from Dynamic Disorder

Shubham Kumar[1], Zhiye Tang[1,2] and Shinji Saito[1,2,*]

[1]Institute for Molecular Science, Myodaiji, Okazaki, Aichi, 444-8585, Japan

[2]The Graduate University for Advanced Studies (SOKENDAI), Myodaiji, Okazaki, Aichi, 444-8585, Japan

*Author for correspondence: shinji@ims.ac.jp

**This PDF file includes:**

Figure S1 to S30

Tables S1 to S7

**Table S1:** The force field parameters for the BKS potential.

| Interaction ($\alpha$–$\beta$) | $A_{\alpha\beta}$ (eV) | $B_{\alpha\beta}$ (Å$^{-1}$) | $C_{\alpha\beta}$ (eV Å$^6$) |
|---|---|---|---|
| **Si-Si** | 0.0 | — | 0.0 |
| **Si-O** | 18003.7572 | 4.87318 | 133.5381 |
| **O-O** | 1388.7730 | 2.76000 | 175.0000 |

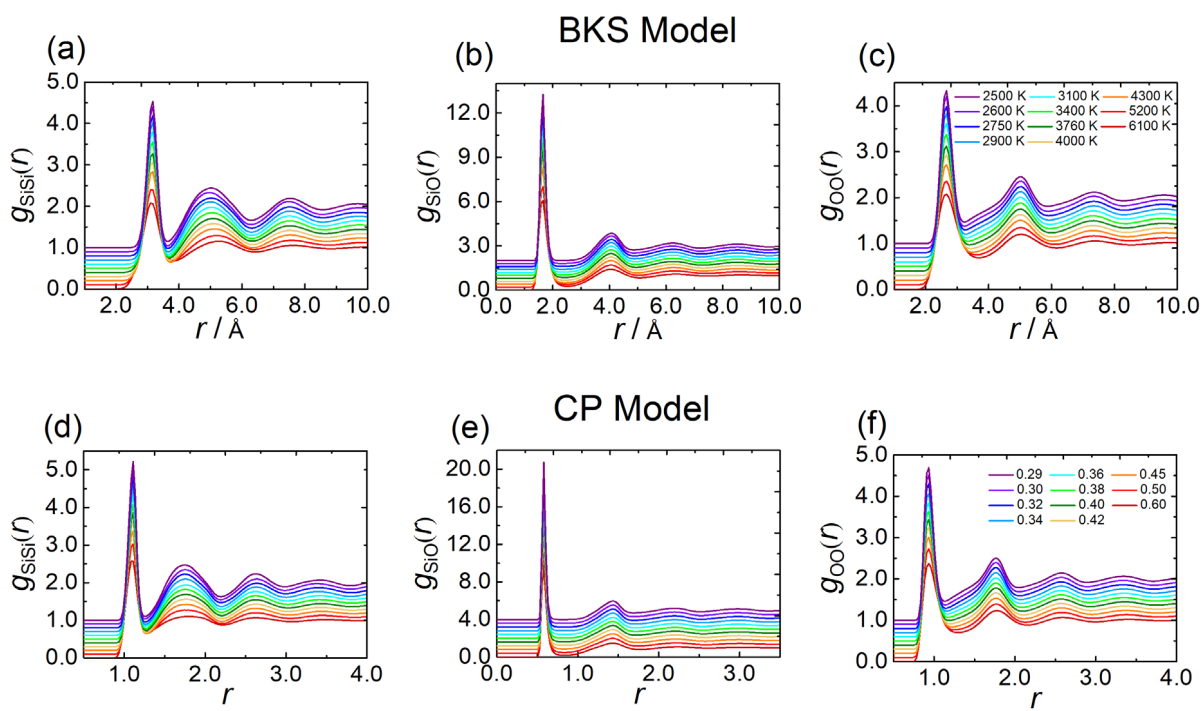

**Fig. S1.** Radial distribution functions (RDFs) of silica for the BKS [(a–c)] and CP [(d–f)] models. Panels (a, d) show $g_{SiSi}(r)$, panels (b, e) show $g_{SiO}(r)$, and panels (c, f) show $g_{OO}(r)$. Legends in panels (c) and (f) list the temperatures for the BKS and CP models, respectively. Curves are vertically offset for clarity: by 0.1 in panels (a, c, d, f), by 0.2 in panel (b), and by 0.4 in panel (e). Both models exhibit well-defined nearest-neighbor peaks with modest temperature dependence, confirming stable short-range order.

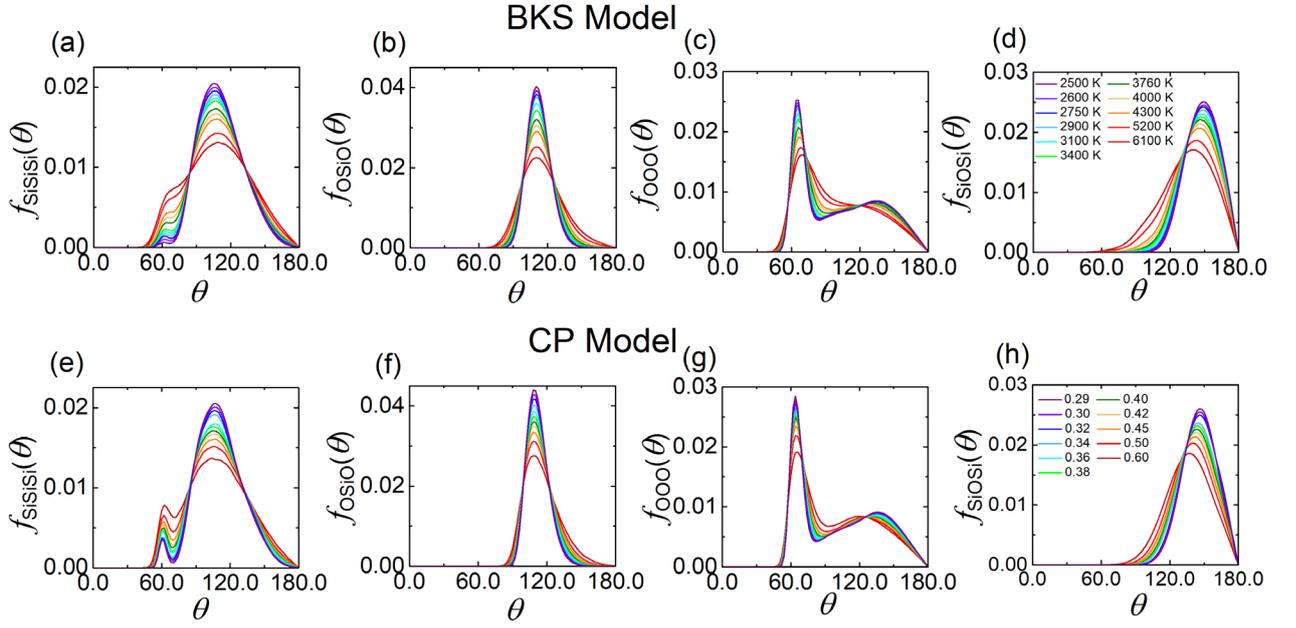

**Fig. S2.** Angular distribution functions of silica for the BKS [(a–d)] and CP [(e–h)] models. Panels (a, e) show $f_{\mathrm{SiSiSi}}(\theta)$, panels (b, f) show $f_{\mathrm{OSiO}}(\theta)$, panels (c, g) show $f_{\mathrm{OOO}}(\theta)$, and panels (d, h) show $f_{\mathrm{SiOSi}}(\theta)$. Legends in panels (d) and (h) list the temperatures for the BKS and CP models, respectively. Both models display well-defined peaks reflecting tetrahedral and network-related angular correlations, with only modest changes upon cooling.

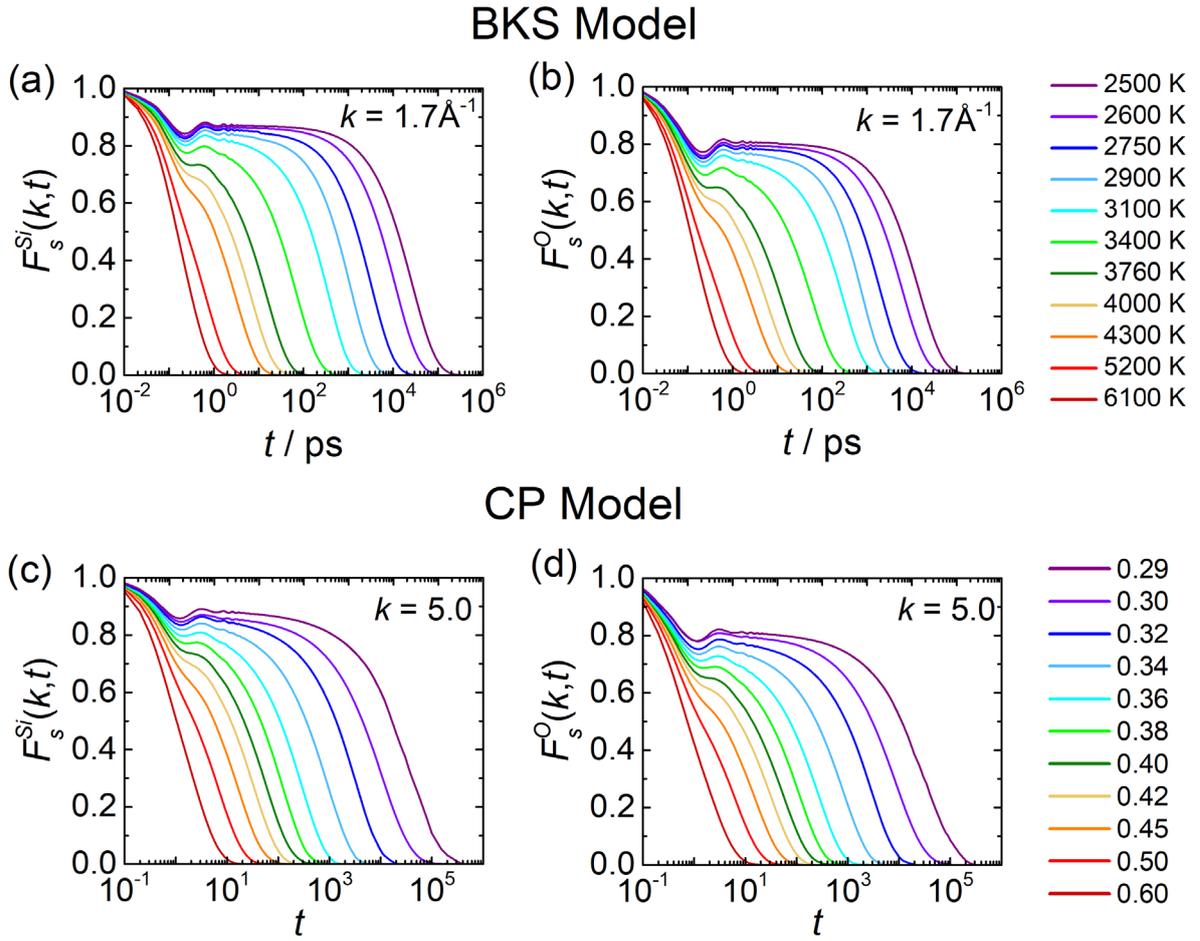

**Fig. S3.** (a,b) Self-intermediate scattering function $F_s(k,t)$ for silicon (a) and oxygen (b) atoms in silica melts simulated with the BKS model. The wavevector is chosen at $k = 1.7\,\text{Å}^{-1}$, corresponding to the first sharp peak of the static structure factor $S(k)$. (c,d) Self-intermediate scattering function $F_s(k,t)$ for silicon (c) and oxygen (d) atoms in silica melts simulated with the CP model. The wavevector is chosen at $k = 5.0$ (in reduced units).

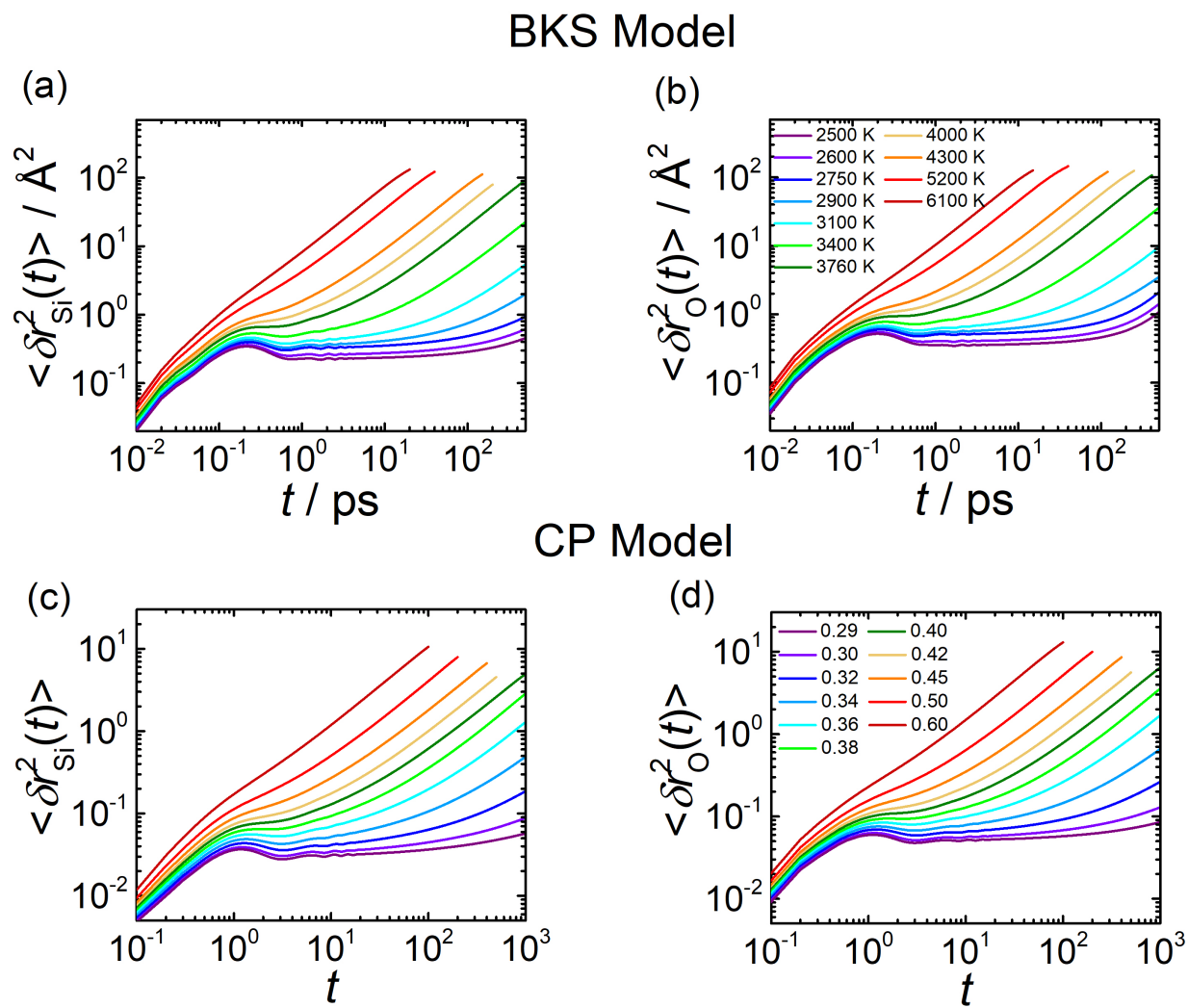

**Fig. S4.** Time evolution of the mean-squared displacement of silicon and oxygen atoms in silica melts. Panels (a, b) show results for the CP model in the range of reduced temperatures $T$ = 0.29-0.60 (legend in (b)). Panels (c, d) show results for the BKS model for $T$ = 2500-6100 K (legend in (d)).

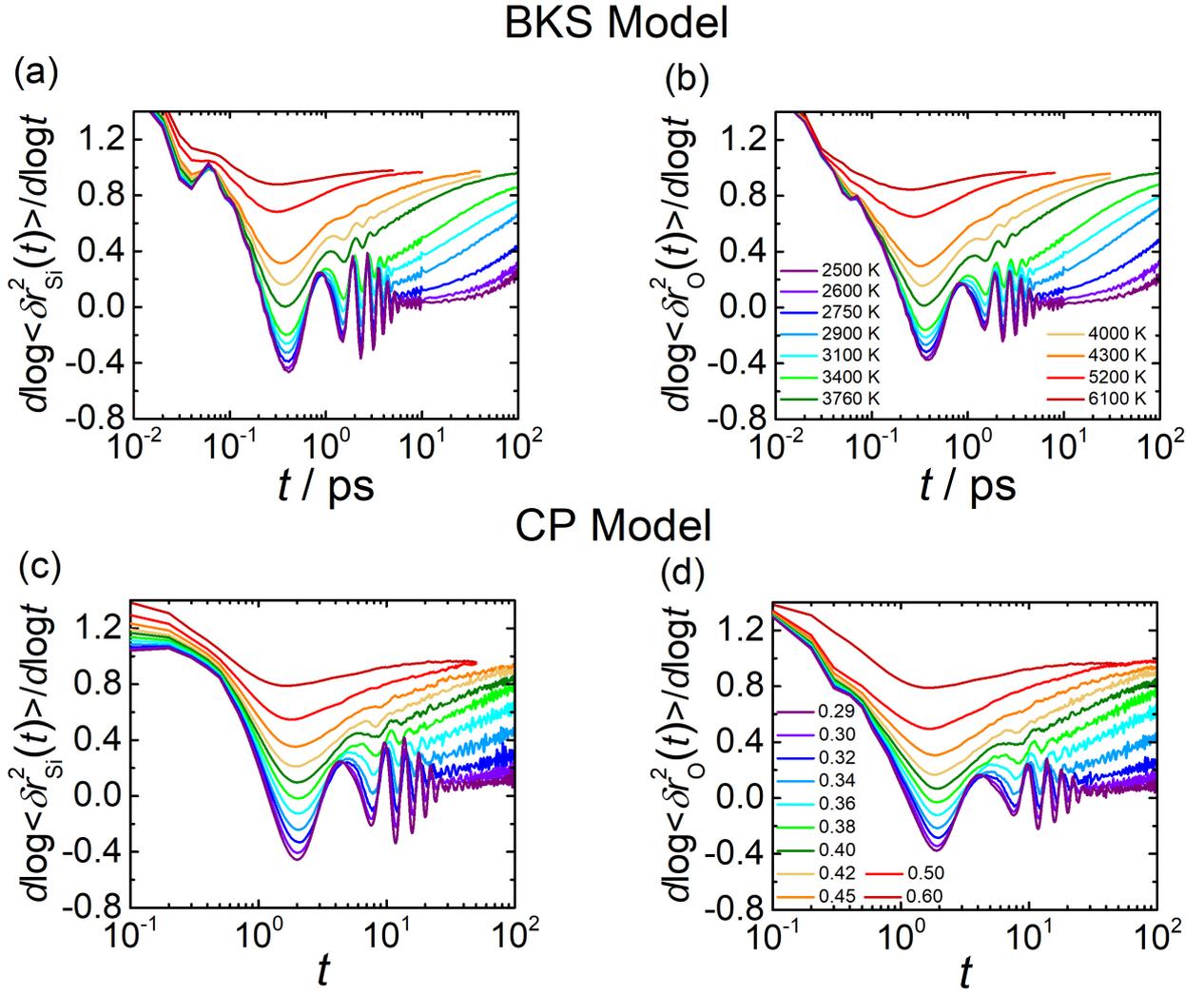

**Fig. S5.** Temperature dependence of the logarithmic derivative of the mean-squared displacement of silicon and oxygen atoms in silica melts. Panels (a, b) show results for the CP model at reduced temperatures $T = 0.29$-$0.60$ (legend in (b)), and panels (c, d) for the BKS model at $T = 2500$-$6100$ K (legend in (d)).

**Table S2.** Activation energies (eV) obtained from Arrhenius fits of the relaxation times $\tau_\alpha(k)$ and diffusion coefficients for silicon and oxygen atoms in silica melts.

**(a) BKS model** ($k_1 = 1.7$ Å$^{-1}$, $k_2 = 2.8$ Å$^{-1}$)

| Species | $E_a^{\tau_\alpha}(k_1)$ [eV] | $E_a^{\tau_\alpha}(k_2)$ [eV] | $E_a^{D}(k_2)$ [eV] |
|---|---|---|---|
| Si | 4.98 | 5.54 | 4.60 |
| O | 4.68 | 5.20 | 4.32 |

**(b) CP model** ($k_1 = 5.0$, $k_2 = 8.0$, reduced units)

| Species | $E_a^{\tau_\alpha}(k_1)$ [eV] | $E_a^{\tau_\alpha}(k_2)$ [eV] | $E_a^{D}(k_2)$ [eV] |
|---|---|---|---|
| Si | 4.07 | 4.57 | 3.72 |
| O | 3.80 | 4.27 | 3.47 |

**Table S3.** Time window $\Delta t$ used to calculate the hop function for silicon and oxygen atoms in silica melts at different temperatures.

| $T$ (BKS, K) | $\Delta t_{Si}$ (BKS, ps) | $\Delta t_O$ (BKS, ps) | $T$ (CP) | $\Delta t_{Si}$ (CP) | $\Delta t_O$ (CP) |
|---|---|---|---|---|---|
| 6100 | 0.26 | 0.22 | 0.60 | 1.60 | 1.60 |
| 5200 | 0.30 | 0.26 | 0.50 | 1.80 | 1.70 |
| 4300 | 0.34 | 0.30 | 0.45 | 1.90 | 1.80 |
| 4000 | 0.36 | 0.32 | 0.42 | 1.90 | 1.80 |
| 3760 | 0.37 | 0.34 | 0.40 | 2.00 | 1.90 |
| 3400 | 0.38 | 0.36 | 0.38 | 2.00 | 1.90 |
| 3100 | 0.38 | 0.36 | 0.36 | 2.00 | 1.90 |
| 2900 | 0.38 | 0.36 | 0.34 | 2.00 | 1.90 |
| 2750 | 0.40 | 0.36 | 0.32 | 2.10 | 2.00 |
| 2600 | 0.41 | 0.38 | 0.30 | 2.10 | 2.00 |
| 2500 | 0.41 | 0.38 | 0.29 | 2.10 | 1.90 |

Note: CP values are expressed in reduced units.

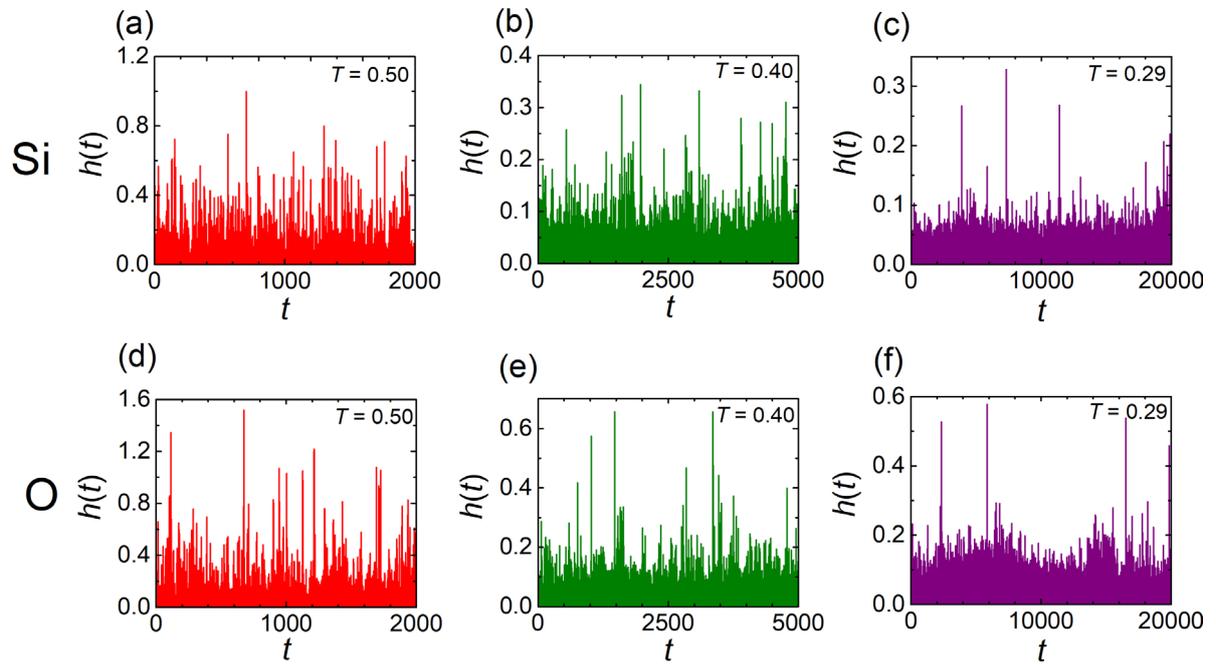

**Fig. S6.** Time evolution of the hop function $h(t)$ for individual Si (top row) and O (bottom row) atoms in the CP model at $T = 0.50$, $0.40$, and $0.29$.

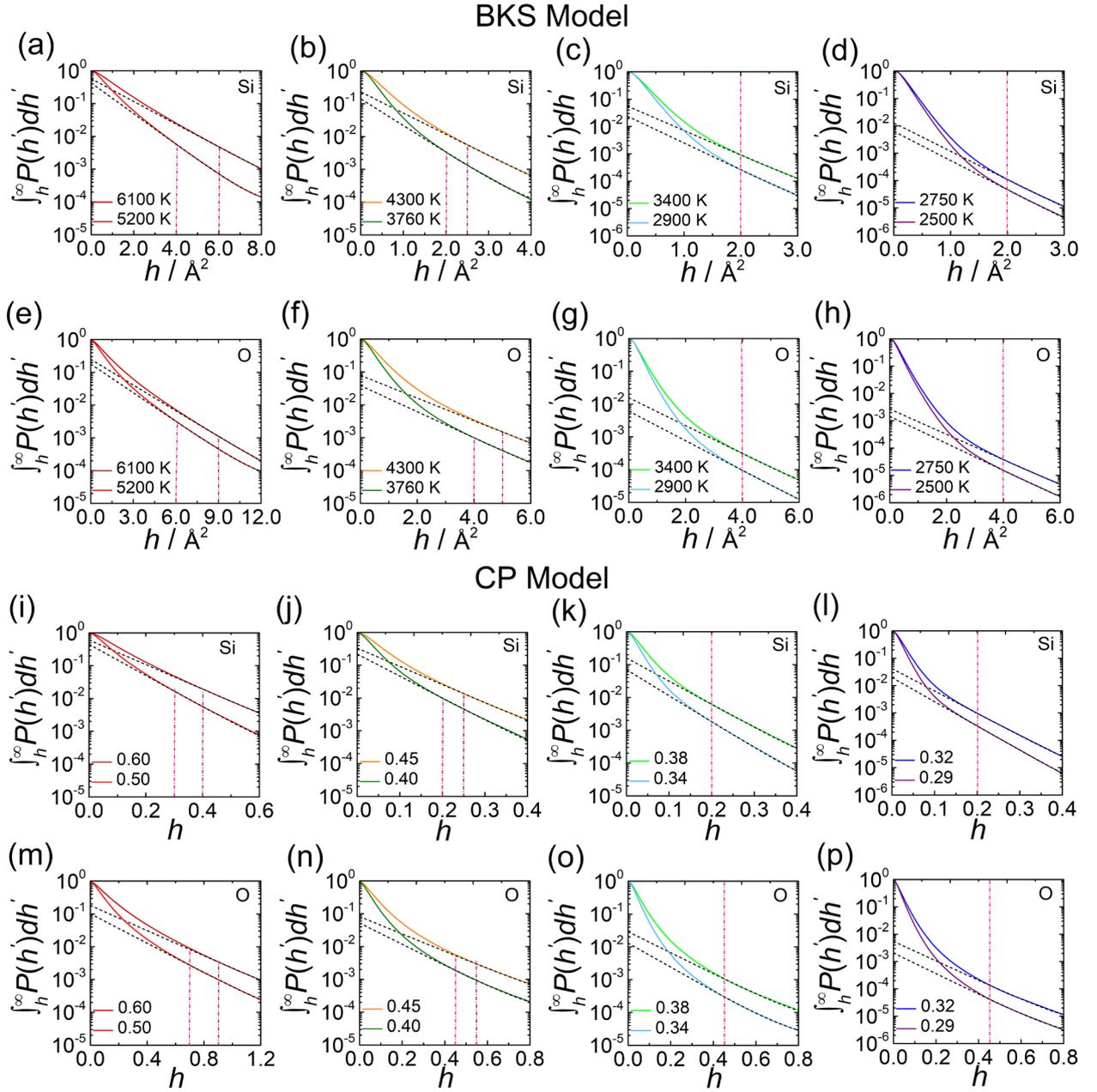

**Fig. S7.** Cumulative probability distributions of the hop function $h$ for silicon and oxygen atoms in silica melts at different temperatures. Panels (a–d) and (e–h) show results for silicon and oxygen atoms, respectively, in the BKS model over the temperature range $T = 6100\text{-}2500$ K. Panels (i–l) and (m–p) show results for silicon and oxygen atoms, respectively, in the CP model for $T = 0.60\text{-}0.29$. Vertical dashed lines indicate the threshold values $h^*$ used for distinguishing cage and jump states.

**Table S4.** Hop thresholds $h^*$ used to distinguish cage and jump states of silicon and oxygen atoms in silica melts.

| $T$(BKS, K) | $h^*_{Si}$ (BKS, Å²) | $h^*_{O}$ (BKS, Å²) | $T$(CP) | $h^*_{Si}$ (CP) | $h^*_{O}$ (CP) |
|---|---|---|---|---|---|
| 6100 | 6.0 | 9.0 | 0.60 | 0.40 | 0.90 |
| 5200 | 4.0 | 6.0 | 0.50 | 0.30 | 0.70 |
| 4300 | 2.5 | 5.0 | 0.45 | 0.25 | 0.55 |
| ≤ 4000 | 2.0 | 4.0 | ≤ 0.42 | 0.20 | 0.45 |

Note: CP values are expressed in reduced units.

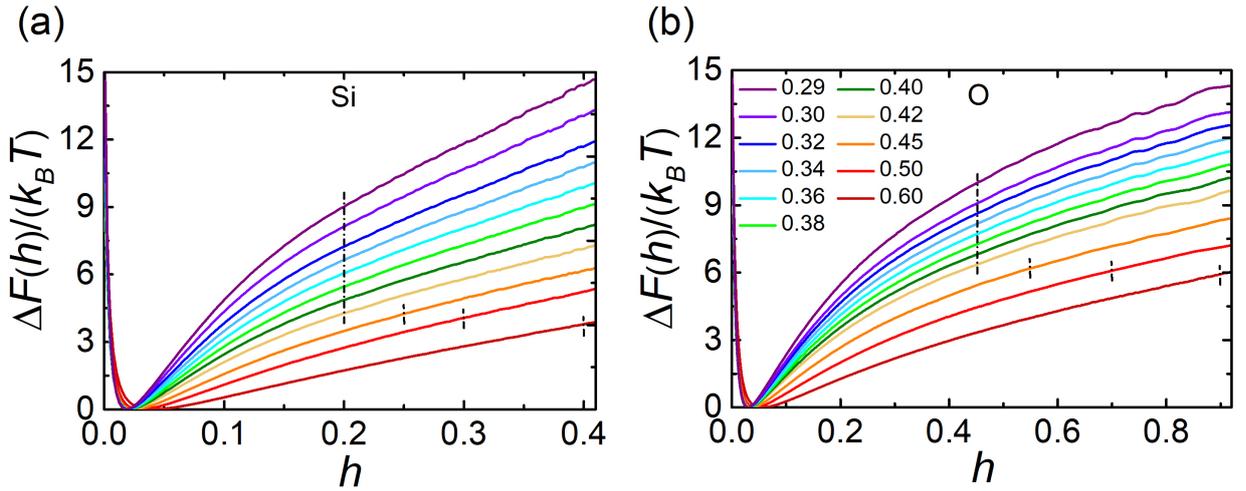

**FIG. S8.** Temperature dependence of the free-energy profiles $\Delta F(h)/(k_B T)$ for (a) Si and (b) O in the CP model, showing the progressive growth of the barrier height with cooling (legend in (b)). Vertical black dashed–dotted lines indicate the hopping threshold $h^*$ at the corresponding temperatures. For silicon, the barrier height increases from $\Delta F(h^*)/(k_B T) = 4.27$ at $T = 0.42$ (~1.05 eV) to 9.01 at $T = 0.29$ (~1.53 eV), corresponding to a growth of about 0.48 eV. For oxygen, the barrier increases from 6.34 at $T = 0.42$ (~1.56) to 9.96 at $T = 0.29$ (~1.70 eV), corresponding to a smaller increase of about 0.14 eV.

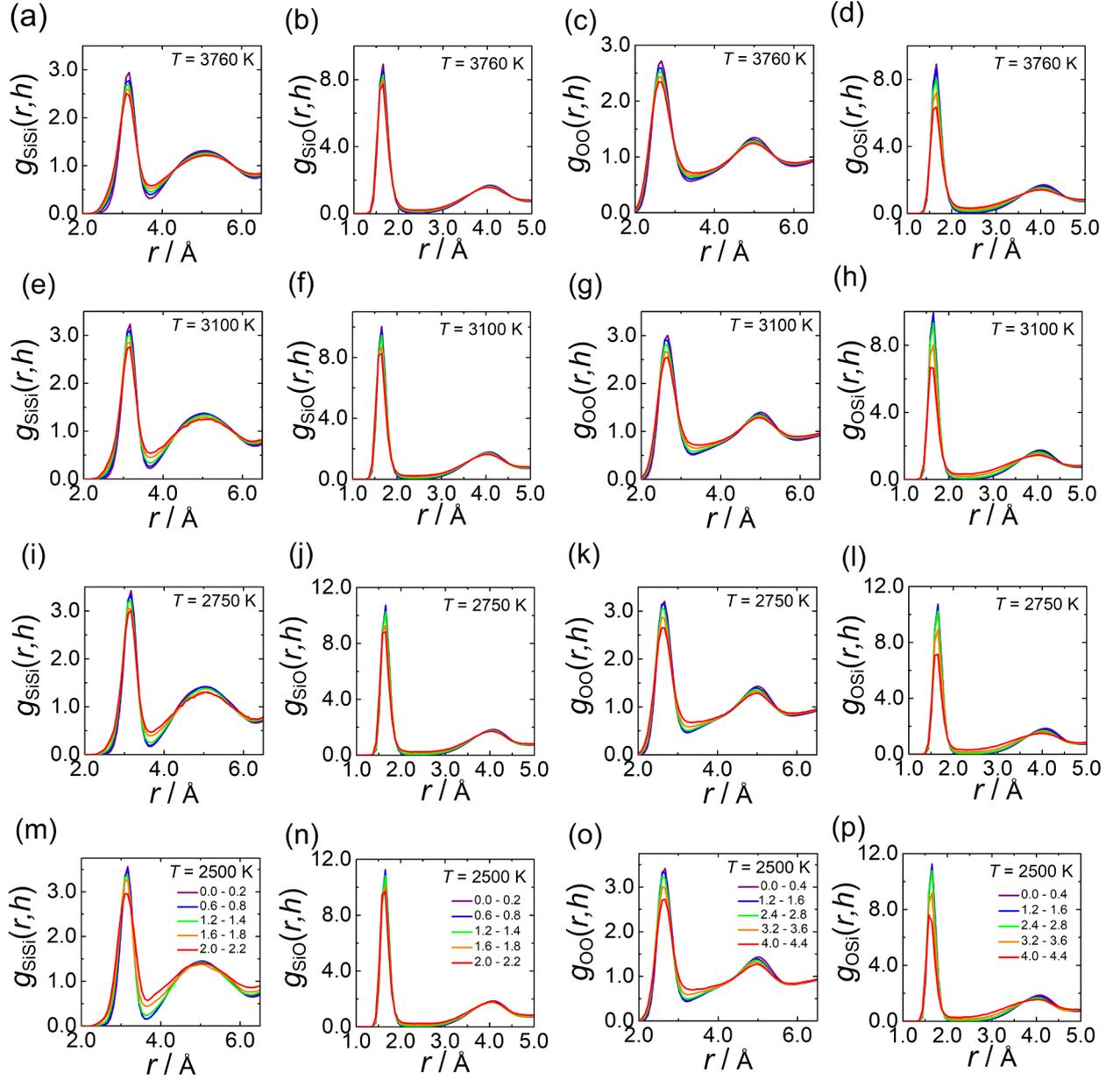

**Fig. S9.** Hop-dependence of the partial radial distribution functions $g_{\alpha\beta}(r, h)$ of silica (BKS model) at representative temperatures $T = 3760, 3100, 2750,$ and $2500$ K. Rows correspond to temperatures (labeled in the figure), and columns to different atomic pairs: $g_{SiSi}$, $g_{SiO}$, $g_{OO}$, and $g_{OSi}$. Legends are shown only in the panels at $T = 2500$ K. The hop-range colors are kept consistent across temperatures for each type of $g_{\alpha\beta}(r, h)$.

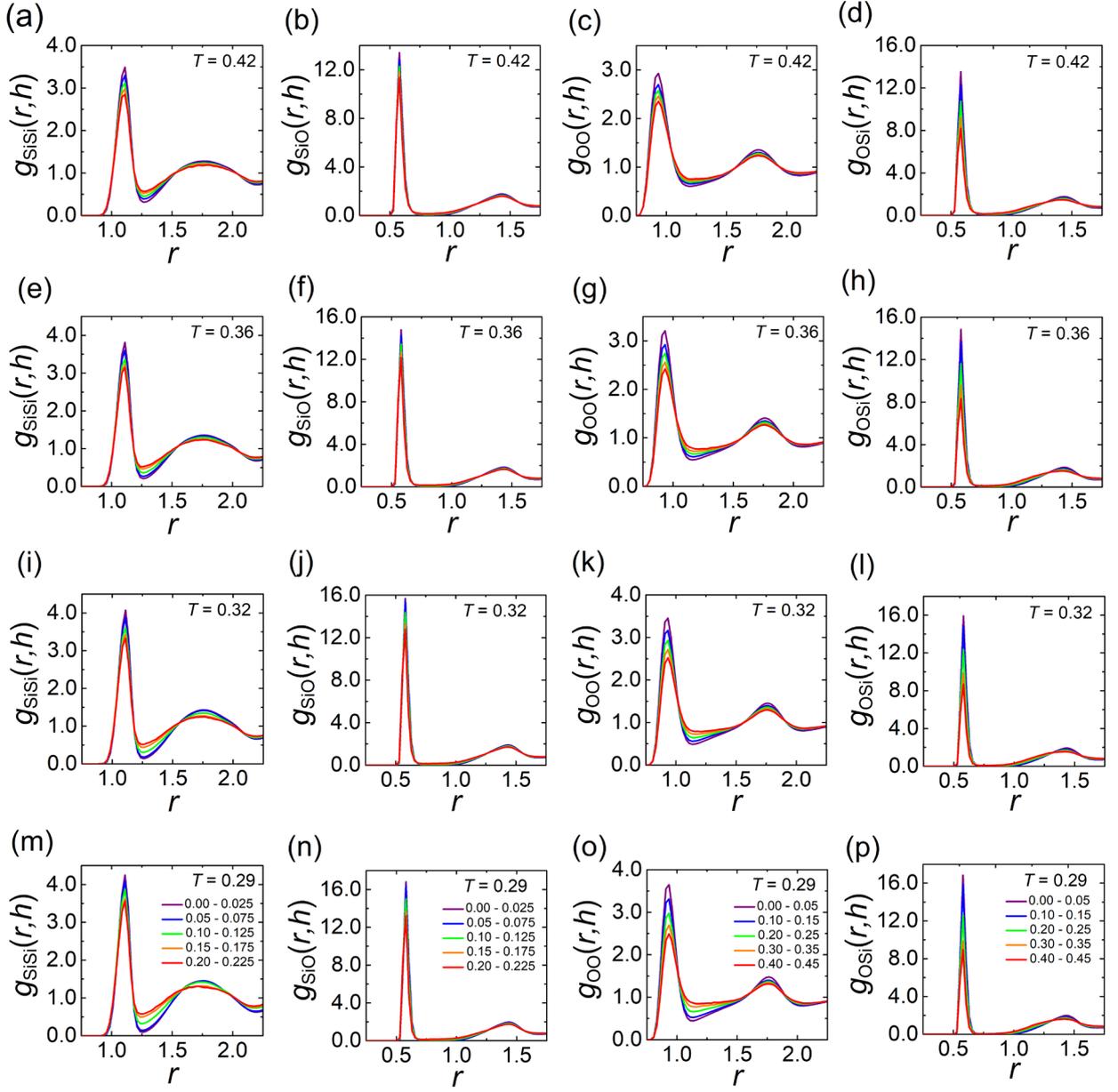

**Fig. S10.** Hop-dependence of the partial radial distribution functions $g_{\alpha\beta}(r, h)$ of silica (CP model) at representative reduced temperatures $T = 0.42$, $0.36$, $0.32$, and $0.29$. Rows correspond to temperatures (labeled in the figure), and columns to different atomic pairs: $g_{SiSi}$, $g_{SiO}$, $g_{OO}$, and $g_{OSi}$. Legends are shown only in the panels at $T = 0.29$. The hop-range colors are kept consistent across temperatures for each type of $g_{\alpha\beta}(r, h)$.

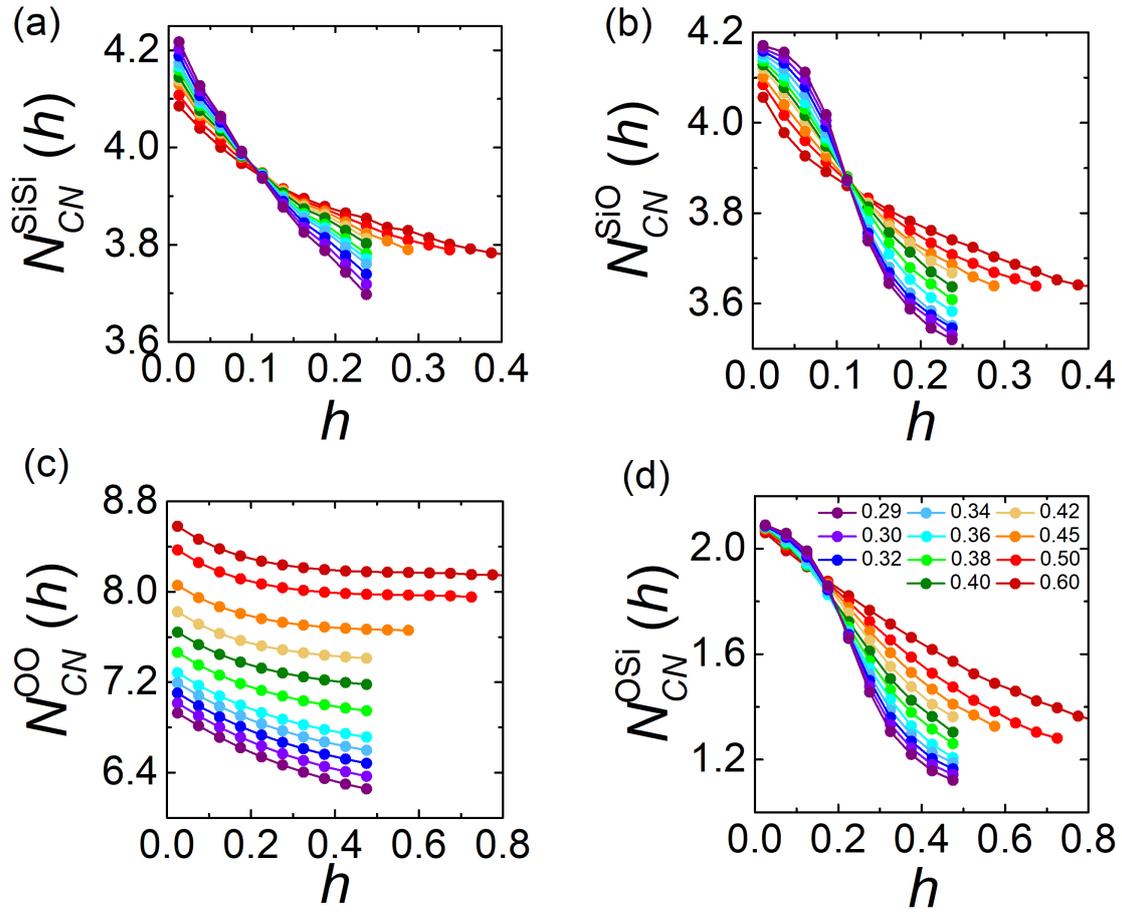

**Fig. S11.** Temperature dependence of the coordination numbers (a) $N_{CN}^{SiSi}$, (b) $N_{CN}^{SiO}$ for jumping Si, and (c) $N_{CN}^{OO}$, (d) $N_{CN}^{OSi}$ for jumping O atom in the CP model, showing a progressive loss of neighbors with increasing $h$ (legend in (j)).

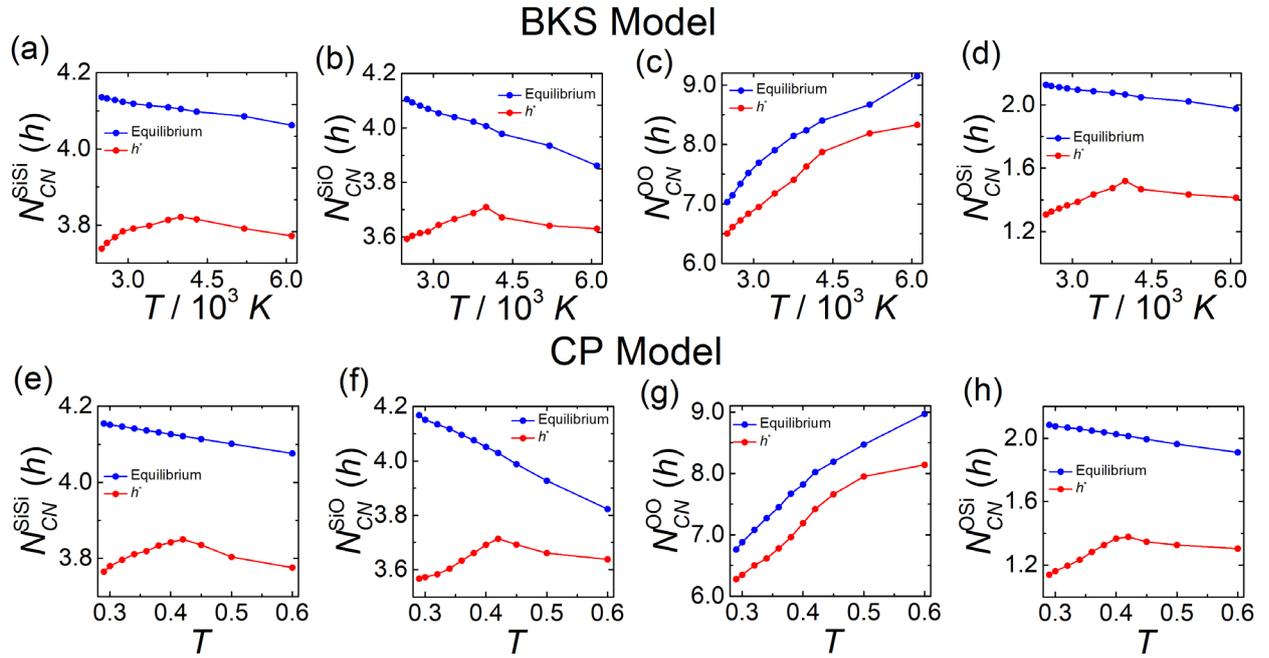

**Fig. S12.** Comparison of coordination numbers $N_{CN}^{\alpha\beta}$ for jumping Si and O atoms at equilibrium (blue) and at the jump threshold $h^*$ (red), shown as a function of temperature for the BKS model (a–d) and the CP model (e–h).

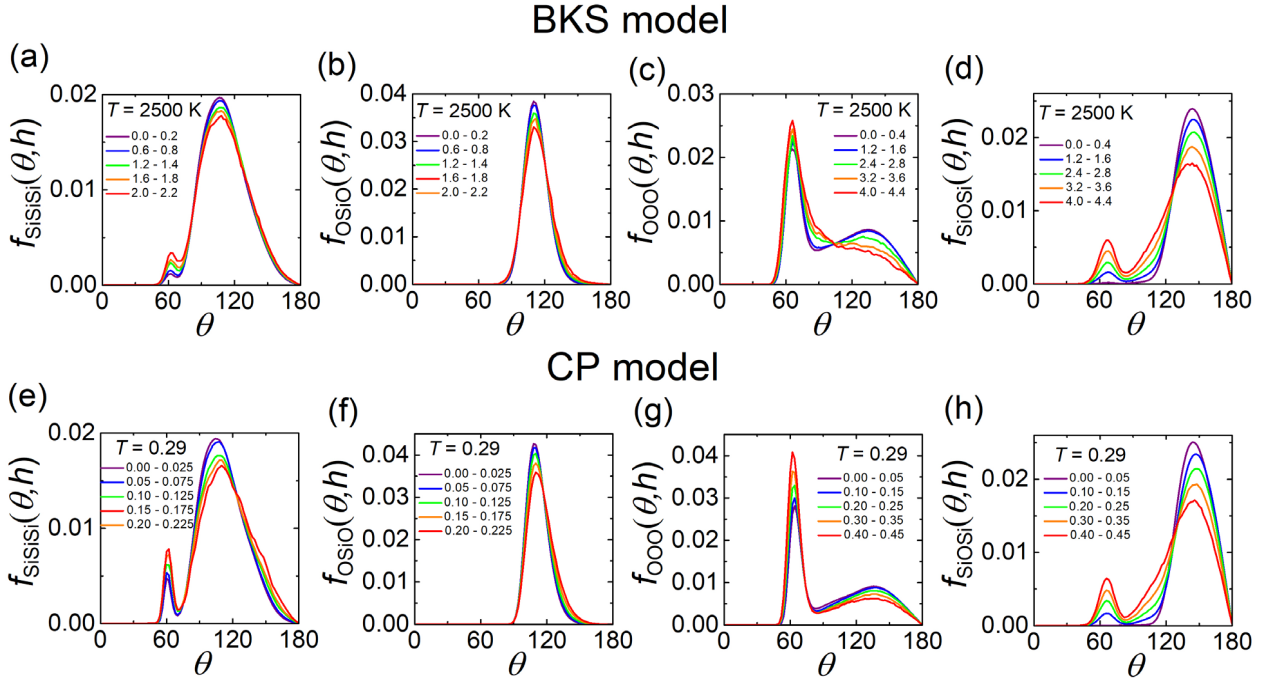

**Fig. S13.** Hop dependence of angular distribution functions $f_{SiSiSi}(\theta)$, $f_{OSiO}(\theta)$, $f_{OOO}(\theta)$, and $f_{SiOSi}(\theta)$ at the lowest temperature for the BKS (a–d, 2500 K) and CP (e–h, $T = 0.29$) models. $f_{SiOSi}(\theta)$ develops a distinct secondary lobe, indicating Si–O–Si bending during hopping of oxygen atoms.

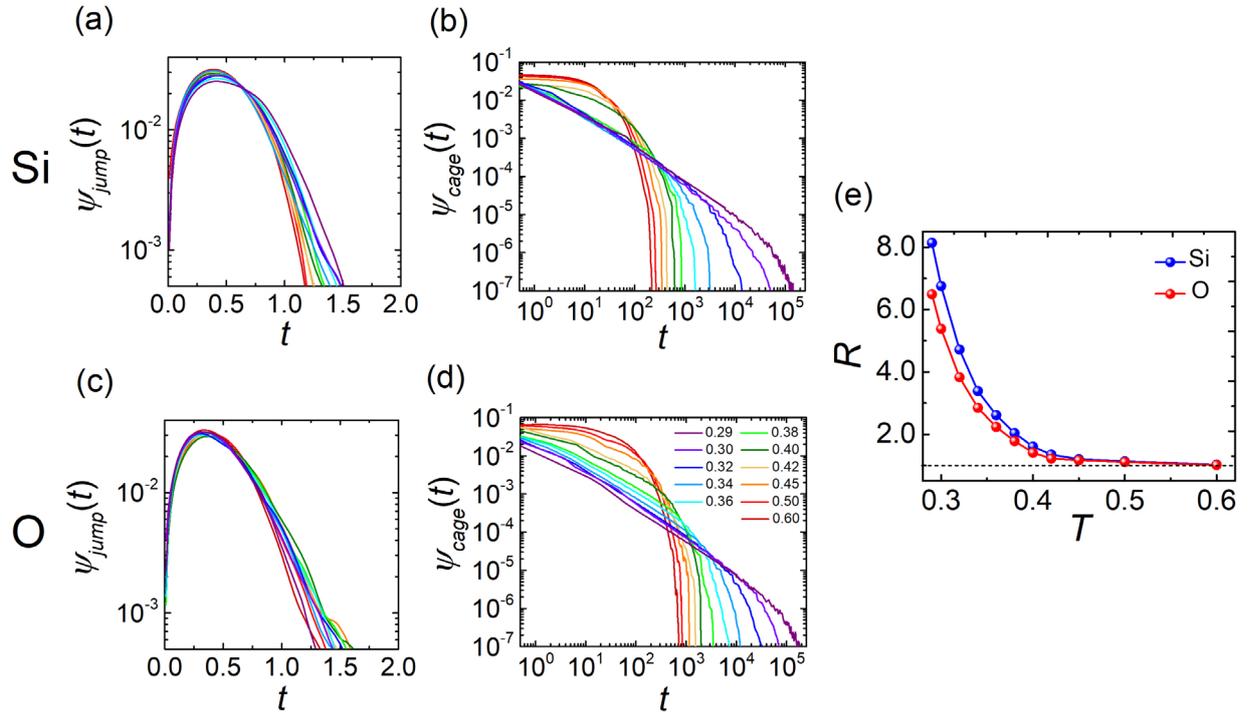

**Fig. S14.** Temperature dependence of (a), (c) jump-time distributions $\psi_{jump}(t)$ for Si and O, respectively, and (b), (d) the corresponding cage-time distributions $\psi_{cage}(t)$ in the CP model (legend in (d)). (e) Temperature dependence of the randomness parameter $R$ for Si and O, quantifying deviations from Poisson statistics.

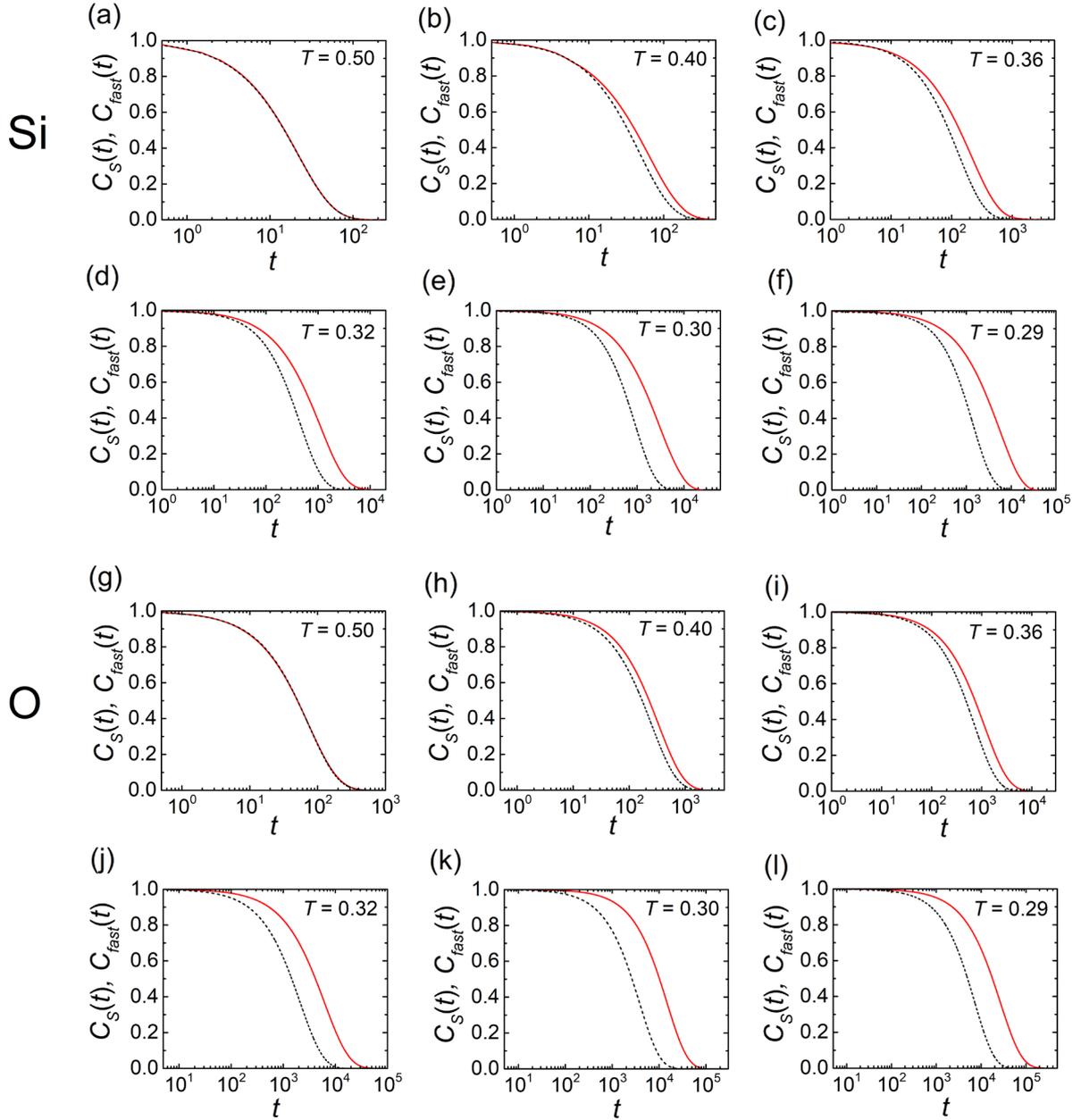

**Fig. S15.** Survival probability of the cage state, $C_S(t)$ (red), and its fast-fluctuation limit, $C_{fast}(t)$ (dashed black), for Si at (a) $T = 0.50$, (b) 0.40, (c) 0.36, (d) 0.32, (e) 0.30, and (f) 0.29 in the CP model. Panels (g-l) show the corresponding results for O at the same temperatures. At high temperatures, $C_S(t)$ closely follows $C_{fast}(t)$, consistent with Poisson statistics. However, upon cooling, systematic deviations emerge, indicating the breakdown of the fast-fluctuation approximation.

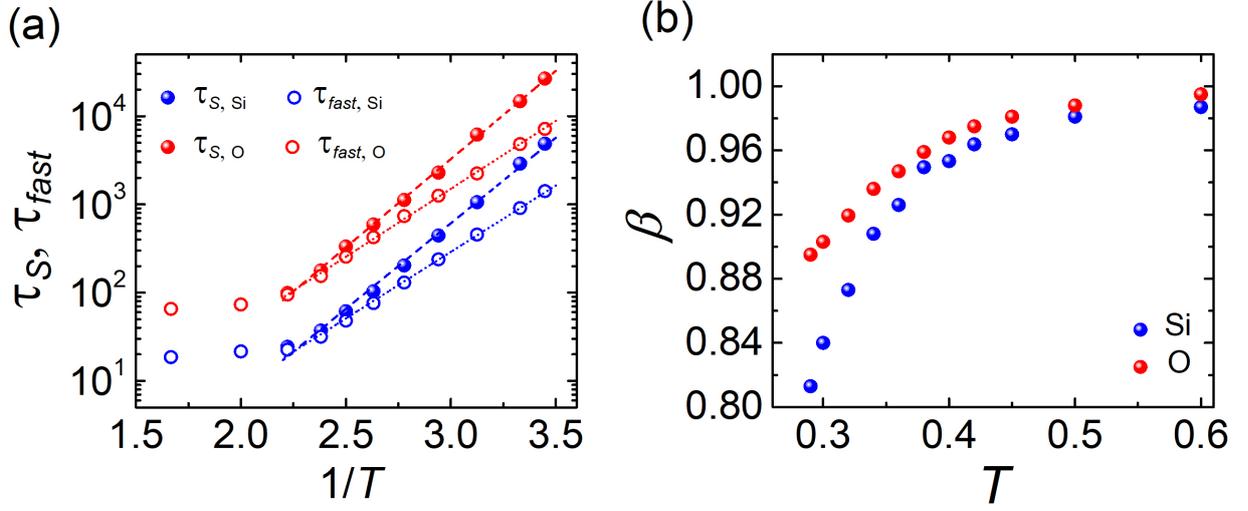

**Fig. S16.** (a) Temperature dependence of the survival time $\tau_S$ compared with the fast-fluctuation timescale $\tau_{fast}$ for the CP model. Both quantities exhibit Arrhenius behavior below $T = 0.45$, as indicated by the dashed and dotted lines. (b) Stretching exponent $\beta$ as a function of temperature. As in the BKS model (Fig. 6), $\tau_S$ exceeds $\tau_{fast}$ upon cooling, and $\beta$ decreases from near unity, indicating the emergence of temporal heterogeneity in the jump dynamics.

**Table S5.** Activation energies (eV) obtained from Arrhenius fits of the $\tau_S$ and $\tau_{fast}$ for silicon and oxygen atoms in silica melts.

**(a) BKS model**

| Species | $E_a^{\tau_S}$ [eV] | $E_a^{\tau_{fast}}$ [eV] |
|---|---|---|
| **Si** | 4.34 | 3.77 |
| **O** | 4.12 | 3.51 |

**(b) CP model**

| Species | $E_a^{\tau_S}$ [eV] | $E_a^{\tau_{fast}}$ [eV] |
|---|---|---|
| **Si** | 3.07 | 2.57 |
| **O** | 2.90 | 2.34 |

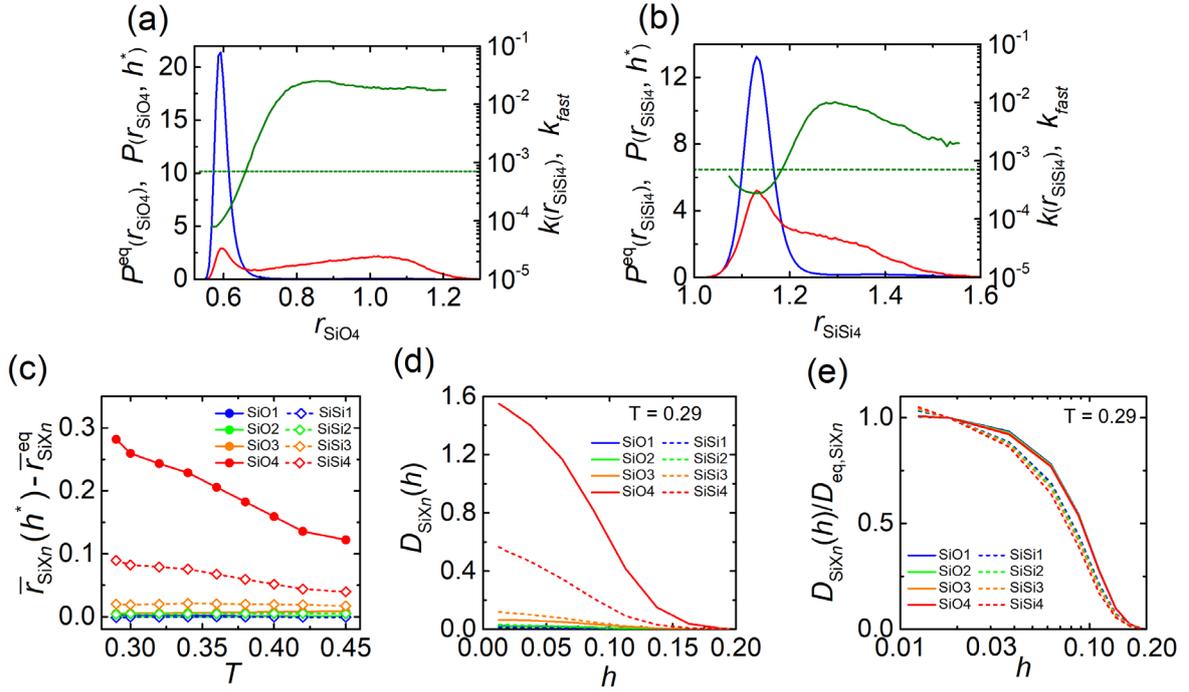

**Fig. S17.** (a) Distributions of the distance between a Si atom and its fourth-nearest O neighbor, $r_{SiO4}$, in equilibrium (blue) and for a jumping Si at the hop threshold $h^*$ (red) at $T = 0.29$ using the CP model. The green curve and green dashed line represent $k(r_{SiO4})$ and $k_{fast}$, respectively. (b) Distributions of the distance between a Si atom and its fourth-nearest Si neighbor, $r_{SiSi4}$, in equilibrium (blue) and for a jumping Si at the hop threshold $h^*$ (red) at $T = 0.29$. The green curve and green dashed line represent $k(r_{SiSi4})$ and $k_{fast}$, respectively. (c) Temperature dependence of the difference between the average distances at equilibrium and at $h^*$ for SiXn. (d) KL divergence $D_{SiXn}(h)$ and (e) scaled KL divergence $D_{SiXn}(h)/D_{eq,SiXn}$ at $T = 0.29$. In (c)-(e), $X \in \{O, Si\}$ and $n \leq 4$.

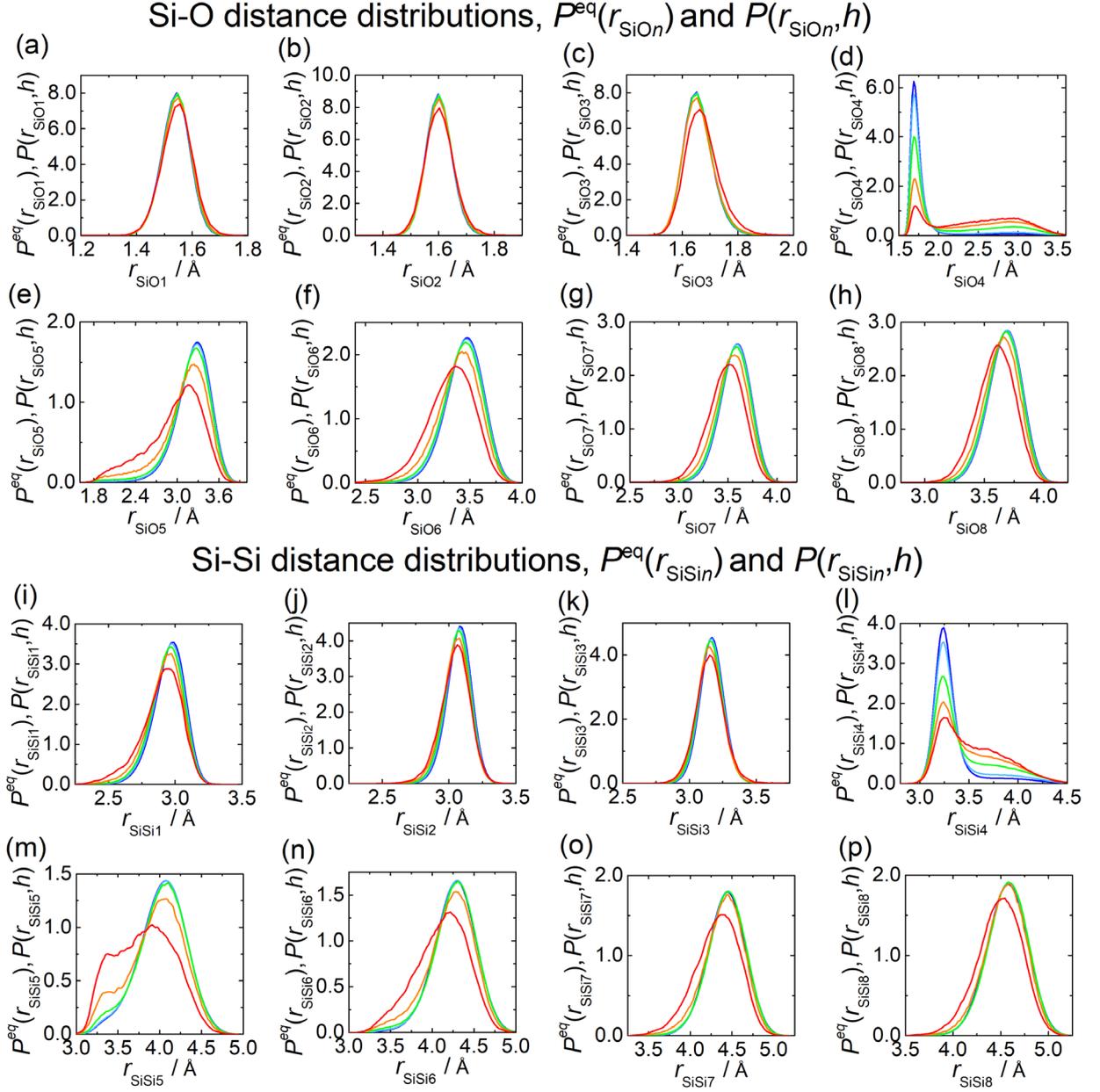

**Figure S18.** Distributions of the distances between a jumping silicon atom with $h$ and its $n$th nearest oxygen neighbors ($1 \leq n \leq 8$) in silica at $T = 2500$ K using the BKS model [(a-h)]. Distributions of the distances between a jumping silicon atom with $h$ and its $n$th nearest silicon neighbors ($1 \leq n \leq 8$) [(i-p)]. The blue curves represent the equilibrium distributions $P^{eq}(r_{SiOn})$ and $P^{eq}(r_{SiSin})$. The light blue, green, orange, and red curves represent the distributions of jumping atoms with $h$ in the ranges (0.2-0.4), (0.8-1.0), (1.4-1.6), and at $h^*$, respectively.

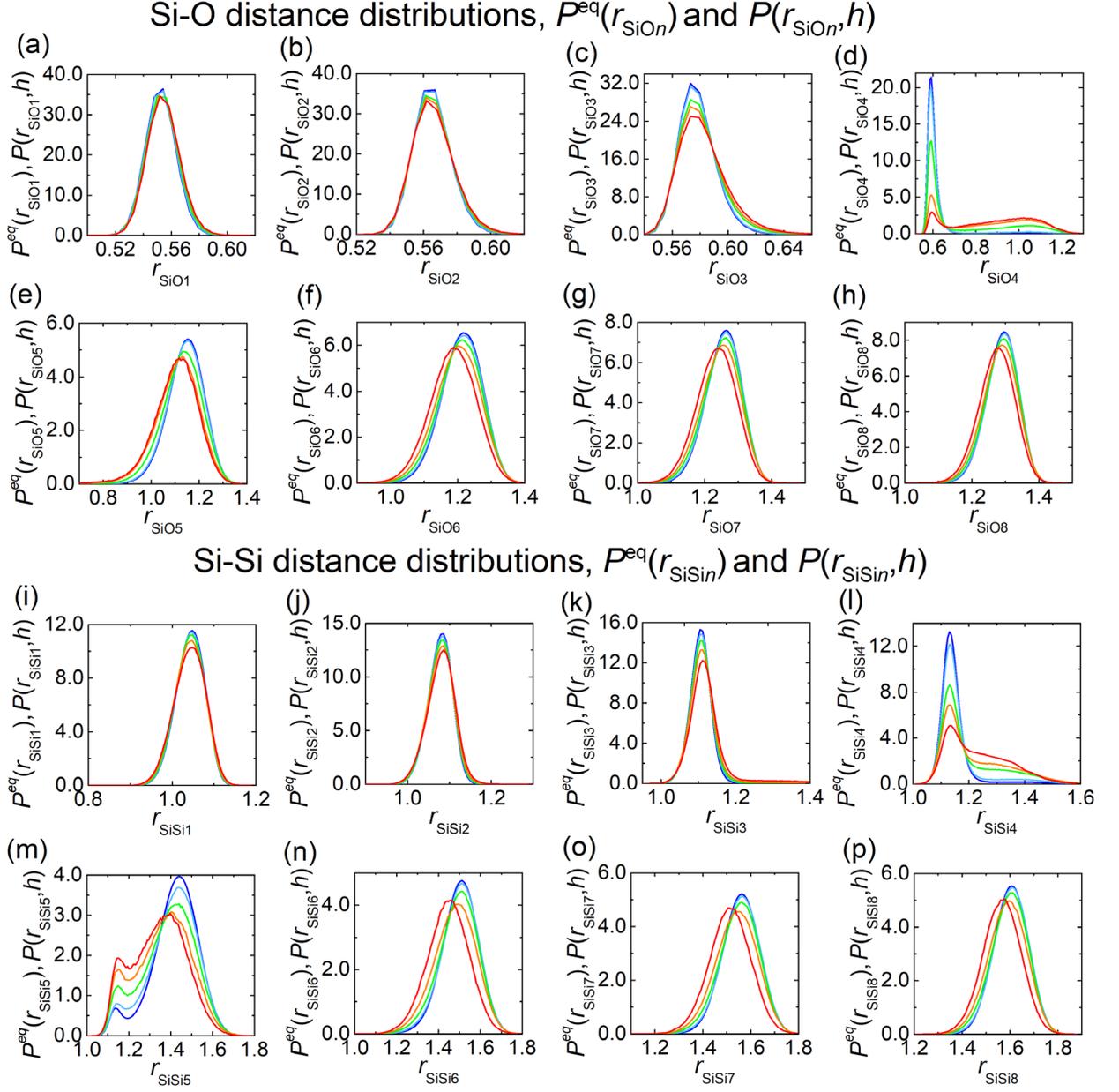

**Figure S19.** Distributions of the distances between a jumping silicon atom with $h$ and its $n$th nearest oxygen neighbors ($1 \leq n \leq 8$) in silica at $T = 0.29$ using the CP model [(a-h)]. Distributions of the distances between a jumping silicon atom with $h$ and its $n$th nearest silicon neighbors ($1 \leq n \leq 8$) [(i-p)]. The blue curves represent the equilibrium distributions $P^{eq}(r_{SiOn})$ and $P^{eq}(r_{SiSin})$. The light blue, green, orange, and red curves represent the distributions of jumping atoms with $h$ in the ranges (0.05-0.075), (0.10-0.125), (0.15-0.175), and at $h^*$, respectively.

**Table S6.** Ratios of the silicon jump rate at the average distance $\bar{r}_{SiXn}(h^*)$ to that at the equilibrium average $\bar{r}_{SiXn}^{eq}$, for X = O, Si and $1 \leq n \leq 4$.

**(a) BKS model**

| T(BKS, K) | SiOn | | | | SiSin | | | |
|---|---|---|---|---|---|---|---|---|
| | 1 | 2 | 3 | 4 | 1 | 2 | 3 | 4 |
| **4300** | 1.04 | 1.07 | 1.09 | 1.65 | 1.06 | 1.07 | 1.05 | 1.14 |
| **4000** | 1.09 | 1.10 | 1.15 | 2.84 | 1.08 | 1.09 | 1.08 | 1.52 |
| **3760** | 1.14 | 1.15 | 1.21 | 4.37 | 1.09 | 1.12 | 1.14 | 2.11 |
| **3400** | 1.17 | 1.18 | 1.27 | 7.87 | 1.12 | 1.17 | 1.19 | 2.62 |
| **3100** | 1.20 | 1.27 | 1.32 | 11.53 | 1.16 | 1.22 | 1.27 | 3.21 |
| **2900** | 1.26 | 1.33 | 1.40 | 16.54 | 1.22 | 1.26 | 1.43 | 4.12 |
| **2750** | 1.27 | 1.36 | 1.47 | 20.36 | 1.27 | 1.34 | 1.56 | 4.82 |
| **2600** | 1.32 | 1.39 | 1.51 | 26.98 | 1.28 | 1.38 | 1.64 | 5.68 |
| **2500** | 1.36 | 1.43 | 1.58 | 36.52 | 1.32 | 1.48 | 1.76 | 6.76 |

**(b) CP model**

| T(CP) | SiOn | | | | SiSin | | | |
|---|---|---|---|---|---|---|---|---|
| | 1 | 2 | 3 | 4 | 1 | 2 | 3 | 4 |
| **0.45** | 1.01 | 1.03 | 1.02 | 2.25 | 1.01 | 1.01 | 1.03 | 1.22 |
| **0.42** | 1.01 | 1.04 | 1.03 | 4.19 | 1.01 | 1.02 | 1.05 | 1.90 |
| **0.40** | 1.02 | 1.03 | 1.03 | 10.86 | 1.02 | 1.02 | 1.04 | 2.56 |
| **0.38** | 1.04 | 1.07 | 1.06 | 17.78 | 1.04 | 1.05 | 1.11 | 3.69 |
| **0.36** | 1.05 | 1.09 | 1.11 | 25.82 | 1.08 | 1.09 | 1.25 | 4.74 |
| **0.34** | 1.07 | 1.14 | 1.14 | 34.88 | 1.18 | 1.17 | 1.36 | 5.68 |
| **0.32** | 1.09 | 1.19 | 1.19 | 65.63 | 1.23 | 1.22 | 1.43 | 8.11 |
| **0.30** | 1.10 | 1.22 | 1.24 | 123.77 | 1.25 | 1.28 | 1.55 | 11.41 |
| **0.29** | 1.12 | 1.24 | 1.36 | 173.86 | 1.26 | 1.34 | 1.62 | 18.41 |

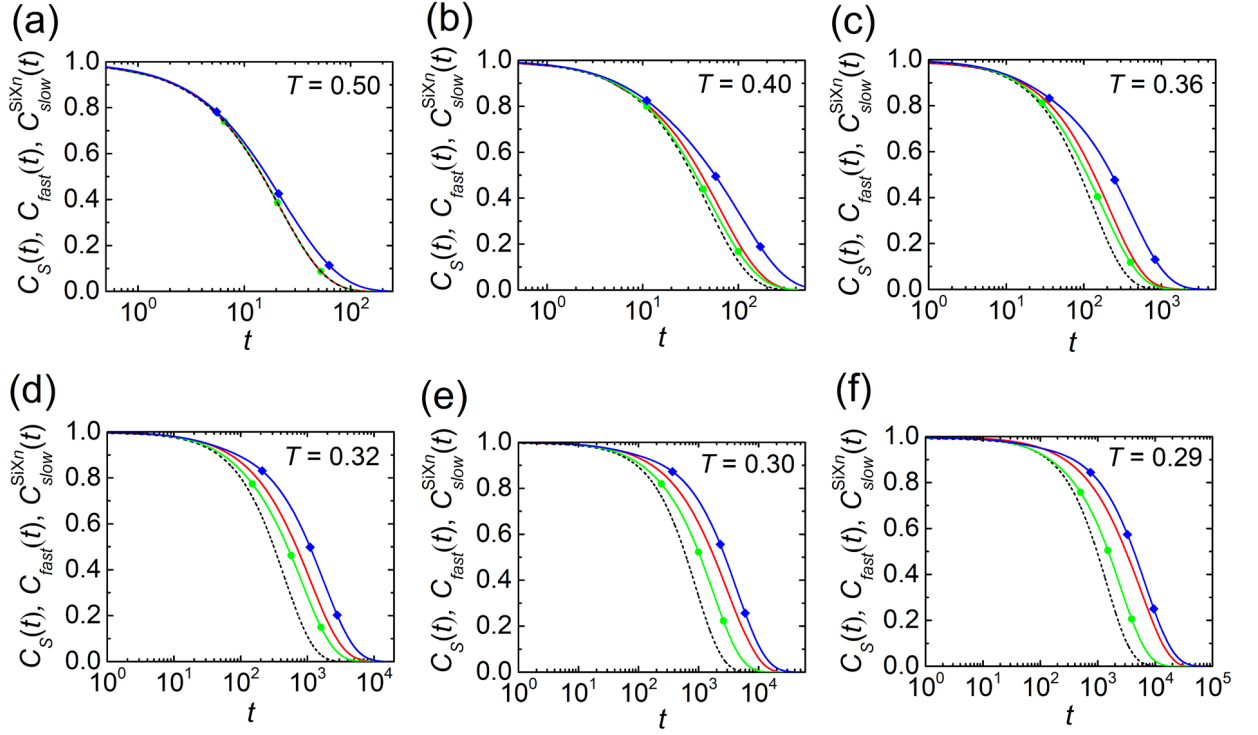

**Fig. S20.** Survival probability for the Si cage state, $C_S(t)$, its fast-fluctuation limit, $C_{fast}(t)$, and the slow-fluctuation limits $C_{slow}^{SiXn}(t)$ (with X∈{O,Si}) in the CP model at (a) $T = 0.50$, (b) 0.40, (c) 0.36, (d) 0.32, (e) 0.30, and (f) 0.29. The red curve, black dashed curve, blue curve with squares, and green curve with circles represent $C_S(t)$, $C_{fast}(t)$, $C_{slow}^{SiO4}(t)$, and $C_{slow}^{SiSi4}(t)$, respectively.

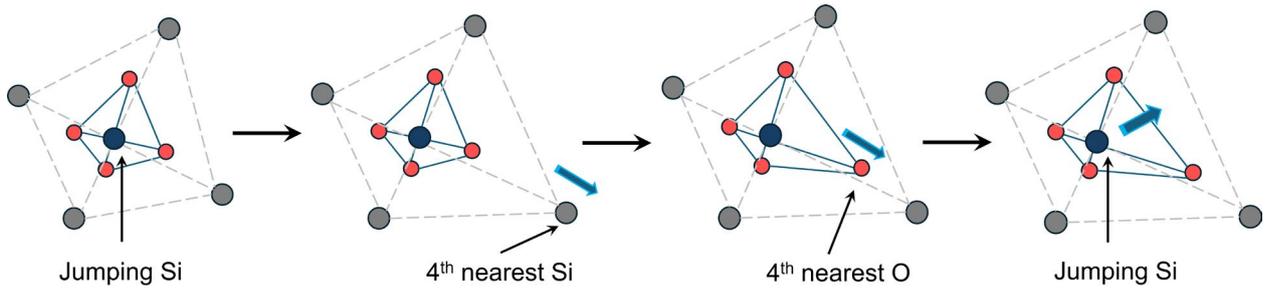

**Fig. 21.** Illustration of a Si jump event in silica. The jumping Si is shown in deep blue, oxygen atoms in red, and other Si atoms in grey. Blue arrows indicate atomic displacements. The sequence from left to right highlights the cooperative motion of the 4th nearest Si and O neighbors that facilitates the jump of the central Si.

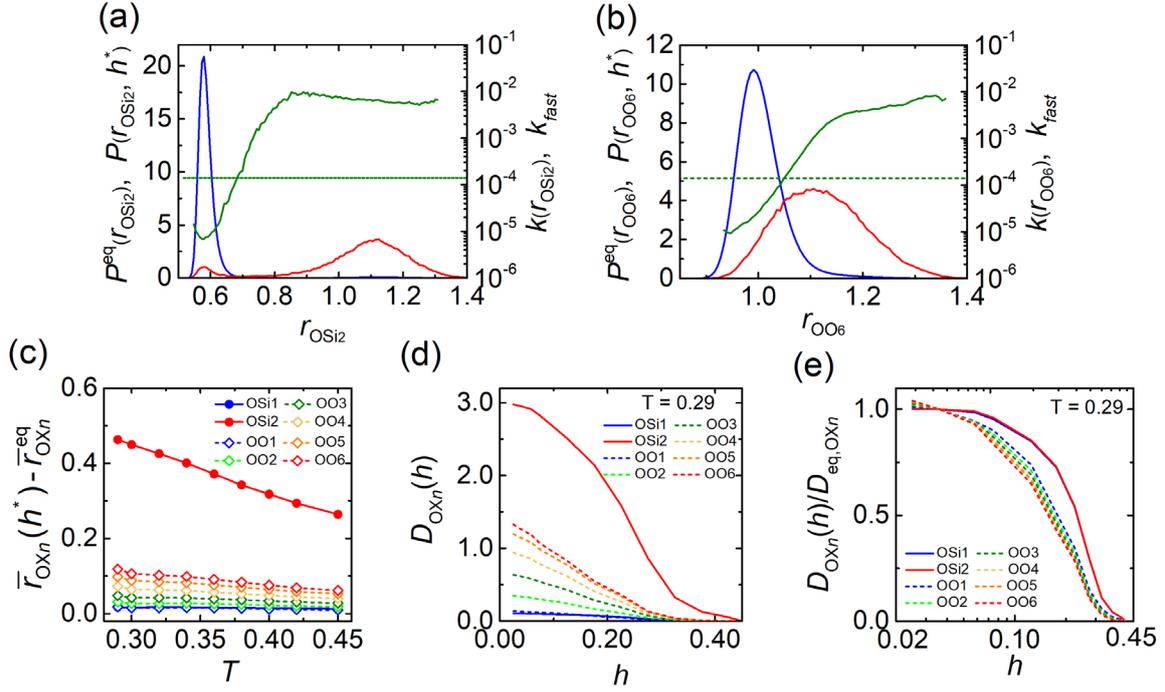

**Fig. S22.** (a) Distributions of the distance between an O atom and its second-nearest Si neighbor, $r_{OSi2}$, in equilibrium (blue) and for a jumping O at the hop threshold $h^*$ (red) at $T = 0.29$. The green curve and green dashed line represent $k(r_{OSi2})$ and $k_{fast}$, respectively. (b) Distributions of the distance between an O atom and its sixth-nearest O neighbor, $r_{OO6}$, in equilibrium (blue) and for a jumping O at the hop threshold $h^*$ (red) at $T = 0.29$. The green curve and green dashed line represent $k(r_{OO6})$ and $k_{fast}$, respectively. (c) Temperature dependence of the difference between the average distances at equilibrium and at $h^*$ for OX$n$. (d) KL divergence $D_{OXn}(h)$ and (e) scaled KL divergence $D_{OXn}(h)/D_{eq,OXn}$ at $T = 0.29$. In (c)-(e), $X \in \{Si, O\}$; for $X = Si$, $n \leq 2$; for $X = O$, $n \leq 6$.

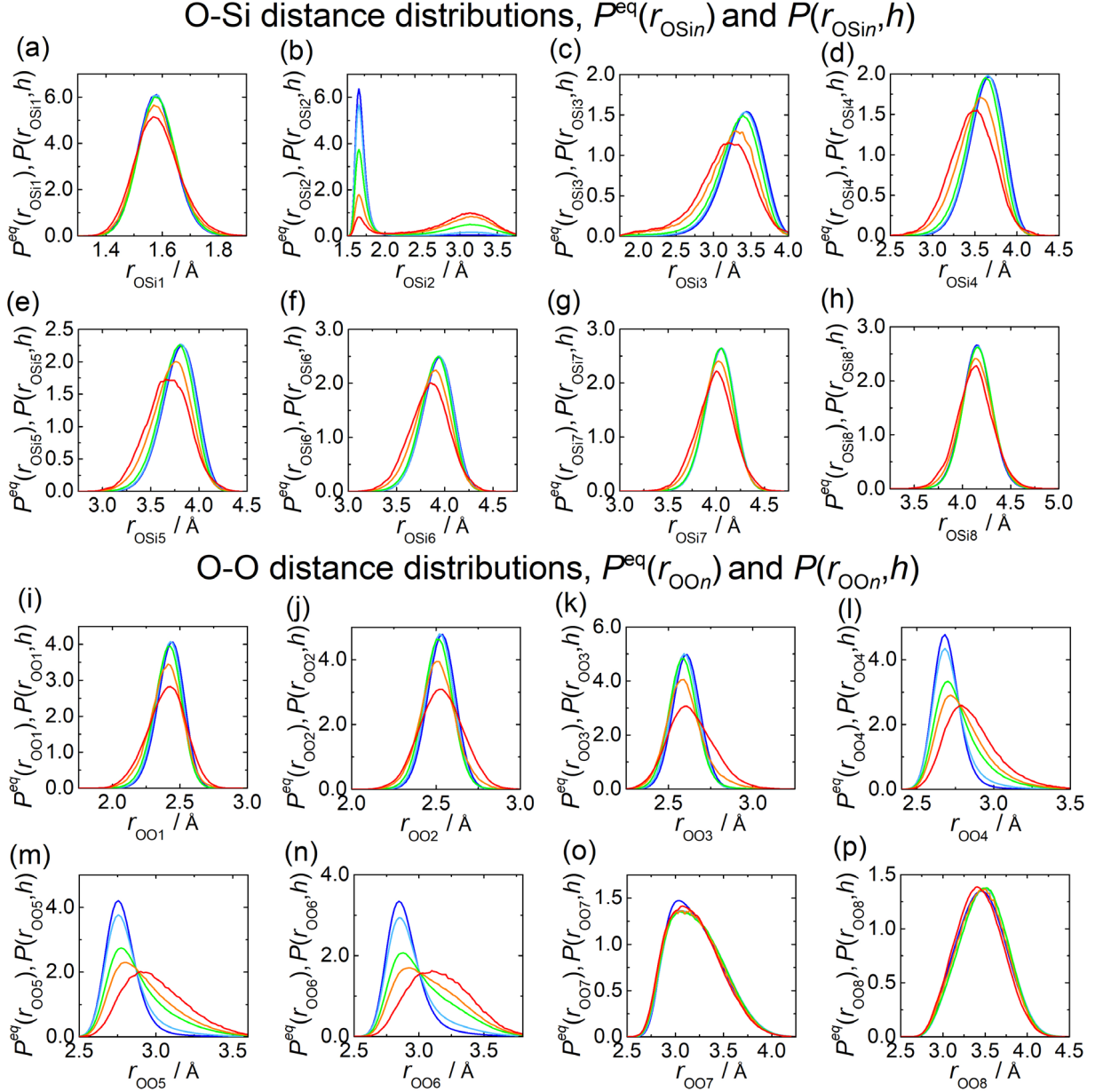

**Figure S23.** Distributions of the distances between a jumping oxygen atom with $h$ and its $n$th nearest silicon neighbors ($1 \leq n \leq 8$) in silica at $T = 2500$ K using the BKS model [(a-h)]. Distributions of the distances between a jumping oxygen atom with $h$ and its $n$th nearest oxygen neighbors ($1 \leq n \leq 8$) [(i-p)]. The blue curves represent the equilibrium distributions $P^{\mathrm{eq}}(r_{\mathrm{OSi}n})$ and $P^{\mathrm{eq}}(r_{\mathrm{OO}n})$. The light blue, green, orange, and red curves represent the distributions of jumping atoms with $h$ in the ranges (0.4-0.8), (1.6-2.0), (2.8-3.2), and at $h^*$, respectively.

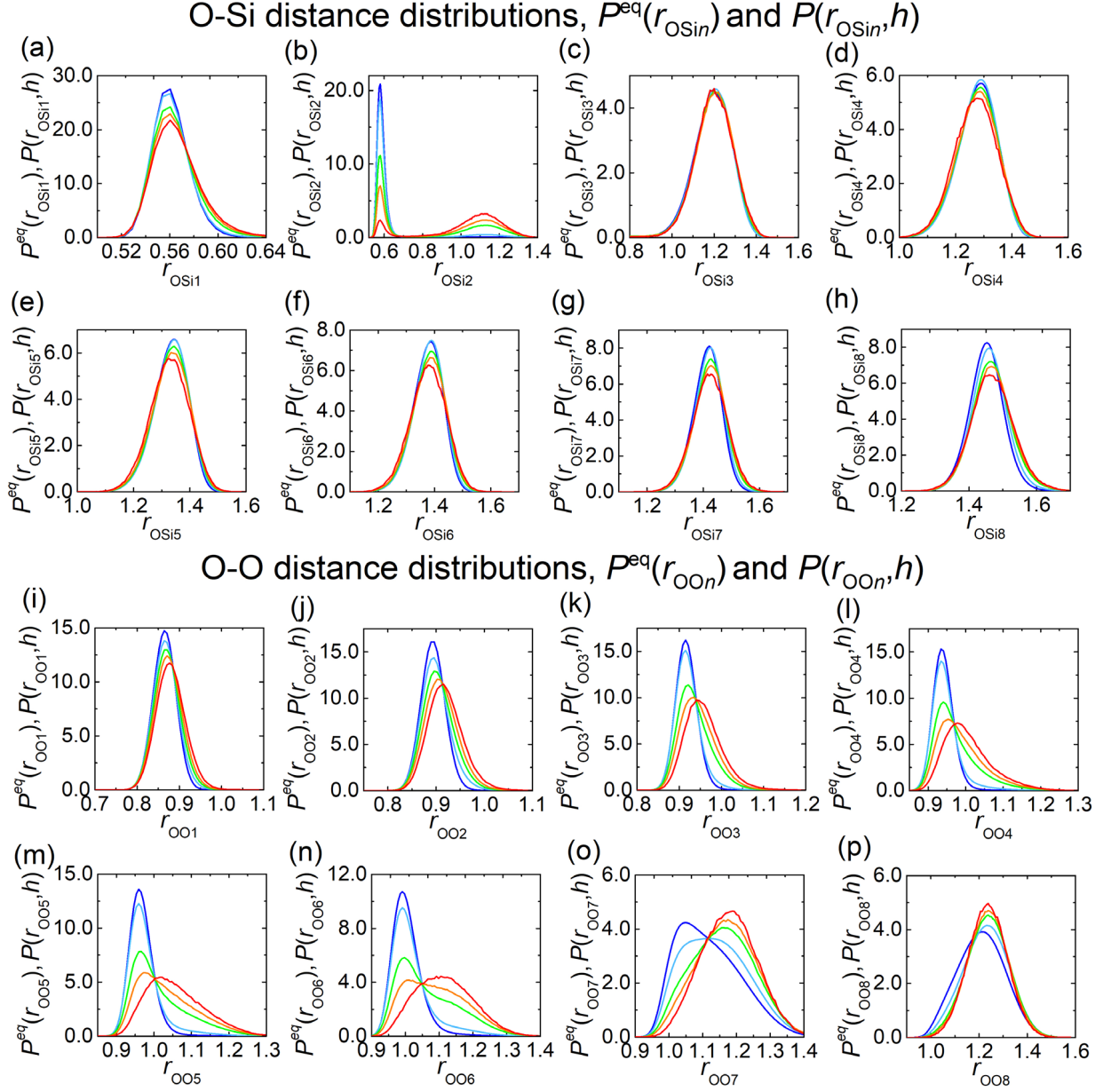

**Figure S24.** Distributions of the distances between a jumping oxygen atom with $h$ and its $n$th nearest silicon neighbors ($1 \leq n \leq 8$) in silica at $T = 0.29$ using the CP model [(a-h)]. Distributions of the distances between a jumping oxygen atom with $h$ and its $n$th nearest oxygen neighbors ($1 \leq n \leq 8$) [(i-p)]. The blue curves represent the equilibrium distributions $P^{eq}(r_{OSin})$ and $P^{eq}(r_{OOn})$. The light blue, green, orange, and red curves represent the distributions of jumping atoms with $h$ in the ranges (0.10-0.15), (0.20-0.25), (0.30-0.35), and at $h^*$, respectively.

**Table S7.** Ratios of the oxygen jump rate at the average distance $\bar{r}_{OX_n}(h^*)$ to that at the equilibrium average $\bar{r}_{OX_n}^{eq}$, for X = Si ($n \leq 2$) and X = O ($n \leq 6$).

**(a) BKS model**

| T(BKS, K) | OSi$n$ | | OO$n$ | | | | | |
|---|---|---|---|---|---|---|---|---|
| | 1 | 2 | 1 | 2 | 3 | 4 | 5 | 6 |
| **4300** | 1.01 | 4.21 | 1.02 | 1.04 | 1.05 | 1.08 | 1.11 | 1.21 |
| **4000** | 1.02 | 6.10 | 1.05 | 1.09 | 1.07 | 1.15 | 1.52 | 2.13 |
| **3760** | 1.05 | 11.12 | 1.07 | 1.15 | 1.19 | 1.28 | 2.21 | 3.34 |
| **3400** | 1.07 | 15.51 | 1.13 | 1.21 | 1.25 | 1.42 | 3.78 | 5.06 |
| **3100** | 1.11 | 24.21 | 1.19 | 1.27 | 1.37 | 1.61 | 5.12 | 7.11 |
| **2900** | 1.14 | 37.33 | 1.23 | 1.34 | 1.52 | 1.92 | 7.21 | 9.52 |
| **2750** | 1.19 | 53.98 | 1.26 | 1.39 | 1.74 | 2.54 | 9.61 | 11.81 |
| **2600** | 1.26 | 81.36 | 1.31 | 1.45 | 1.95 | 4.58 | 13.56 | 16.23 |
| **2500** | 1.33 | 145.58 | 1.34 | 1.52 | 2.32 | 7.48 | 21.67 | 26.76 |

**(b) CP model**

| T(CP) | OSi$n$ | | OO$n$ | | | | | |
|---|---|---|---|---|---|---|---|---|
| | 1 | 2 | 1 | 2 | 3 | 4 | 5 | 6 |
| **0.45** | 1.11 | 6.75 | 1.04 | 1.06 | 1.11 | 1.67 | 1.98 | 2.21 |
| **0.42** | 1.09 | 16.21 | 1.07 | 1.08 | 1.31 | 1.79 | 3.05 | 3.82 |
| **0.40** | 1.12 | 26.52 | 1.09 | 1.09 | 1.45 | 2.42 | 5.12 | 5.91 |
| **0.38** | 1.16 | 41.17 | 1.15 | 1.16 | 1.76 | 3.17 | 8.19 | 9.76 |
| **0.36** | 1.23 | 72.86 | 1.19 | 1.29 | 2.01 | 4.25 | 11.17 | 13.33 |
| **0.34** | 1.27 | 125.90 | 1.23 | 1.46 | 2.23 | 6.31 | 14.22 | 18.21 |
| **0.32** | 1.39 | 356.34 | 1.27 | 1.68 | 3.37 | 9.54 | 22.16 | 27.86 |
| **0.30** | 1.46 | 524.86 | 1.29 | 1.97 | 5.28 | 13.38 | 36.61 | 43.52 |
| **0.29** | 1.58 | 790.67 | 1.32 | 2.23 | 9.20 | 28.78 | 64.23 | 78.69 |

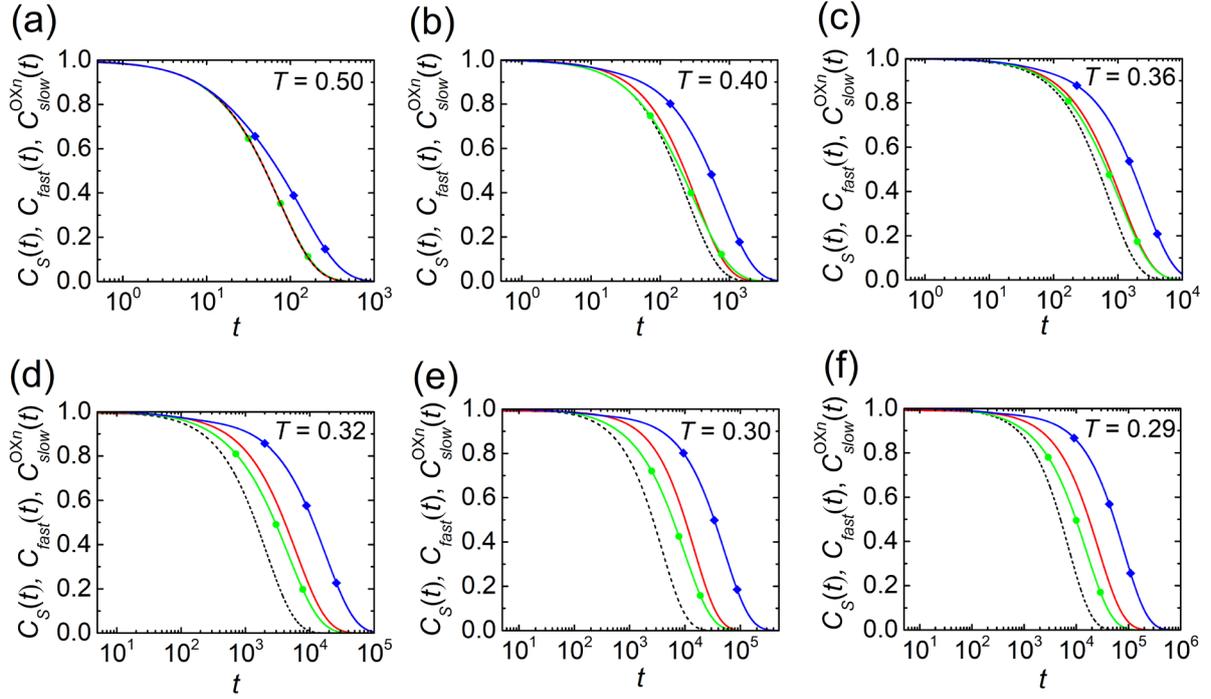

**Fig. S25.** Survival probability for the O cage state, $C_S(t)$, its fast-fluctuation limit, $C_{fast}(t)$, and the slow-fluctuation limits $C_{slow}^{OXn}(t)$ (with X∈{Si,O}) for O-centered substates at (a) $T = 0.50$, (b) 0.40, (c) 0.36, (d) 0.32, (e) 0.30, and (f) 0.29. The red curve, black dashed curve, blue curve with squares, and green curve with circles represent $C_S(t)$, $C_{fast}(t)$, $C_{slow}^{OSi2}(t)$, and $C_{slow}^{OO6}(t)$, respectively.

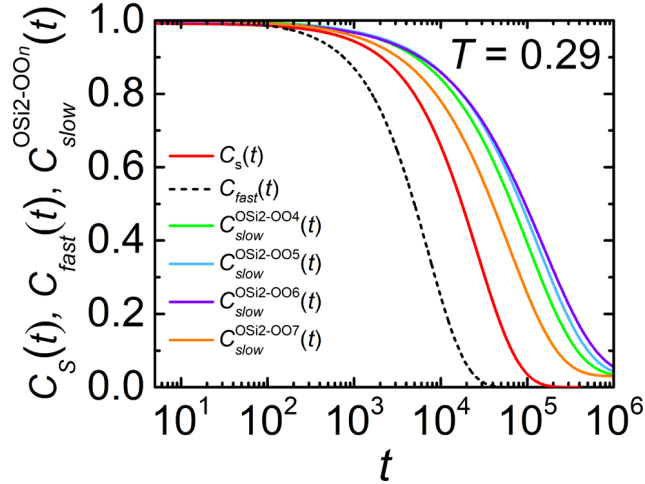

**Fig. S26.** Comparison of the survival probability for the O cage state, $C_S(t)$, with its fast fluctuation limit $C_{fast}(t)$ and slow-fluctuation limits constructed from the joint variables $C_{slow}^{(OSi2-OOn)}(t)$ at $T = 0.29$. The solid red and dashed black curves represent $C_S(t)$ and $C_{fast}(t)$, respectively. The green, light blue, violet, and orange curves represent the results for $n = 4$, 5, 6, and 7, respectively.

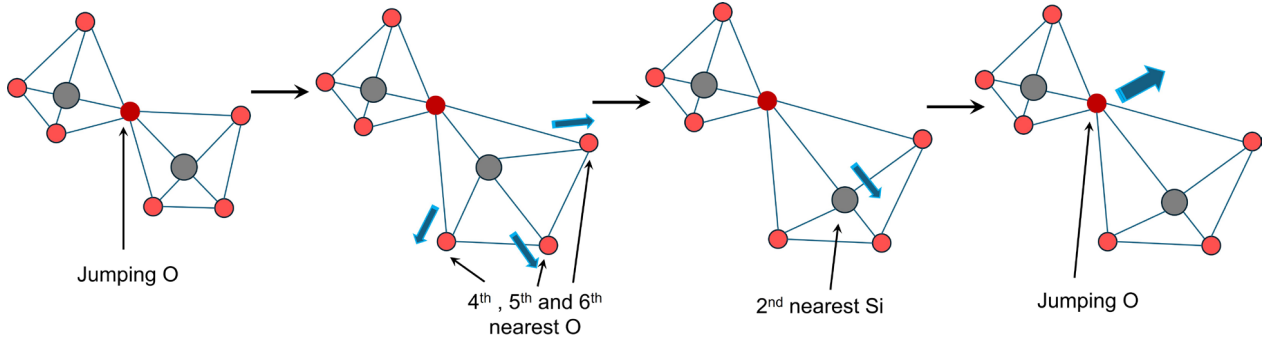

**Fig. S27.** Illustration of an O jump event in silica. The jumping O is shown in brown-red, Si atoms in grey, and other O atoms in red. Blue arrows indicate atomic displacements. The sequence from left to right highlights the cooperative motion of the 4th, 5th, and 6th nearest O neighbors, together with the 2nd nearest Si, that facilitates the jump of the central O.

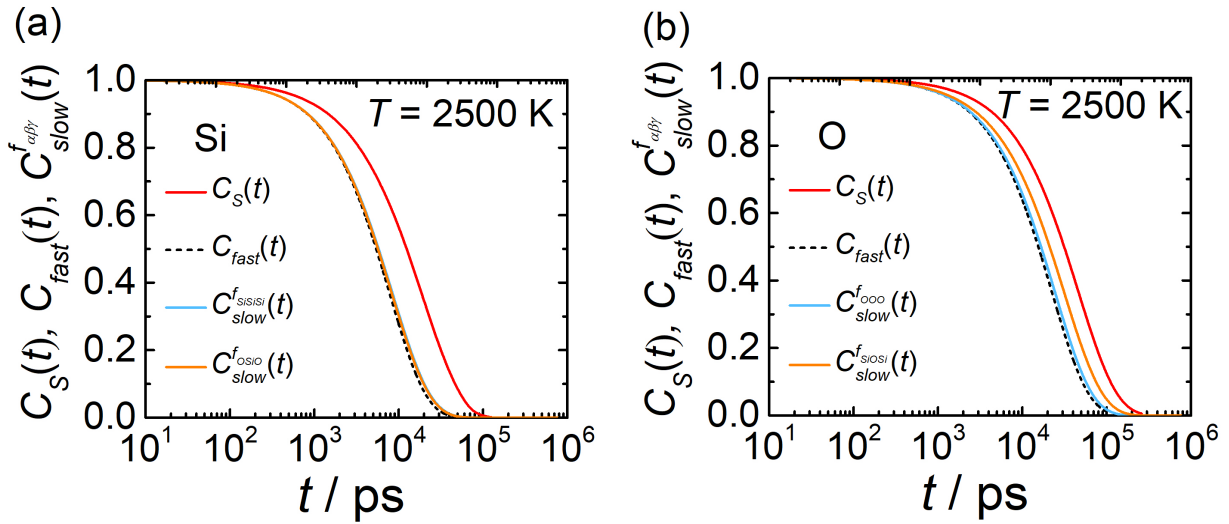

**Fig. S28.** Comparison of the survival probability $C_S(t)$, its fast fluctuation limit $C_{fast}(t)$, and the slow-fluctuation limits constructed from hop-dependent angular distribution functions $C_{slow}^{f_{\alpha\beta\gamma}}(t)$ at $T = 2500$ K for the BKS model. (a) For Si, using $f_{SiSiSi}(\theta)$ and $f_{OSiO}(\theta)$. (b) For O, using $f_{OOO}(\theta)$ and $f_{SiOSi}(\theta)$.

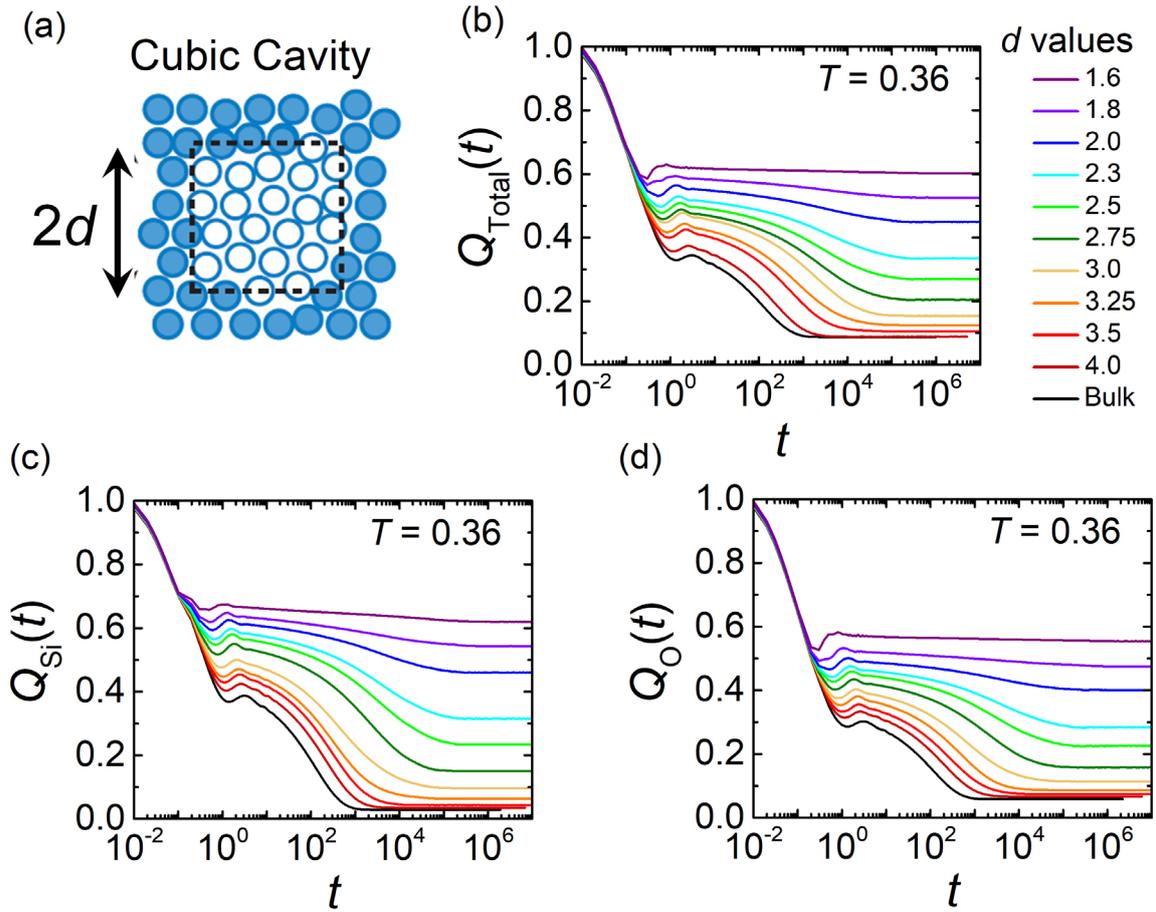

**Fig. 29.** (a) Schematic of the cubic cavity geometry with cavity size $2d$. (b) Time dependence of the total overlap function $Q_{\text{Total}}(t)$ at $T=0.36$ using CP model for different cavity sizes $d$, compared with the bulk (black). (c) Species-resolved overlap function for silicon atoms, $Q_{\text{Si}}(t)$. (d) Species-resolved overlap function for oxygen atoms, $Q_{\text{O}}(t)$. With decreasing cavity size, the long-time plateau of the overlap increases, indicating enhanced confinement effects, with a more pronounced sensitivity for silicon than for oxygen.

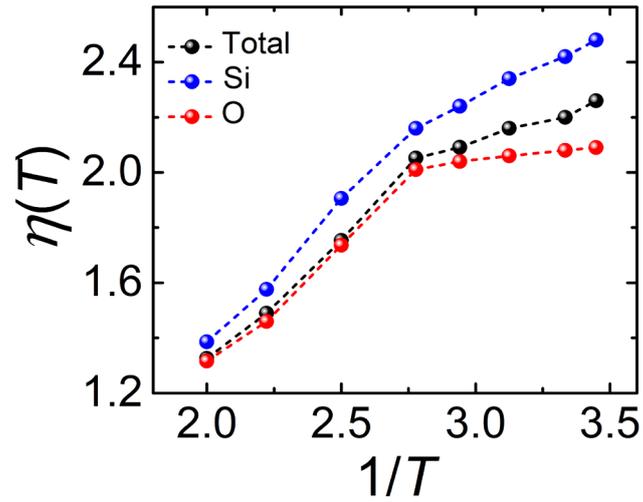

**FIG. S30.** Temperature dependence of the stretching exponent $\eta(T)$ obtained from fits of the excess overlap in the cubic cavity analysis [Eq. (21)] for the total, silicon, and oxygen components. While $\eta$ for silicon and the total system increases steadily upon cooling, oxygen shows little to no growth at lower temperatures, reflecting its comparatively localized dynamics.